\documentclass{aa}
\usepackage{graphicx}
\usepackage{color}
\usepackage{lscape}
\usepackage{txfonts}
\usepackage{natbib}
\def\l{$\lambda$}
\def\mbh{$M_{\rm BH}$\/}
\def\lledd{$L_\mathrm{bol}/L_{\rm Edd}$\/}

\def\rfe{$R_{\rm FeII}$}

\def\feiiq{{\rm Fe{\sc ii}$\lambda$4570}\/}

\def\msol{M$_\odot$\/}

\def\ltsima{$\; \buildrel < \over \sim \;$}
\def\simlt{\lower.5ex\hbox{\ltsima}}            % < over MMM
\def\gtsima{$\; \buildrel > \over \sim \;$}

\def\simgt{\lower.5ex\hbox{\gtsima}}            % > over MMM

\def\ha{{\sc H}$\alpha$}
\def\lya{{ Ly}$\alpha$}
\def\civ{{\sc{Civ}}$\lambda$1549\/}

\def\cm3{cm$^{-3}$\/}
\def\hb{{\sc{H}}$\beta$\/}

\def\hbbc{{\sc{H}}$\beta_{\rm BC}$\/}
\def\hbvbc{{\sc{H}}$\beta_{\rm VBC}$\/}
\def\hbnc{{\sc{H}}$\beta_{\rm NC}$\/}
\def\mgii{{Mg\sc{ii}}$\lambda$2800\/}

\def\oiiiopt{{\sc{[Oiii]}}$\lambda\lambda$4959,5007\/}
\def\o4363{{\sc{[Oiii]}}$\lambda$4363\/}

\def\feiiopt{{Fe\sc{ii}}$_{\rm opt}$\/}
\def\feii{{Fe\sc{ii}}\/}
\def\fe{{\sc{Fe}}\/}

\def\heii{{\sc{Heii}}$\lambda$4686\/}
\def\fe76087{{\sc [Fe vii]}$\lambda$6087\/}
\def\oiii{{\sc [Oiii]}$\lambda$5007}

\def\kms{km~s$^{-1}$}
\def\rk{$R_{\rm K}$\/}
\def\ergss{erg s$^{-1}$\/}

\def\rk{{$R{\rm _K}$}\/}
\def\heii{{{\sc H}e{\sc ii}}$\lambda$4686\/}

\def\ledd{$L_{\rm Edd}$\/}
\def\coneq{$c(\frac{1}{4})$\/}
\def\cthreeq{$c(\frac{3}{4})$\/}
\usepackage[stable]{footmisc}
\begin{document}

\title{VLT/ISAAC Spectra of the H$\beta$ Region in Intermediate-Redshift
Quasars\thanks{Based on observations made with ESO Telescopes at the  Paranal Observatory under programme ID  073.B--0398(A) and
075.B--0171(A).}}
   \subtitle{III. H$\beta$ Broad Line Profile Analysis and Inferences about BLR Structure}
   \author{
         P. Marziani\inst{1}, J. W. Sulentic\inst{2},
           G. M. Stirpe \inst{3},  S. Zamfir\inst{2}, and M.Calvani\inst{1}}

   \offprints{P. Marziani}

\institute{ INAF, Osservatorio Astronomico di Padova,
              Vicolo dell' Osservatorio 5, 35122 Padova, Italy\\
              \email{paola.marziani@oapd.inaf.it; massimo.calvani@oapd.inaf.it}
\and
              Department of Physics and Astronomy, University of
              Alabama, Tuscaloosa, AL 35487, USA\\
              \email{jsulenti@bama.ua.edu,zamfi001@bama.ua.edu}
\and
               INAF, Osservatorio Astronomico di Bologna,
               Via Ranzani 1, 40127   Bologna, Italy\\
               \email{giovanna.stirpe@oabo.inaf.it}
             }
\date{Received ; accepted }
\abstract {} {We present new VLT ISAAC spectra  for 30 quasars,
which we combine with previous data to yield a sample of 53
intermediate redshift ($z \approx$  0.9 -- 3.0) sources. The sample is used to
explore properties of prominent lines in the \hb\ spectral region of
these very luminous quasars.}
{We compare this data with two large low redshift ($z<$0.8)
samples in a search for trends over almost 6dex in source 
luminosity.}
{We find two major trends: (1) a systematic increase of minimum
FWHM \hb\ with luminosity (discussed in a previous
paper). This lower FWHM envelope is best fit assuming that the
narrowest sources radiate near the Eddington limit, show line
emission from a virialized cloud distribution, and obey a well
defined broad line region size vs. luminosity relation. (2) A
systematic decrease of equivalent width \oiiiopt\ (from W$\approx$15
to $\sim$1\AA) with increasing source bolometric luminosity (from $\log L_\mathrm{bol} \approx $ 43 to  $\log L_\mathrm{bol} \approx$ 49). 
Further identified trends require discrimination between so-called
Population A and B sources. We generate median composite spectra in
six luminosity bins to maximize S/N. Pop. A sources show reasonably
symmetric Lorentzian \hb\ profiles at all luminosities while Pop. B
sources require two component fits involving  an unshifted broad 
and a redshifted very broad  component. Very broad \hb\ increases in
strength with increasing $\log L_\mathrm{bol}$\  while the broad component
remains constant resulting in an apparent ``Baldwin effect'' with equivalent width
decreasing from $W\sim$80 to $\sim$20 \AA\ over our sample
luminosity range. The roughly constant equivalent width  shown by the \hb\ very broad component
implies production in optically-thick, photoionized  gas. The onset of the
redshifted very broad component appears to be a critical change that
occurs near the Pop. A-B boundary at FWHM \hb\ $\approx$ 4000 \kms
which we relate to a critical Eddington ratio ($\approx 0.2
\pm$0.1). %Assuming that only BLR \hb\ in pop. sources provides a
%``virial'' measure reduces black hole mass estimates and increases
%Eddington ratio estimates for these sources (including virtually all
%radio-lou%d quasars. 
}{}
%contributes $\approx $40\%\
%of the total \hbbc\ flux at low L increasing to $\sim$75\%\ in the
%high VLT ISSAC Pop. B sources. %After VBLR correction unshifted (assumed
%virial) BLR \hbbc\ in Pop. B sources shows an essentially constant
%FWHM$\approx$ 4000 \kms.
\authorrunning{Marziani et al. }
\titlerunning{\hb\ in Intermediate $z$\ Quasars: III. Profile Analysis}
\maketitle

\section{Introduction}

Much remains to be defined in quasar research even if one restricts
attention to the broad emission lines often used to define them.
Their broad line spectra show considerable diversity complicating
attempts to generate composite spectra and making estimates of
intrinsic properties such as black hole mass unreliable. In
addition, the  relationship between broad line emitting active galactic nuclei
(Type 1 AGN) and  various classes of sources that do not show broad
lines (e.g. Type 2 AGN, LINERs, Blazars, NLRGs) is still uncertain.
It is difficult to imagine how advances in physical understanding
can come until source phenomenology is clarified. We have focused on
clarifying the phenomenology of Type 1 sources because of their
relatively unambiguous broad line signature. Leaving aside
reverberation studies making use of source variability,
observational advances can be expected to come from two areas of
spectroscopic investigation: (1) inter-line comparisons (from \lya\
to \ha) in terms of relative intensity and profile shape
\citep[e.g., ][]{shangetal07}; (2) moderate dispersion and high S/N
studies of the \hb\ spectral region in quasars with $z \geq 1.0$.
There is a long history of infrared spectroscopy of the Balmer lines
in high redshift AGN \citep[e.g.,
][]{kuhretal84,espeyetal89,carswelletal91,hilletal93,bakeretal94,elstonetal94,evansetal98,murayamaetal98,murayamaetal99} but only
recently has it become possible to obtain spectra with resolution
and S/N comparable to those for optical spectra of low $z$ sources.
IR spectroscopy of \hb\ at $z \ga $1.0 enables one to use the same
line (\hb), the same rest frame determination (\oiiiopt\ or narrow component of \hb)
and the same reduction procedures for estimating the central black
hole mass (\mbh). Even so observations of the most numerous and most
luminous quasars at $z \approx$ 2 can only be made when the optical
lines are redshifted into one of the IR bands of high atmospheric
transmission.

Spectrographs on HST  have provided UV coverage of higher ionization
broad lines in low-$z$\ Type 1 sources (Seyfert 1 and quasars) with
$\sim$150 sources  having useful measures of \civ\
\citep[e.g.,][]{brothertonetal94,marzianietal96,baskinlaor05,sulenticetal07}.
Almost all of these sources now have matching optical spectroscopic
measures of the \hb\ region \citep[e.g.,
][]{marzianietal03a,baskinlaor05,shangetal07}. Continuing with a
litany of difficulties we note that even some of these UV
observations show marginal S/N while  many do not cover the full
range from Ly$\alpha$ to 3500 \AA. Optical coverage is usually not synoptical. 
This makes detailed
intercomparisons of the strongest high and low ionization broad
lines possible for only a few tens of objects at best. The lack of
spectra covering the rest frame from Ly$\alpha$\ to H$\alpha$\ is
unfortunate since line intensity and profile ratios provide a wealth
of physical constraints. The ability to compare lines from ions of
widely different ionization potentials has helped to elucidate two main
emitting regions within the broad line region (BLR): (a) one
responsible for the production of low-ionization lines (LIL) like
\feii, \hb\ as well as \mgii\ and (b) one that emits mainly lines
(\civ, \heii) from ions of high ionization potential \citep[HIL; e.g.,
][]{collinetal88}. Meaningful line profile studies begin with
observations of  \civ and \hb\ as the most typical HIL and LIL.

Our attempt at clarifying type 1 AGN phenomenology involves the
Eigenvector 1 formalism of \citet{borosongreen92} and was later
expanded into 4 dimensions (4DE1). The latter focused on four
parameters including  measures of  \hb\ and \feii\
\citep{sulenticetal00a,sulenticetal00b,sulenticetal07}. Changes
in the width and relative strength of  \hb\ and \feii\ lines
appear to be primarily related to Eddington ratio convolved with
source orientation \citep{shangetal03,marzianietal03b,yipetal04,boroson05,collinetal06}.
4DE1 studies also introduced the concept of two quasar populations A
and B that maximize phenomenological differences and possibly
identify sources with higher and lower Eddington ratios
respectively. The width of broad \hb\ (FWHM or 2$^\mathrm{nd}$ profile
moment $\sigma$) have been widely used as measures of velocity
dispersion in the line emitting gas \citep[see
e.g.][]{petersonetal04,sulenticetal06,vestergaardpeterson06},
allowing what are thought to be the most reliable estimates of black
hole mass in low $z$\ quasars. This approach has motivated us to seek
similar measures in sources with the highest possible redshift
rather than using other lines as \hb\ surrogates.

In this paper we leave aside inter-line comparisons and focus on
the \hb\ spectral range in low and in intermediate-to-high $z$\
quasars ($z \ga$ 1). Near IR spectra of the latter were obtained
with the infrared spectrometer ISAAC on ESO VLT Unit 1. The high S/N and resolution
of the new observations allow  a meaningful comparison with
measures obtained for low $z$\ sources. Low $z$\ measures come from two
samples: (1) an  atlas of bright sources with
z$\la$ 0.8 \citep[][hereafter, ATLAS sample]{marzianietal03a}  and (2)
a magnitude-limited sample of 321 SDSS quasars \citep[{\it m}$_\mathrm{g} <$17.0, $z
<$0.7 from ][hereafter SDSS sample]{zamfiretal08}.

%Most importantly the new
%data also allow us to search for trends between emission line
%properties and source luminosity $L_\mathrm{bol}$\ and/or redshift $z$.

In two previous papers  we focused on a search for luminosity
effects involving \hb, \feiiopt, and \oiiiopt\ \citep[][hereafter
Paper I]{sulenticetal04} as well as on the use of \hb\ as a virial
black hole mass estimator \citep[][hereafter Paper
II]{sulenticetal06}. This paper presents observations and reductions
for 30 additional sources ($z= 1.08- 3.09$; \S \ref{obs}).  We
briefly discuss  the  \hb, \feii\ and narrow line measures (\S
\ref{imm}) and then show that the addition of 27 new sources to the
VLT sample reenforces the luminosity trends described in Paper I (\S
\ref{further}). We then consider the distribution of VLT sources in
the optical plane of 4DE1. Median 4DE1 and luminosity binned
composite spectra are discussed in the context of the 4DE1
Population A-B concept \cite[][]{sulenticetal00a,sulenticetal08}. 
In \S \ref{disc} we use the binned composite
spectra to decompose \hb\ and make estimates of black hole mass
\mbh\ and Eddington ratio \lledd\ for Pop. A and B sources.

%469 for g<17.5

\section{Observations and Data Reduction}\label{obs}

Spectra for intermediate-high redshift sources were obtained
between 07/2004 and 07/2006 in service mode with the infrared
spectrometer ISAAC mounted on VLT1 (ANTU) at the European Southern
Observatory.  Table \ref{tab:obs} summarizes the new
observations. The basic format is given below the Table, following
almost exactly Tab.~1 of Paper~II.  In addition to the format
provided below  Table \ref{tab:obs} we note that Col. (3) lists
the blue apparent magnitude taken from Hamburg-ESO survey papers
\citep{reimersetal96,wisotzkietal00} while  Col. (4) lists the
source redshift $z$\ computed as described in section \ref{red}.
Col. (8) lists the radio-loudness parameter \rk\ defined as the
ratio between the specific flux at 6 cm and 4400 \AA\ in the rest
frame. We applied a $k-$correction for both the radio and optical
data. In the case of the radio data: $f_\mathrm{e,6cm} =
f_{\nu,\mathrm{obs}} \cdot \left[ \nu_\mathrm{obs}(1+z)/
\nu_\mathrm{6 cm} \right] ^\alpha$ where the radio spectral-index
is $\alpha =0.5$. Only 14 of the sources have radio detections;
1.4GHz upper limits were derived for undetected sources from the
NVSS detection threshold  or the Sydney University Molonglo Sky
Survey \citep[SUMSS;][]{bocketal99}. Two sources are formally
radio-loud following our definitions \citep{zamfiretal08}.

Each spectrum corresponds to a wavelength range (IR windows sZ, J,
sH, sK; Col. (10) of Tab. \ref{tab:obs}) that covers all or part
the region involving \hb\, \feiiq\ and/or Fe{\sc
ii}$\lambda$5130. Reduction of quasar spectra and standard stars
followed exactly the same procedures described in Papers I and II.
Wavelength calibration yielded rms residuals of 0.4, 0.6 and
0.9~\AA\ in the sZ, J and sH windows, respectively. Absolute flux
measures will be inaccurate because atmospheric seeing almost
always exceeded the slit width ($\approx$ 0''.6) resulting in
significant light loss. Small offsets were present in the
wavelength calibration because the arc lamp frames were obtained
in daytime and therefore usually after grism movement. A
correction for these shifts was obtained by measuring the
centroids of 2--3 OH sky lines against the arc calibration and
calculating the average difference.

\subsection{Sample Considerations}

All 53 sources  in the VLT-ISAAC sample were selected from the
Hamburg-ESO (HE) quasar survey which was a color-selected and
magnitude-limited ($m_{\mathrm B} \approx$ 17.5) quasar survey
\citep{wisotzkietal00}. Thirty-one new spectra are presented for
30 sources in \S \ref{fig:spectra} with 27 sources not previously
observed. They are merged with Paper I and II measures to
yield a total sample of 53 sources. Two of the new observations
involve previously observed sources that were discussed in Paper
I. The independent pairs of measures provide a valuable
consistency check and are treated as independent data points in
our statistical analysis of \S \ref{further}. Two new spectra
obtained for HE 1505+0202 are listed in Table \ref{tab:obs} with
average values used in the analysis. HE 0353-3919 (Paper I) has
been excluded from quantitative analysis because its \hb\
profile is compromised by a gap in coverage between the sZ and Z
bands (see Fig. 2 of Paper I). The resultant sample of 52 sources
(and 54 data points) is henceforth referred to as the ISAAC
sample.

The samples of  \citet[][ATLAS]{marzianietal03a}  and
\citet[][SDSS]{zamfiretal08}  are used  as low-$z$\ spectroscopic
comparison samples. SDSS is more complete, involving the 321
brightest $g$-band selected ($\leq$17.0) in SDSS DR5 with $z
\leq$0.7 while the ATLAS sample involves a more heterogeneous
selection of 215 quasars largely brighter than $m_\mathrm{B}$ =
16.5. We make wide use of the SDSS sample in this study because of
its higher level of completeness, larger number of sources, and
because results from the ATLAS sample were already presented in
Paper I and II. Both the SDSS and ATLAS data enable: (1) accurate
measurements of \hb\ and \feiiopt\ emission, (2) decomposition of
the broad \hb\ profile, and (3) measures of the narrow \hb\
component (\hbnc) and \oiiiopt\ lines. 
The SDSS sample  suffers a
strong Malmquist bias. The ATLAS sample shows a more uniform
luminosity distribution making a correlation analysis less biased.
It will be used in the analysis of luminosity trends involving FWHM
\hb\ (\S \ref{disc}).

\subsection{Data Analysis \label{red}}

Analysis of ISAAC spectra made use of  {\tt IRAF} tasks to
accomplish continuum and \feiiopt\ modeling as well as
subtraction. The {\tt specfit} task was employed to make a
reliable model of the \hb\ spectral region. The
simultaneous fitting of continuum and \feii\   should
 be less dependent on subjective evaluations by the
observer. We assume that the continuum underlying the \hb\
spectral region is a power law of variable slope. Continuum
subtraction has severe limitations due to the small bandwidth of
our spectra  and to (unknown) internal reddening effects. However,
{\tt specfit} usually found a plausible  continuum (see Fig.
\ref{fig:spectra}). Estimation of continuum subtraction
uncertainty due to S/N involves choosing continuum fluxes at about
--3$ \sigma $ (minimum) and +3$ \sigma $ (maximum) levels where
$\sigma$ is the standard deviation of the most likely continuum
choice. Uncertainties of continuum placement were empirically
defined from the difference between extreme high/low continua and
the most probable one derived by {\tt specfit}.

\feiiopt\ emission was modeled using a scaled and broadened template
as previously employed by \citet{marzianietal03a}. An important
change was introduced  to the \feii\ template emission underlying
\hb\ as described below (\S \ref{template}). We stress again that a
notable advantage of {\tt specfit} is that the scaled and broadened
template can be simultaneously fit over $\approx 1000$~\AA\ with a
power-law continuum, the \hb\ profile and the narrow lines (\hbnc,
\oiiiopt, \ion{Fe}{vii}\l5160, \ion{Fe}{vii}\l5177). The strong
\feii\ blue blend (4450--4680 \AA: \feiiq) was measured as an estimator of the \feiiopt\ strength. The Gaussian broadening factor from the best fitting template yields an
estimate of FWHM \feii. A careful estimate of minimum and maximum
plausible broadening factors was made to derive  $\pm 3\sigma$\
uncertainties.  The blue side of the spectrum including
\feiiq\ is missing or only marginally covered in several sources. In
these cases the best template fit was achieved for the red blend
(Fe{\sc ii}$\lambda$5130 in the range 5200--5600 \AA) and \feiiq\
was estimated assuming a fixed ratio between the red and blue
blends.  Fig. \ref{fig:hbeta} shows the estimated \feii\ emission
(green lines).

The \hb\ profile  was modelled with: (1a) a core Lorentzian component plus a
weaker/broader Gaussian  on the blue (usually) or on the red side of
the core or, alternatively, (1b) the sum of two Gaussians one always very broad
and showing non-negligible velocity shift relative to a narrower
core component; (2) a narrow component (\hbnc).  We do not attribute any physical meaning to the ``blind" decompositions performed with {\tt specfit} in the case of
the double Gaussian fits for individual sources. The two types of
fit provide the simplest {\em empirical} description of broad \hb\
profiles in the majority of sources. The \hb\ profile was fit with a
high order spline function ({\tt IRAF} task {\tt sfit}) in a few
cases where the model fits showed significant deviations from the
actual profile. Spline functions do not yield a model fit but only
an empirical fit that smooths noise and reproduces the main features
and inflections in \hb. The spline fit approach was the one used in
our previous work \citep{marzianietal03a} where sources with FWHM
\hb\ $\leq$ and $\geq$ 4000 \kms\ favored Lorentzian and double
Gaussian fits respectively. { Systematic differences in flux and
$W$\ measurements as well as in line profile parameters between the
empirical and {\tt specfit} approach  occur since the extended wings
of Gaussian and especially Lorentzian profiles lead  to a lower
continuum placement. The effect however is within the estimated
measurement uncertainty. }

The \oiiiopt\ lines were also simultaneously modelled with the
following conditions strictly enforced during the fitting procedure:
(1) the flux ratio between \oiii\ and {\sc [Oiii]}$\lambda$4959
$\approx$3, (2) both lines should show identical profiles. It was
often necessary to include a ``semi-broad" and/or blueshifted
component to model the \oiiiopt\ profile since evidence for a strong
blue asymmetry was observed  in several cases. This approach yields
a satisfactory reproduction of the observed \oiiiopt\ profiles
within our resolution  and S/N limits. It is again an empirical
method to properly estimate the total \oiiiopt\ strength and
describe the line profile shape. We do not make any assumptions
about the actual nature of \oiiiopt\ emission which will be the
subject of a separate paper.

\subsection{The \feiiopt\ template}
\label{template}

Considerable attention has been given to theoretical and empirical
estimation of \feiiopt\ emission in the spectral region of \hb,
for a number of scientific (i.e., obtain a diagnostic of the BLR)
and technical goals. Theoretical calculations of \feii\ emission,
assuming the predominance of photoionization in the Broad Line
Region have greatly improved in the last decade
\citep{sigutpradhan03,sigutetal04,verneretal99} with a 371-level model of the
Fe$^+$\ ion presently included in the photionization code {\tt
CLOUDY} v. 07.01 \citep{ferlandetal98}. A detailed analysis of the
\feii\ emission in I Zw 1 \citep{veroncettyetal04} revealed rich
and complex emission from several line systems, each associated
with a well-defined redshift and FWHM. In the case of sources with
larger FWHM and data with intermediate resolution as well as
average S/N an analysis as detailed as that made for I Zw 1
\citep{veroncettyetal04} cannot be done. The \feiiopt\ template
approach works satisfactorily for the heterogeneous variety of
type-1 AGN \citep{borosongreen92,marzianietal03a,zamfiretal08} with the probable
exception of some outliers and extremely strong \feii\ emitters.

%The contribution of the host galaxy and the AGN
%thermal disk emission (modelled locally by a power-law)  reach
%parity near 4800--5000 \AA\ with slight changes in the relative
%strengths of these continuum components likely responsible for the
%mismatch of template fits to the \feiiopt\ blends on the red and
%blue sides of \hb.

There is however  a major problem associated with the empirical
template: it is difficult to estimate the \feii\ emission  under
the \hb\ line.  Our empirical template differs from that of \citet{borosongreen92}
because of lower \feii\ emission under \hb\ and from
\citet{veroncettyetal04} for the opposite reason. The new analysis
of the I Zw 1 spectrum by \citet{veroncettyetal04}, coupled with
photoionization models, suggests that emission under \hb\ should
not be very strong. Therefore,  the strength of the \feii\ lines
underlying \hb\ has been computed with an appropriate
photoionization model. In the context of photoionization the
present understanding of the \feii\ emitting region suggests low
ionization and high electron density
\citep{verneretal99,bruhweilerverner08}. We assumed an ionization
parameter $\log \Gamma = -2.25$, and electron density [cm$^{-3}$]
$\log n_\mathrm{e} = 12.75$. This model provides an intensity
ratio between \feii\ m29 and \hb\ consistent with that measured
in I Zw1. These improvements in the template do not significantly
affect our \rfe = F(\feiiq)/F(\hb) $\approx$ W(\feiiq)/W(\hb)
measurements. Even in the strongest \feii\ emitters (\rfe $\ga 1$)
the change in \hb\ intensity is much lower than the associated
uncertainty. The effect on  \hb\ profile is to lower the \feiiopt\
subtraction on the red side of \hb\ making it stronger. The change
with respect to the previous template  is negligible in the case of
weak emission but may not be so if \feii\ is strong and relatively
narrow (see \S \ref{e1}).  A blueshifted, broad component ascribed
to a wind \citep{sulenticetal02,marzianietal08} is affected
qualitatively by the new \feiiopt\ subtraction, although there are
several  examples of individual
ISAAC sources (HE0248$-$3628, HE0512$-$3329 in Paper II) that
unambiguously confirm the existence of such a component which is obvious in
Ly$\alpha$ and especially  \civ\ for several low-$z$\ sources \citep{marzianietal08}.

\section{Reduced Spectra and Line Measures}
\label{imm}

\subsection{Redshift Determination and Rest Frame Corrections}

Rest frame determination is important in quasar spectroscopy
because both broad and narrow emission lines are known to show
velocity displacements relative to one another. We followed our
standard procedure \citep{marzianietal03a} and adopted the \hb\
peak redshift ($\lambda = 4861.33$~\AA) as rest frame for our new
sources. The \oiiiopt\ lines ($\lambda \approx $ 4958.9, 5006.85 \AA) were only
used when they gave the same value as \hb\ (see Col. (6) of
Tab.~\ref{tab:obs}). They sometimes show a blueshift relative to
the \hb\ peak  and are designated as ``blue outliers" in extreme
cases \citep{zamanovetal02,marzianietal03b,aokietal05,boroson05,komossaetal08}.  Blue
outliers usually show FWHM \hb $<$\ 2000 \kms\ but caution should be used
when  adopting \oiii\ or \hb\ for rest frame determination
\citep[see also][ who suggest a systematic \feiiopt\ {\em redshift} with respect to the quasar rest frame]{huetal08a}. The adopted
$z$\ estimates were used to de-redshift the spectra while the {\tt
IRAF dopcor} task applied a $(1+z)^3$\ correction to convert
observed specific fluxes into rest frame values.
Fig.~\ref{fig:spectra} shows the flux and wavelength calibrated
spectra.

\subsection{Optical \feii\  Lines}

Fig.~\ref{fig:hbeta} shows the continuum subtracted spectra (left
panels) with the adopted \feiiopt\ template model superimposed.
Tab.~\ref{tab:broad} provides flux and equivalent width of \hb,
as well as flux and FWHM of \feiiq. The data format is explained
in the footnotes of each table. Care should be applied in using
the FWHM values for \feiiq. These values were obtained through a
$\chi^2$\ minimization in the {\tt specfit} procedure and
therefore should at least be free of any subjective bias.
Uncertainties were estimated by superimposing a sequence of
templates with different FWHM values on the optimal \feiiq\
spectrum and estimating the $\pm 3\sigma$ confidence levels where
FWHM \feii\ became unacceptably large or small. We find that
FWHM(\feiiq) is systematically smaller than FWHM(\hb) and that
the median value is $\approx$ 4000 \kms. Uncertainties of
individual \feii\ measurements are quite large allowing only
statistical inferences to be made.

\subsection{Narrow Lines}

Table \ref{tab:narrow} reports the  flux and equivalent widths
values for  \hbnc\ and \oiii. \hbnc\ is weak/marginal in most
sources and undetected in 4 sources.
%This is not surprising given
%the weakness of \oiiiopt\ emission in most of these AGN. \hb\
%in sources with  weaker \oiiiopt\ emission also show Lorentzian
%profiles which reduces the inflection between broad and narrow
%emission components. A large number of lower luminosity sources in
%the low $z$\ comparison samples show Gaussian profiles with a
%clear inflection between broad and narrow emission components.
The
total flux values for \oiii\ often include narrow and semi-broad
components which are required to adequately fit the \oiiiopt\ line
profiles using {\tt specfit}. Differences between the \hbnc\ and
\oiiiopt\ profiles indicate that crude intensity ratios \oiii/\hbnc\
may not be meaningful. The ratio \oiii/\hbnc\ has a clear meaning
only if there is a redshift and FWHM consistency between \oiiiopt\
and \hb.

\subsection{\hb\ Line Profiles}

%\citep[]{marzianietal96,marzianietal03a,sulenticetal04}

Measurements of the broad \hb\ profile including FWHM and other
important line parameters such as asymmetry index, kurtosis and line
centroid at various fractional intensities were obtained using a
{\tt FORTRAN} program developed for the purpose. These parameters
are the same as defined in Paper I and are given in Tables
\ref{tab:profs} and \ref{tab:centroids}. The right-hand panels of
Fig.~\ref{fig:hbeta} emphasize the shape of \hb\ for each source.
Table \ref{tab:centroids} lists measurements of the \hbbc\ centroid
at various fractional intensities (in \kms). Each line measurement
in Tables \ref{tab:profs} and \ref{tab:centroids} is followed in the
next column by its appropriate uncertainty at the $2 \sigma$\
confidence level. The uncertainties on line profile parameters were
estimated by changing the fractional height by $\pm 0.05$. All of
these measures are affected by the compositeness of the broad \hb\
profile which frequently shows unambiguous evidence for at least two
distinct components: a broad component  proper (\hbbc) and very
broad component (\hbvbc\ presumably from a  very broad line region, VBLR). In
the following, we will continue to refer to \hb\ as the total broad
emission excluding \hbnc, and to \hbbc\ as to the \hb\ emission
after \hbvbc\ (and of course also \hbnc) removal.

%Cols. (2) and (4) of Tab. \ref{tab:profs}
%give Full Width at Zero Intensity (FWZI) and FWHM. Col. (6) gives
%the asymmetry index (AI) as defined in \citet{sulenticetal04}.
%Col. (8) lists kurtosis values.

\section{Results}
\label{further}
\subsection{Luminosity Effects}
\label{lumeff}

\paragraph{\hb\ --} Fig.~\ref{fig:lumeff} shows the distribution
of \hb\ profile measures as a function of bolometric luminosity
where $L_\mathrm{bol} \approx 10 \cdot \lambda L_{\lambda}$
($\lambda$ = 5100 \AA) for the SDSS data. Bolometric luminosity was
computed from $z$ and $m_\mathrm{B}$\ for the ISAAC sample because
of the spectra uncertain flux scale. Figure \ref{fig:lumeff}a (upper left
panel) shows the distribution of FWHM(\hb) measures as a function of
log $L_\mathrm{bol}$\ where low $z$\ measures  come from the SDSS
sample \citep{zamfiretal08} and ISAAC sources are indicated by
larger filled circles. The only obvious, well defined FWHM trend
found over $\sim$5 decades of source luminosity involves an increase
in the minimum FWHM \hb\ with luminosity as previously reported in
Paper I. Fig.\ref{fig:lumeff}a shows a systematic increase with
$\log L_\mathrm{bol}$ of mean/median FWHM \hb\ from mean/median
3200/2550 \kms\ to 5075/4900 \kms\ which is likely driven by the
minimum FWHM trend. No trend is observed, for example, among sources
with FWHM $>$ 4000 \kms. The excess of sources with FWHM $>$
8000 \kms\ in the range $\log  L_\mathrm{bol}$ = 45.5 -- 46.5 simply
reflects Malmquist bias amplified by the increase of sources above
$\log L_\mathrm{bol}$ $\approx$  45.5 due to the onset of strong
source evolution in the range $z =$ 0.5 -- 0.7. The correlation of
FWHM \hb\ with source luminosity is weak (c.f. Paper I) but
statistically significant for the increased samples of this paper,
with Pearsson's correlation coefficient $r \approx 0.26$ ($P \la
10^{-6}$).

Figure \ref{fig:lumeff}b (upper right) shows the distribution of
asymmetry index (A.I.) as a function of luminosity. An excess of red
asymmetric profiles is seen at all luminosities but does not show up
well in the mean or median values which lie in the range 0.03 --
0.06 because of the strong SDSS source concentration between A.I. =
$\pm$ 0.05. The ISAAC sample shows the highest (reddest) mean and
median values with A.I. = 0.11 and 0.09, respectively. If we focus
on the sources with more extreme A.I. values then we find the
largest difference between the ``strips'' $-0.4 < $A.I.$ < - 0.2$\ and $0.2 <$ A.I.$ < 0.4$:  red
asymmetries outnumber blue ones by a factor of seven. Only 17 SDSS
sources (zero ISAAC) show blue asymmetries greater than --0.2 while
$\approx$85 show red asymmetries greater than +0.2 (17 ISAAC). No
luminosity correlation is detected with the red asymmetric excess
visible at all luminosities.

Figures \ref{fig:lumeff}c (lower left) and \ref{fig:lumeff}d (lower right) show the
distributions of  \hbbc\ ``peak'' \cthreeq\ and ``base'' \coneq\
profile velocity displacements (line shifts), respectively. \coneq\
shows a strong source concentration around zero (unshifted profile
base) but a significant excess of redshifts at all luminosities.
There is evidence for a trend with mean/median values increasing
from 56/57 \kms\ in the lowest luminosity decade to 671/292 \kms\
for the ISAAC sample. Thirteen sources (7 ISAAC) show base
redshifts between 2000--4000 \kms\ while no sources show blueshifts
in that range. Clearly the red excess in Figures \ref{fig:lumeff}b
and \ref{fig:lumeff}d are related. If we distinguish between radio quiet (RQ)
and radio loud (RL) sources (Zamfir et al., {\em in preparation})
we find the striking absence of a RL source
concentration near zero. Radio loud sources span the full range
of observed red and blueshifts however with a preference for
redshifts. The concentration around zero is a pure RQ effect.

%The
%fraction of RL sources increases in each luminosity bin of Figures
%\ref{fig:lumeff} with no RL observed below log $L_\mathrm{bol}
%\approx$ 44.0.

The centroid \cthreeq\  shows only a small number of sources with
shifts above $\pm$1000 \kms\ (4 red and 7 blue and 0 ISAAC). The
approximately 2 for 1 preference for blueshifts is confirmed in
the range \cthreeq\ = 600 --1000 \kms\ with 8 red and 15 blueshifted
sources. Mean values are slightly blue for lower luminosities (--20
to --40 \kms\ for $\log L_\mathrm{bol} \la {46}$) and become slightly red in the highest luminosity
SDSS and ISAAC bins (+35 \kms). The simplest accretion disk models
predict redshifted profile bases and blueshifted peaks
\citep{chenetal89,eracleoushalpern03}
leaving open the possibility that an
underlying disk signature is present in all sources. Later
discussion however does not support that conclusion for redshifted
and red asymmetric profiles. The absence of strong luminosity
correlations in Figure \ref{fig:lumeff} is confirmed by Pearson
and Spearman correlation coefficients.

%\citep{sulenticetal90}

\paragraph{\oiiiopt --} Perhaps the most striking difference between the
SDSS and ISAAC samples involves the relative weakness of narrow line
emission in $\approx$ 80\%\  of the latter. Table \ref{tab:narrow}
shows only three ISAAC sources with $W$(\oiii) $\ga$ 13 \AA. We
previously found \citep{sulenticetal00b} mean values for low $z$\
sources in the range $W$(\oiii)= 23 -- 27 \AA\ with standard
deviations for various subsamples in the range 10 -- 30 \AA. Only
one source in the ISAAC sample shows $W$(\oiii) greater than the
mean of the low $z$\ sample: HE 0109-3518 with $W$(\oiii) $\approx$
38.5 \AA, an extremely large value especially considering the high
 luminosity of this $z \approx$ 2.4 quasar. This is the
ISAAC source (Figure \ref{fig:spectra}) that most closely resembles
a  low $z$\ quasar. Twelve blue outliers identified by
\citet{marzianietal03b} show $W$(\oiii) $\approx$ 7.5\AA\ which is
the same as the mean value for the sources listed in Table
\ref{tab:narrow}. There is a high fraction of sources in the ISAAC
sample that may show a systematic blueshift with respect to \hbnc,
making several ISAAC sources similar to the blue outliers in terms
of the \oiiiopt\ properties. The issue will be however discussed in
a separate paper.

%Also, in
%some of the NLSy1 sources with narrowest broad line FWHM the
%\oiii\ line shows a significant blueshift (blue outliers)
%\citep{zamanovetal02,marzianietal03b,aokietal05,boroson05}.
%Despite this rejection \oiii\
%occasionally intrudes into 4DE1 formalism where
%in the high and low redshift samples \oiii\  shows a blue wing or asymmetry.
%A
%high s/n optical spectrum \citep{arp84} shows a blue asymmetric
%\civ\ profile typical of a pop. A quasar.

\subsection{4DE1 Parameters}
\label{e1} A 4DE1 spectroscopic parameter space previously defined
\citep{sulenticetal00b,sulenticetal07} involves measures of
FWHM(\hb), \rfe\  \civ\ shift, and soft-X ray spectral slope
\citep[see also][]{borosongreen92}. We focus here on the optical
plane of 4DE1 where, as far as we can tell, none of the four
principal parameters show a correlation with optical or UV
luminosity out to $z \approx$ 0.8. Fig. \ref{fig:4de1} shows the
SDSS and ISAAC source distributions in the 4DE1 optical plane (FWHM
\hb\ vs. \rfe). Low-$z$\ quasars are distributed along a curved
sequence with sources showing {\em both} large FWHM \hb\ and \rfe\
apparently forbidden. The distribution of SDSS sources matches
closely the one found in the ATLAS sample. The zone of occupation
extends from the upper left where FWHM(\hb) is large, \feiiopt\ is
least prominent and $\approx$50\%\ of the sources are RL. At the
opposite (lower right) end we find sources with the narrowest FWHM
\hb\ profiles, strong \feiiopt\ and $\sim$1\%\ radio loud sources.
The centroid of the SDSS source distribution lies near FWHM \hb =
3800 -- 4000 \kms\ and \rfe = 0.3 -- 0.4.

We  use the 4DE1 optical plane to see if the higher luminosity ISAAC
sample shows any difference in source distribution that might hint
at a luminosity trend. ISAAC sources are  shown as large dots with
error bars in Fig. \ref{fig:4de1}. The largest distribution
differences involve: (a) the absence of ISAAC sources below FWHM
$\approx$ 3000 \kms\ and (b) an apparent displacement of  almost
half of the ISAAC sample towards larger values of FWHM \hbbc\ and
\rfe: on average by $\approx$1 -- 2000 \kms\ in FWHM \hbbc\ and by
0.1--0.2 in \rfe. Difference (a) involves the previously discussed
increase in minimum FWHM \hbbc\ with luminosity and can be described
as a  zone of avoidance for high luminosity sources in the 4DE1
optical plane. Difference (b) requires more detailed study in the
next section.

\subsection{4DE1 Spectral Types and Composite Profiles}

Further analysis of 4DE1 and source luminosity trends are best
accomplished using composite spectra. This avoids confusion
introduced by details in individual spectra and optimizes 
S/N. Composite spectra in the context of the optical 4DE1 plane have
special value because \feiiopt\ prominence is one of the key
parameters and FWHM \hb\ is arguably the most reliable virial
estimator of black hole mass. Fig.~\ref{fig:4de1} is subdivided into
bins of $\Delta$FWHM(\hb) $\approx 4000$~\kms\ and $\Delta$ \rfe
$\approx 0.5$, following \citet{sulenticetal02}.

Spectral types (bins) B1$^+$, B4 and A3 each contain a  single ISAAC
source. A reexamination of the single B4 source HE1505+0212 suggests
that \feiiq\ emission is unusually strong with no reasonable
continuum adjustment capable of moving it further to the left than
bin B3. This source is either a genuine outlier or an extreme Pop. A
source (following the luminosity-dependent boundary between Pop. A
and B discussed in \S \ref{refinement}). Many ISAAC sources are
located in bin B2 while  SDSS sources are apparently rarer there and
much more concentrated toward the lower left corner of the bin. \rfe\ is relatively insensitive to small
changes in the adopted continuum leading us to conclude that the
displacement of ISAAC sources towards higher \rfe\ cannot only be
attributed to an effect of the data reduction. This supports the
reality of distribution difference (b) discussed above.

\subsubsection{Decomposition and Fitting of  ISAAC Spectra in the Optical Plane of 4DE1}

Median composite ISAAC spectra were computed for sources
in 4DE1 bins A1, A2, B1, B2 (Fig. \ref{fig:4bin}) using calibrated and de-redshifted
spectra normalized to specific flux of unity at 5100 \AA. The
first two composites show broader FWHM(\hb) and the second two
show stronger \rfe\ compared to ATLAS \citep{sulenticetal02} or
SDSS composites for the same bins.

Pop. A sources can usually be modelled with a single unshifted
Lorentzian component. This is also true for bin A1 and A2 ISAAC
sources. Pop. B sources usually require a double Gaussian
\citep[unshifted broad \hbbc\ and redshifted very broad \hbvbc\
components:][]{sulenticetal02}. \hb\ profiles in bin B1 and B2
sources are best fit with a double Gaussian model implying that both
unshifted \hbbc\ and redshifted \hbvbc\  components seen in low $z$\
quasars are also present in high luminosity sources (Fig.
\ref{fig:4binan}). The decomposition is not always unique and it can
therefore be difficult to estimate the relative contributions of the
two components. An important clue about the proper decomposition
comes from profile variations in the source PG1416--129 \citep[][see
also 3C 206 in \citet{corbinsmith00}]{sulenticetal00c} where the \hbbc\ almost disappeared following a continuum decline while the
\hbvbc\ showed less variation and was revealed much more
clearly as very broad and redshifted. The highest S/N inflected
spectra in the ISAAC sample (e.g. HE0109--3518) also clearly favor
this decomposition which is adopted throughout our analysis.

\subsection{Luminosity Binned Spectra}

Comparing mean or median spectra for random samples of quasars
observed in different luminosity ranges is unlikely to yield
meaningful results. Any difference would be obscured by the
intrinsic dispersion of spectral properties at any given apparent
luminosity. As stressed in much of our previous work \citep[see
especially the review by][]{sulenticetal00a}, and illustrated again
in Fig. \ref{fig:4bin}, all quasar spectra are not self-similar.
Figure \ref{fig:lumeff}a reenforces this point in one dimension by
showing that FWHM \hb\ spans a very large range from $\approx$ 500
\kms\ to $\approx $ 20000 \kms. We suggest that a {\it minimum
requirement} for physically meaningful comparisons among low-$z$\
samples involves distinguishing between 4DE1 Pop.~A  and B sources.
More refined comparisons  becomes  possible by
generating composite spectra for individual Pop. A and B bins: e.g.
A1, A2, A3, \ldots and B1, B2, B1$^+$, etc. The population A -- B
concept was perhaps the most controversial part of our 4DE1
formalism where sources were arbitrarily divided into population A
and B based on FWHM \hb\ profile measures above and below $\approx$
4000 \kms\ respectively. However we find a multitude of differences
when we distinguish sources this way \citep{sulenticetal07}. Most RL
sources belong to Pop. B, and about 25-30\%\ of RQ sources also
belong to Pop. B, but the Pop. A/B distinction appears to be more
effective than more traditional RQ vs. RL comparisons in terms of
BLR structure.

Fig. \ref{fig:lbin} shows median composite spectra in six luminosity
intervals using 321 SDSS and 53 ISAAC sources. Composites were
generated separately for Pop. A and B sources in decades of $\log
L_\mathrm{bol}$ = 43 -- 44 to 48 -- 49. All ISAAC sources lie in the
interval $\log L_\mathrm{bol}$ = 47--49  with a small number
creating the highest $L_\mathrm{bol}$ bin (48--49).  Pop A and B
quasars show parallel trends with increasing luminosity: \oiiiopt\
and \hbbc\ systematically decrease in equivalent width  but \hb\
at the same time increasingly becomes redward asymmetric. The strong decrease in
narrow line strength is expected from standard nebular theory
\citep[see e.g.][with the caveats of
\citet{sulenticetal07}]{netzer90}. The Eigenvector 2
anti-correlation between strength of \heii\ and luminosity
\citep{borosongreen92} is also seen. No obvious trend involving
\feiiopt\ is seen which reflects the orthogonality of luminosity
with respect to Eigenvector 1 measures using principal component
analysis (PCA). The ISAAC sample median spectra are most similar to
the highest luminosity SDSS bin composite. The change in \hb\
towards high luminosity appears to be concentrated in the strength
of the red side of the line which was already anticipated as a red
asymmetry and line base \coneq\ red shift in Figures
\ref{fig:lumeff}b and \ref{fig:lumeff}c.

\subsubsection{Decomposition of Luminosity Binned
Spectra}

The next logical step is to decompose and quantify individual lines
and line components in different Pop. A and B luminosity bins.
Earlier attempts revealed one of the most obvious differences among
low redshift \hb\ profiles  and provided one of the motivations for
the population A--B concept
\citep{sulenticetal02,marzianietal03b,sulenticetal07}.  Fig.
\ref{fig:dcom} shows Pop A and B median spectra binned in luminosity
after continuum subtraction. Tab. \ref{tab:decn} provides line
equivalent width measures and FWHM \feii\ values for various
spectral bins discussed in the text. Tab. \ref{tab:pro2} tabulates
line profile measures  and Tab. \ref{tab:cen2} provides \hb\
centroid measures and estimated uncertainties for the same bins.
Formats appear as footnotes below each table. The first seven lines
present various 4DE1 bin values for ISAAC sources while the
remainder provide luminosity bin values following the 
designations in Fig. \ref{fig:4de1}. Measures in these tables
reflect the results of simultaneous fits to all lines and components
in each median composite spectrum. Attempts at a sequential
line-by-line decomposition yielded very similar results. Numbers
quoted in the discussion of Figure \ref{fig:lumeff} are averages or
medians based on individual source measures while Tables
\ref{tab:decn}, \ref{tab:pro2}, \ref{tab:cen2} give values derived
from composite spectra. There are no
significant discrepancies between measures estimated in these two
ways.
%Measures for luminosity bins in Tables
%\%ref{tab:decn}, \ref{tab:pro2}, \ref{tab:cen2} are given .
Source codes for $L_\mathrm{bol}$\ bins use the first two digits of
the $L_\mathrm{bol}$\ bin followed by A or B for the population (e.g. values for
line 43A correspond to measures of the Pop. A luminosity composite
for log $L_\mathrm{bol}$ = 43 -- 44). Narrow line changes with
luminosity will be discussed in a separate paper so that we can
concentrate on the properties of \hbbc\ and \hbvbc. FWHM \hb\ in Pop. A
sources shows a systematic increase with $L_\mathrm{bol}$ from 2000 \kms\
to 4100 \kms\ across the six luminosity bins likely driven by the
increase in minimum FWHM \hb\ with $\log L_\mathrm{bol}$ shown in Figure
\ref{fig:lumeff}a.

%Note that the final Pop. A bin shows FWHM H$\beta>$ 4000 \kms\
%because it was computed using a revised (luminosity dependent) Pop.
%A-B boundary discussed in the next section.

Pop. B  shows no clear FWHM trend with $\log L_\mathrm {bol}$\
but we find that $W$(\hb)  decreases systematically with increasing
luminosity. Pop. A does not show an obvious trend although  $W$(\hb)
is smallest in the most luminous Pop. A bins. The Pop. B ``Baldwin Effect'' shows $W$(\hb) decreasing from 143 \AA\ to 82 \AA\ with increasing  $\log L_\mathrm
{bol}$. Pop. B sources show 6--7 times larger asymmetry indices at
the highest source luminosities as well as evidence for a redshift
near the base (\coneq) of the line. The mean velocity redshift
increases by 9 times from the lowest to highest luminosity bin.
Fig. \ref{fig:hist} shows the increase with source luminosity of the
fraction of \coneq\ redshifts for Pop. B sources. The excess flux in
the red wing of \hb\ is revealed in Fig. \ref{fig:dcom}
decompositions to be a real \hb\ flux excess and not due to
\feiiopt\ contamination \citep[see also][who found a similar result]{netzertrakhtenbrot07}.

Conversely \feiiopt\ subtraction reveals that \hb\ shows little
evidence for profile shifts and asymmetries in Pop. A sources. The
apparent Pop. A red asymmetry appears to increase with luminosity in
Figure \ref{fig:lbin}; however Figure \ref{fig:dcom} suggests that it
is mainly due to \feiiopt\ and \oiiiopt\ contamination on the red
side of \hb. Table \ref{tab:decn} shows that  $W$(\feiiq)  does
not increase with luminosity but \hbbc\ decreases somewhat making
red \feiiopt\ contamination appear to grow in strength, even if some redward, very broad emission cannot be excluded at a low flux level in the highest luminosity bins.

The quantification in Tables \ref{tab:decn}, \ref{tab:pro2},
\ref{tab:cen2}, and the two components required to model Pop. B
sources, reenforce past evidence
\citep{corbin95,sulenticetal00c,sulenticetal02} for the existence of
a redshifted very broad line  component in \hb. The \hbvbc\
is the most obvious  profile difference between Pop. A and B, as
well as between high and low luminosity sources in Pop. B. As noted
earlier both RL and RQ sources are found in Pop. B and both Pop. B
RL and Pop. B RQ sources show the \hbvbc. This provides a
strong spectroscopic argument in favor of the importance of the Pop.
A-B concept compared to comparisons on the basis of radio loudness.
The effect of the VBLR component is to produce a broader, more red
asymmetric and redshifted \hb\ profile with the amplitude of these
effects dependent on the relative strengths of the BLR and VBLR
components. HE 0109--3518 in Fig. \ref{fig:spectra} is an example of
a source with a weak/moderate VBLR component while HE 1039--0724
shows one of the strongest VBLR components in the ISAAC sample. Tab.
\ref{tab:pro2}, \ref{tab:cen2}, Fig. \ref{fig:hist}, and the profile
decompositions in Figure \ref{fig:dcom} show evidence for a uniform
increase of  \hbvbc\ strength with  luminosity.

\subsection{Luminosity Dependence of Spectral Types}

The left-hand panel of Fig.~\ref{fig:lb} shows the bolometric
luminosity distributions for ISAAC sources showing  \hb\ profile
best fit with single Lorentzian and double Gaussian models.
Ambiguous sources are excluded. A K-S test confirms that the two
source luminosity distributions are not significantly different. The
right panel shows the distribution of profile fits in an FWHM \hb\
vs. bolometric luminosity plane with Lorentzian, double-Gaussian and
ambiguous sources denoted L, G and U respectively. No strong
luminosity preference is seen for L or G sources. The difference
between FWHM distributions for L and G profiles is highly
significant with a K-S test suggesting that  $P \la 10^{-8}$ for
sources with different profile shapes to be are drawn from the same
parent population. This can be taken as an independent confirmation
of the Pop. A-B dichotomy.

\subsubsection{Spectral Types: A Refinement}
\label{refinement}

The Pop. A-B  FWHM(\hbbc) = 4000 \kms\ boundary was adopted using a
low $z$\ and low luminosity sample. There is no strong evidence that
it has a direct physical meaning even if it is better motivated
empirically than the limit at FWHM(\hb) $= 2000$~\kms\ used to
separate ``Narrow Line'' Seyfert 1s (NLSy1s) from the rest of broad
line AGN.
%We previously suggested that the Pop A-B FWHM limit boundary might be luminosity
%dependent and indicative of a fixed or critical Eddington ratio.

Fig.\ref{fig:kasp} presents a reprise of Figure \ref{fig:lumeff}a
with a much expanded low redshift sample including: (a) SDSS DR5
quasars brighter than g=17.0 within z=0.7, (b) FIRST radio detected
SDSS DR5 quasars between g=17.0-17.5 within z=0.7
\citep{zamfiretal08} and (c) the ATLAS  sample of low redshift
quasars \citep{marzianietal03a}. This provides a somewhat better
mapping of the low $z$\ source distribution in the FWHM -- $\log
L_\mathrm{bol}$ plane.  The SDSS quasar sample does not include
sources with FWHM \hb $<$1000 \kms\ which were classified in SDSS as
galaxies. The narrowest of these ``broad line'' sources   can be
confirmed as Type 1 AGN (NLSy1) by the presence of \feiiq\ emission
in their spectra. Among the 400+ highest S/N SDSS quasars that we
have studied in detail \citep{zamfiretal08} all but five show
detectable \feiiopt\ emission suggesting that the presence of
optical \feiiopt\ emission is an ubiquitous property of Type 1 AGN.
A recent survey \citep{zhouetal06} of NLSy1s using SDSS DR3 provides
a representative census of these sources. They found $41$\ sources
($z\la $ 0.7) brighter than $g = 17.5$\ between FWHM \hb = 450 --
2000 \kms\ and these are indicated in Figure \ref{fig:kasp} with
filled triangles. The old Pop. A-B and NLSy1-rest of Type-1 AGN
``boundaries'' at 4000 and 2000 \kms, respectively are indicated.
This figure provides us with a better defined lower edge to the FWHM
\hb\ distribution. We superpose curves corresponding to exponents
a=0.52 \citep{bentzetal06} and  0.67 (Paper II) in the Kaspi
relation which will be considered in the discussion.
Fig.\ref{fig:kasp} also shows a revised Pop. A-B boundary following
the form of an a = 0.67 Kaspi relation as done in Paper I.

Figures~\ref{fig:lb} and \ref{fig:kasp} suggest a modified 4DE1 Pop.
A/B boundary is required for high luminosity quasars. We define
modified 4DE1 spectral types MA1, MA2, MB1, MB2, etc. where the
boundary between the MA and MB Populations is now
luminosity-dependent and given by FWHM$_\mathrm{AB} \approx 3500+625
\cdot L_\mathrm{bol}^{0.165}$\ (a = 0.67; analogous to Paper I).
Composite line \hb\ profiles for the revised 4DE1 bins are similar
to those obtained with the fixed FWHM definition
(Fig.~\ref{fig:4bin} and Fig. \ref{fig:4binan}) and are therefore
not shown.

In order to decide whether the fixed (4000 \kms) or luminosity
dependent boundary is more significant we defined three samples: (1)
the old Pop.~A, (2) an intermediate population between 4000 \kms\
and new luminosity-dependent limit and (3) a modified Pop.~B MB
$\equiv  B \setminus M$. We can also consider a modified Pop.~A, MA
$ \equiv A \cup M $. The \hb\ spectral region before and after
continuum subtraction for the median spectra of A, M and MB are
shown in Fig.~\ref{fig:medianmod} and Fig. \ref{fig:medianmodan}
respectively. The M median most closely resembles the Pop.~A median
rather than MB (or B).  Fig.\ref{fig:medianmod} and
\ref{fig:medianmodan}  show that the M \hb\ spectral region
resembles the one of Pop. A, and that the \hb\ fit is well fit by a
Lorentzian function, as it is the case for Pop. A sources. This
suggests that the luminosity dependent limit FWHM$_\mathrm{AB}(L)$\
might be more appropriate. The distribution of G and L sources in
the FWHM(\hbbc) vs.~ luminosity plane (Fig.~\ref{fig:lb}) also
supports this result. Although individual fits should be treated
with caution we see that the shapes are predominantly Lorentzian in
the Pop.~A and M regions while they are Gaussian in the MB area of
the plane.

\subsection{Black Hole Mass and Eddington Ratio}
\label{bhmass}

FWHM(\hb) has been used as a  virial estimator of \mbh\
\citep{vestergaardpeterson06} at low redshift. The problem is what
to use above redshift $z \approx$ 0.9. The use of \civ\ as an \hb\
surrogate involves problems that seriously reduce the reliability of
high z \mbh\ estimates \citep{sulenticetal07,netzeretal07}. We have
chosen instead to follow \hb\ out to the highest possible redshifts
via infrared spectroscopy. We use the updated relationships linking
FWHM(\hbbc), specific luminosity at 5100 \AA\ and BLR radial
distance $R_\mathrm{BLR}$ from \citet[][their Eq.
(5)]{vestergaardpeterson06}. We do not enter into the caveats of
black hole mass and Eddington ratio determinations \citep[see
e.g.][]{collinetal06,sulenticetal06,marconietal08,shenetal08}
remarking only that \mbh\ values are  uncertain by a factor 2--3 at 1$\sigma$ confidence level if derived from single-epoch observations. It is also possible to infer
bolometric luminosity from the specific continuum flux assuming a
constant bolometric correction. One can therefore also estimate the
luminosity-to-mass ratio ($L_\mathrm{bol}$/\mbh). This approach is
very crude \citep[see e.g.][]{hopkinsetal07,kellyetal08} but a
bolometric correction is still relatively stable across RQ and most
RL sources with the obvious exception of core-dominated (beamed)
sources.

The distribution of ISAAC quasars with L- and G--type fits as a
function of \mbh\ and \lledd\ are shown in the left- and right-hand
panels of Fig.~\ref{fig:distrlm} respectively. Fig.~\ref{fig:prof}
shows all ISAAC sources (including U  \hbbc\ profiles) in the
\lledd\ vs. \mbh\ plane. We derive quite large masses in the range
log \mbh=8.5-10.0 with a range of log \lledd = $-1.0$\ to 0.0. L
(assumed Pop. A) and G (assumed Pop. B) profiles favor lower/higher
log \mbh\ and higher/lower log \lledd\ values, respectively, as
previously found for our ATLAS sample \citep[][]{marzianietal03b}.
The difference in the distribution is significant for both
\mbh\ and \lledd, with a  low probability that sources with L- and
G-type profiles come from the same parent distribution of \lledd\
and \mbh: K-S tests indicate a probability $\la 10^{-3}$. However,
there is a range of \lledd ($-0.5 \div$ --0.3) where L-- and G--
type sources are almost equally represented. The absence of a sharp
\lledd\ boundary can be understood in terms of error effects as well
as of  systemic biases. Typical uncertainty in $\log$\lledd\ is
$\approx$ 0.1, if one assumes that the bolometric luminosity is
uncertain by 50\%, and that the FWHM(\hb) has typical error of 10\%.
However, this estimate neglects the scatter in the size-luminosity
relationship for the BLR. In addition, orientation effects -- which
are established among  radio loud quasars \citep[see
e.g.,][]{sulenticetal03,rokakietal03} -- may well raise the \lledd\
of several sources of Pop. B. They are not considered in the \lledd\
and \mbh\ estimates presented here. Moreover, the magnitude limit of
our ISAAC sample makes possible that we can detect {\em only} Pop. B
sources with the largest \lledd\ values (cf. Fig. 7 of Paper II). 

 Fig. \ref{fig:aic14} shows the dependence of \coneq\ and A.I.
on \mbh\ and \lledd. Significant correlations between \mbh\ and both
A.I. and \coneq\ are found while no significant correlations are
found involving \lledd. The A.I. vs \mbh\ correlation is marginally
significant ($P\approx 4 \cdot 10^{-3}$) if Pop. B sources are
considered alone; in this case \coneq\ and \mbh\ are not
significantly correlated. This is perhaps not surprising considering
the small number of sources, the typical errors, the \mbh\
``segregation" between  Pop. A and B. Median spectra in bins of
\mbh\ and \lledd\ point toward the origin of the correlation with
\mbh\ (Fig. \ref{fig:ml}). The bin labeled as $\log$ \mbh $< 9.5$\
actually covers masses in the range $8.5 \la \log$ \mbh $< 9.5$, and
for $\log$\lledd$\la -0.5$ corresponds to the bin $3.9 \la \log
L_\mathrm{bol}$/ \mbh$\la 4.4$ in Figure 9 of
\citet{marzianietal03b}. The remaining three bins represent the
extension to higher masses and \lledd\ that were not sampled at $z
\la 0.7$. We see  a trend, as  observed at low-$z$, that  for fixed
\lledd the redward component is stronger for larger \mbh. This result is 
discussed in \S \ref{disc:mbh}.

\section{Discussion}
\label{disc}

\subsection{Luminosity Trends}
\label{disc:e1}

One of the motivations for the ISAAC survey was to search for
trends, or even correlations, between emission line properties and
source luminosity. Another motivation was to compare high and low
luminosity source properties in the 4DE1 context. PCA of a bright
quasar sample \citep{borosongreen92} found source luminosity to be
orthogonal to all line measures except $W$(\heii). An \oiii\ measure
was included among the original Eigenvector 1 parameters
\citep{borosongreen92} showing a large scatter likely due to the
strong $L_\mathrm{bol}$\ dependence that we find with ISAAC. We did
not adopt an [OIII] measure in our 4DE1 formalism because we
preferred to avoid narrow line considerations as much as possible
\citep{sulenticetal00b}. Subsequent 4DE1 studies at low $z$ (e.g.,
Paper I and references found no luminosity dependence on broad line
LIL measures. What was thought to be the most significant (inverse)
correlation between luminosity and broad line HIL measures
(especially \civ) \citep{baldwin78} is now known to be present in
quasar samples with small redshift/luminosity dispersion
\citep{xuetal08} and is likely related to intrinsic source
properties such as Eddington ratio \citep{bachevetal04}.

Only one luminosity trend was found over the 5-6 luminosity decades
covered in this study   (without Pop. A/B discrimination). Profile
parameter distributions (Figure \ref{fig:lumeff}) are essentially
the same at all luminosities. The one exception involves an increase
in minimum FWHM \hb\ with $\log L_\mathrm{bol}$ visible in Figure
\ref{fig:lumeff}a (see also Paper I). Low luminosity ($\log
L_\mathrm{bol}$ = 43.0) sources show FWHM \hbbc\ values as low as
500--750 \kms\ while the minimum value at $\log L_\mathrm{bol}$ = 48
appears to be FWHM \hb\ $ \approx $ 3000 \kms. The lower FWHM \hb\ limit
is expected if broad Balmer line emission arises in a  virialized
cloud distribution (or accretion disk) that obeys a Kaspi \citep[see
e.g. ][]{kaspietal05} relation. We can think of the minimum
FWHM-$\log L_\mathrm{bol}$\ trend as the boundary for sources
radiating sub-Eddington.  NLSy1s defined (at low $z$) as sources
with FWHM \hb\ $<$2000 \kms\ are absent from the ISAAC sample.

%This is expected as a selection
%effect in a flux limited sample. The \citet{zhouetal06} NLSy1
%sample suggests that we would have to observe sources 4--5
%magnitudes fainter in the ISAAC $z$\ range before we would have a
%reasonable chance to find NLSy1 with FWHM \hb$\sim$ 1000\kms.

It is unclear if a Kaspi relation ($R_\mathrm{BLR} \propto
L^{-\mathrm{a}}$) is valid above redshifts of a few tenths.  There
has been much discussion about the correct exponent in the Kaspi
relation with values between a = 0.5 -- 0.7 favored
\citep{kaspietal00,kaspietal05,bentzetal06}. Figure \ref{fig:kasp}
provides an empirical representation of the minimum FWHM \hb\ trend.
Note that the expected trend has been quadratically combined with
the instrumental resolution (assumed to be 200 \kms\ at FWHM,
appropriate for the SDSS).  Superposed curves corresponding to a =
0.52 \citep{bentzetal06}, and 0.67 \citep[][Paper II]{kaspietal05}
show that discrimination between these values becomes possible above
$\log L_\mathrm{bol} \approx$ 47. A first impression is that a =
0.52 is disfavored  by the ISAAC data unless a considerable number
of these sources are super-Eddington radiators. Thus values in the
range a = 0.6-0.7 are favored.  However a complicating factor
involves the dependence of FWHM \hb\ on source orientation which is
especially important if one accepts the paradigm of a line emitting
accretion disk. Some low redshift sources with small FWHM \hb\
(extreme NLSy1 sources) show rapid high amplitude X-ray variability
\citep[e. g.][]{brinkmannetal03} or blue shifted narrow lines (blue
outliers) which have been argued to signal near face-on orientation. In a
previous paper \citep{marzianietal03b} we attempted inclination
corrections for blue outliers assuming a $\sin i$\ dependence of
FWHM. Using a tentative correction that increases black hole mass by
$\Delta$\mbh $\approx$ 0.4 we were able to move many apparently
superEddington NLSy1 radiators (sources with $\log L_\mathrm{bol}
\sim$ 46 that lie below the a = 0.5 curve in Figure \ref{fig:kasp})
below \lledd  = 1. It seems unlikely that more than a very few ISAAC
sources involve preferred orientation (5 are nominally
super-Eddington) given the distribution of observed FWHM \hb\ values
even if optical luminosity is orientation dependent. It is also
possible that orientation becomes increasingly less important in
higher $L_\mathrm{bol}$\ sources because of radiative instabiliy in
the accretion disk \citep{blaes07}. A value as large as a = 0.67
would imply that all quasars radiate well below the Eddington limit.
We conclude that there is no strong evidence for a population of super-Eddington
radiators in our low $z$\ and ISAAC samples.

\subsection{4DE1 Populations A and B}

This study suggests that, in order to isolate luminosity effects
and other fundamental trends in quasar samples, one should work in
the context of the 4DE1 formalism and take advantage of the
Population A-B concept. The empirical evidence continues to grow
in favor of two distinct populations of quasars. We recently summarized
the multitude of empirical differences between the two populations
\citep{sulenticetal07,sulenticetal08}.  If luminosity trends exist for
only one of the populations then the lack of a trend in the other might
obscure the effect in a mixed sample such as the one displayed
in Figure \ref{fig:kasp}.  The Pop A -- B boundary was
empirically defined, after noting a discontinuity between the
sources whose FWHM(\hb) lies in the range 2000 -- 4000 \kms and
those of Pop.~B (also because almost all RL sources are pop. B).
The similarity in emission line parameters between those sources
and the NLSy1s then suggested a single population for the entire
0--4000 \kms range. In analogy with the minimum FWHM \hb\ trend
which corresponds to a fixed Eddington ratio  of 1 we can expect
that the limit between Pop.~A and B will also be luminosity
dependent (Paper I).

We have previously suggested that the Pop. A-B FWHM \hbbc\ boundary
might reflect  a critical accretion rate/Eddington ratio {(\lledd
$\sim 0.2 \pm 0.1$)}. We note that a number of recent studies have
independently arrived at a similar ``critical'' FWHM or \lledd\
value \citep{marzianietal03b,collinetal06,bonningetal07,kellyetal08,huetal08b}.
Such a  critical value might signal a significant change in disk
structure/kinematics that is reflected in systematically larger FWHM
\hb, the demise of a \civ\ wind as well as weaker soft X-ray and
\feii\ emission. It is interesting to point out that the thin
accretion disk approximation is expected to break down at \lledd
$\la$ 0.3 with the development of an inner thick structure
\citep{hubenyetal00}. This kind of change might well affect the
spectroscopic signature of the BLR.

%The Pop.A/B boundary is likely to be luminosity dependent, even if i
It is probably na\"{\i}ve to expect a luminosity dependence as
strong as the one predicted by a fixed \lledd\ (Fig.
\ref{fig:kasp}). The Pop. A/B boundary is found to be close to 4000
\kms, following four lines of evidence: (1) the median spectrum for
Pop. A in the luminosity bin 43$\la\log L_\mathrm{bol} \la $44 shows
that the \hb\ is Lorentzian; (2) between  constant FWHM(\hb)=4000
\kms\ and the \lledd = 0.15 line, and for $45 \la\log L_\mathrm{bol}
\la 47$, most ATLAS sources show typical Pop. A properties. A
minority of core-dominated radio-loud sources  are located there
since they are expected to be observed almost pole-on, at minimum
FWHM. If $44 \la\log L_\mathrm{bol} \la 45$\ ATLAS sources are
mainly located within --500 \kms\ of FWHM = 4000 \kms, their \hb\
profiles are similar to but not exactly the same as the ones of Pop. A. In the
present paper we consider the luminosity dependence of the boundary
to follow Paper I assuming a $\approx$ 0.67. It is assumed constant
corresponding to  FWHM$\sim$4000 \kms\  below $\log L \approx 47$.

%The choice of a population A-B boundary then becomes important to
%justify beyond empiricism and

\subsection{An \hbbc\ ``Baldwin Effect'' is Present only in Population B}

The Pop. A/B concept provides a useful alternate way to evaluate
luminosity properties.  When we discriminate by source population we
immediately find possible luminosity trends between FWHM and
$W$(\hb) for Pop. A and Pop. B respectively. The former trend sees a
mean FWHM \hb\ increase with luminosity from $\approx$ 2000 \kms\ up
to $\approx$ 4000 \kms. This is likely driven by the minimum FWHM
\hb\ trend already discussed as well as by Malmquist bias since all
of the highest luminosity sources lie at higher redshifts and are
preferentially selected in a flux limited sample. Assuming an
underlying correlation between black hole mass and source luminosity
we expect systematically larger black hole masses for the ISAAC
sources as is observed.

Pop. B sources show a decrease in $W$(\hb) by almost a factor of two
over our source luminosity range. This \hb\ Baldwin effect is
difficult to interpret because \hb\ is a composite feature
(\hbbc\  + \hbvbc). Tab.\ref{tab:decn} presents an attempt at quantitatively
separating BLR and VBLR components for the six Pop. B luminosity
binned composite spectra. The same can be done for Pop. A but the
BLR component always dominates \hb\ in all luminosity bins. The
BLR component in Pop. B ranges from 60\%\ of \hb\ at low
luminosity to 25\% at $\log L_\mathrm{bol}$ = 47 -- 48. Correction
for the  VBLR component yields a ``super''
Baldwin effect where $W$(\hbbc) decreases by a factor of four
over our luminosity range. At the same time we find considerable
scatter but no obvious trend in the strength of \feii\ emission as
measured in the \feiiq\ blend.  The highest luminosity Pop. B
sources with smallest $W$(\hbbc) show an apparently more prominent \hbvbc\ while its contribution in low luminosity sources seems
modest because \hbbc\ dominates. 
%{\bf A similar result was obtained also by \citet{netzertrakhtenbrot07}
%who found a tendency for the fractional luminosity of the red part of the \hb\ line to increase with continuum luminosity.}

The simplest interpretation of these results involves changes in
physical conditions of the BLR emitting region due to: (1) source
luminosity and (2) the Pop. A-B dichotomy. As alluded to above the
first may involve changes related to radiation pressure dominance in
the accretion disk at higher source luminosity and the second
changes related to a critical Eddington ratio.
%We have already argued for 1-2dex increase in BLR density (and
%decrease in ionization parameter) from Pop. B to pop. A sources
%\citep{marzianietal01}.
The tendency for FWHM(\feii) to follow  FWHM \hbbc\ in BLR-dominated
Pop. A sources can be taken as evidence for an \feii\ -- H$\beta$
BLR commonality. The inferred lower density of the VBLR \citep[e.g.,
][]{sulenticetal00c} in  (less BLR dominated) Pop. B sources
disfavors an \feii/VBLR emission region commonality
\citep{marzianietal08}.  $W$(\hbvbc) is found to be approximately
constant in the six luminosity bins suggesting that it arises
largely in optically thick gas photoionized by the central
continuum. The largest challenge is to explain the fourfold decrease
in  \hbbc\ strength observed only Pop. B sources.  The simplest
interpretation involves a change in Pop. B physical conditions that
suppresses \hbbc. Since \hbbc\ and \feii\ are thought to arise in
the same region this might point toward collisional quenching of
\hb\ BLR emission as originally suggested by \citet{gaskell85}.
Another, even simpler possibility involves different sized annuli
for \feii\ and \hbbc\ emitting regions. If we correct $W$(\hb)
measures for \hbvbc\  and recalculate \rfe\ parameters
we find that many/most Pop. B sources will move towards higher \rfe\
values $\approx$1 in the 4DE1 optical plane.  The similar \rfe\
ratios observed in Pop. A luminosity bins (using \hbbc\ only)
suggest also a similarity in physical conditions. In other words, we
would have a similar low-ionization BLR  in both Pop. A and B
sources plus a second  VBLR  only in Pop. B sources.

%The major 4DE1 implication of the BLR+VBLR decomposition is
%a change in population B source occupation in the optical plane
%(Figure 4). 4DE1 parameter \rfe\ was defined as a ratio of line
%components thought to arise in the same BLR clouds. Removing VBLR
%emission from \rfe will fill in the zone of avoidance in the 4DE1
%optical plane. BLR-VBLR decompositions are too uncertain to permit
%derivation of revised \rfe\ measures on a source by source basis
%however. Tab.\ref{tab:decn} provides revised \rfe\ values for the
%pop B luminosity composites. It is clear that the optical plane of
%4DE1 will no longer be luminosity independent because the highest
%uminosity sources will move farther to the right in the 4DE1
%ptical plane (higher revised \rfe). There will be a trend towards
%higher luminosity sources from left to right in the population B
%art of 4DE1. It will therefore most closely resemble the
%distribution of Pop. A sources except that they show no such
%uminosity trend. Previous optical plane modeling found an L/M
%trend but no direct L trend involving pop. A source occupation
%\citep{marzianietal01}. We conclude that the pop. A-B boundary is
%driven largely by changes in BLR physics/geometry related to the
%VBLR \hbbc\ emission component.

\subsection{Black Hole Mass and Eddington Ratio}
\label{disc:mbh}

Tab. \ref{tab:decn} presents  FWHM \hbbc\  estimates for Pop. B
luminosity bins. The \hbbc\ of Tab. \ref{tab:decn}  is related to
the ``reverberating'' component almost equivalent to the entire
\hb\ profile in Pop. A sources. We note that the decomposition
\hbbc -- \hbvbc\ of Pop. B sources assumes two Gaussians while the intrinsic shapes of
\hbbc\ and \hbvbc, especially in the radial velocity range where
they overlap is, unknown. The origin of the redward asymmetry often
seen in the \hb\ profile is also unclear. It seems too large to be
due to gravitational redshift associated with line emission from
increasingly smaller distances from the central black hole (under
the assumption that the motion is predominantly virial). A likely
possibility is non-virial motion associated with gas infall toward
the central black hole \citep[see e.g. ][who provide renewed support for the hypothesis of infalling emitting regions]{huetal08a,gaskellgoosmann08}, but alternatives like Compton scattering have also been considered
\citep[e.g.][]{kallmankrolik86}. However, if the
VBLR is mainly made of optically thick gas  the radial velocity
range underlying the \hbbc\ is especially ambiguous. Note also that
the width reestimation  is not the same as using the 2$^\mathrm{nd}$\ moment of
the \hb\ profile:  $\sigma$ \citep{petersonetal04,collinetal06}
may have little meaning for a composite line since it is measured on
the whole \hb\ profile. The reestimation leads to a more extreme
reduction than suggested in Paper II; actually, all VBLR corrected
FWHM  values decrease to within 1000 \kms\ of the Pop. A-B
boundary (4--5000 \kms). In fact the highest luminosity bins show
FWHM(\hbbc) within the revised Pop. A range as defined in Figure
\ref{fig:lb}.

An important implication of this correction involves its effect on
\mbh\ and \ledd\ estimation for Pop. B sources.  Pop. B sources
would essentially cease to exist as a separate population in the
context of \mbh\ estimation. All \mbh\ values for Pop B would
decrease from the previous range (Paper II) log \mbh = 7.5 -- 9.5 to
7.0 -- 9.0 which is more similar to the Pop. A range but still
consistent with the view \citep{sulenticetal00a} that Pop. B quasars
are radiating at lower \lledd, as further discussed below.

%These changes represent an
%increase in the reliability of \mbh\ virial estimates based upon
%the use of FWHM \hbbc\ BLR measures in Pop. B sources.
%Since the BLR component of \hbbc\ becomes essentially constant in
%FWHM for Pop. B sources the FWHM \hbbc\ 4DE1 parameter
%\citep{sulenticetalab} must be abandoned or redefined in terms of
%FWHM or $W$  \hbvbc. \rfe as presently defined involves a
%subtle dependence on source luminosity requiring its redefinition
%as well. Further discussion of 4DE1 implications are beyond the
%scope of this paper. The pop. A-B boundary can now be thought of
%as the critical L/L$_{Edd}$ value where the onset of VBLR emission
%becomes important.

Considering the uncertain and somewhat speculative nature of the corrections applied to Pop. B sources, FWHM \hb\ in Pop. A sources appears to provide more
direct and reliable black hole mass estimates yielding a range in
log \mbh\ = 6.0 -- 8.5 at low $z$\ and 9.0 -- 10.0 at $ z \ga$1.0 (see
figures in Paper II). The major source of uncertainty involves the
inclination correction which depends on the BLR geometry. A weak
VBLR component and/or blue asymmetry are present in an uncertain
number of sources which might lead to slight overestimates of \mbh\
when assuming that the virial line component is a symmetric
Lorentzian. Gaussian fits to Pop. A profiles
\citep[e.g.][]{shenetal08} lead to significant overestimates of
\mbh. Median values from our luminosity composite profiles (assuming
mean FWHM \hb\ and $L_\mathrm{bol}$\ for sources in that bin) show a
clear luminosity correlation ranging from $\log$ \mbh $\sim$ 6 -- 10
for the most luminous ISAAC sources. Estimates corrected for \hb\
asymmetries are slightly lower but only for the SDSS bins.

%More symmetric Pop. A profiles and accretion disk
%model considerations may signal a less flattened BLR geometry at
%high luminosity.

{ The \lledd\  values in Table \ref{tab:decn} increase from 0.02 to $\ga$ 0.6 in the highest luminosity bins of Pop. B.  We remark  that the \lledd\ value for the most luminous bin ($\approx$ 1) is especially uncertain because of the limited number of sources and of the large correction amplitude (FWHM(\hb)$\approx$9300 \kms; FWM(\hbbc)$\approx$4300 \kms).  \lledd\ values obtained using FWHM (\hb) without removing \hbvbc\ 
(Table \ref{tab:pro2}) range from $\approx$ 0.02 to 0.27. Similar values are obtained if FWHM(\feiiq) is used as a virial estimator to compute \mbh. The corresponding \lledd\ estimates for Pop. A luminosity composites also show an   increase ranging from \lledd\ $\approx$ 0.2 (lowest luminosity)   to 1.3  in the highest luminosity bin. Ranges for Pop A from Paper II were  \lledd\  = 0.05 -- 1.0 at low $z$\ and 0.5 -- 1.0 at $z \ga$1.0 assuming that sources above \lledd $>$1.0 are the result of face-on orientation.  The luminosity trend is expected for both Pop. A and B since we are sampling preferentially higher \lledd\ radiators with increasing $L_\mathrm{bol}$\ (as shown in Paper II). Without \hbvbc\ correction the difference between the \lledd\ of Pop. A and B  is $\Delta \log$ \lledd $\approx$ 0.6 for the SDSS luminosity bin, decreasing to $\approx$ 0.3 for the ISAAC luminosity bins (47 and 48).  If we consider the width of \hbbc\ only,   $\Delta \log$ \lledd $\approx$ 0.5, even if this difference becomes rather small in the ISAAC bins, probably because of the extreme amplitude of the correction, and of  the caveats listed at the beginning of this section.  At any rate, from the SDSS sample we can conclude that Pop. B appears to be a population of lower Eddington ratio than Pop. A.  } Pop. B is  likely to be an older and more evolved quasar population than
Pop A, since black holes seem to reside in hosts with large
spheroidal components, often elliptical galaxies \citep{wooetal05}.
On the contrary the morphology of NLSy1 hosts  (i.e., in many ways
the ``extreme Pop. A" sources) in the local Universe often involves
high surface brightness star-forming galaxies
\citep{krongoldetal01,ohtaetal07}. They are often barred and/or
remarkably perturbed suggesting that they may be  young systems
sustained by a large flow of matter toward their central black holes
\citep[see e.g. the review by][]{sulenticetal08}.

\subsection{Occupation of Spectral Type B2}

ISAAC sources are found in the B2 bin of the 4DE1 optical plane
(FWHM= 4-8000 \kms\ and \rfe = 0.5 -- 1.0. The low $z$\ ATLAS sample
included only 11 sources ($\sim$5\%) in that bin but the more
complete SDSS sample involves  40 B2 sources ($\sim$10\%). Bin B2
may involve sources with the largest black hole masses (several
times 10$^9$ \msol). The largest masses are expected to fall in the
region of bin B2 \citep{zamanovmarziani02} but such massive black
holes are not observed at low $z$\ (see, for example the \mbh\
distribution as a function of $z$\ in Paper II). In the ISAAC sample
we derive \mbh\ values as high as $\log\hbox{\mbh}\approx 9.7 $
\citep{sulenticetal06}. According to the grid in Fig. 2 of
\citet{zamanovmarziani02} the B2 bin is expected to be occupied by
sources with \mbh~$\sim 5 \times 10^9$ \msol. The expected
$L_\mathrm{bol}/$\mbh\ ratio is modest with \lledd $\sim 0.16$. B2
sources therefore expected to share physical properties that are thought to be
typical of Pop.~B. The \hb\ line profile is well described by a
double Gaussian decomposition also if we restrict our attention to
revised bin MB2. This result is consistent with the conclusion that
most sources in bin B2  are similar to those in B1 although with
stronger \rfe. The \hb\ line profiles imply that both types may fall on
the Pop. B side of a critical \lledd\ that separates  Pop.~A and
Pop.~-B boundary.

\section{Conclusions}

We presented VLT-ISAAC spectroscopic observations for 30
intermediate $z$\ and high-luminosity HE quasars. Combined with
previous data we have a sample of 53 objects ($z$ = 0.9 -- 3.0) which we
compare with two large low-$z$ samples
\citep{marzianietal03a,zamfiretal08}. We find few
correlations/trends between broad emission line properties and
source luminosity. We previously proposed an empirical limit at
FWHM(\hb)~$\approx 4000$~\kms\ to distinguish between two
populations A-B of quasars with very different spectroscopic
properties. Sources with FWHM(\hb) $<$ 4000 \kms\ show Lorentzian
\hb\ profiles while those above this limit are best fit with double
Gaussian models. This low $z$/ low luminosity dichotomy is also
found for the high luminosity ISAAC sources. Elementary \lledd\
computations suggest that  the dichotomy may be explained if a
critical  \lledd\ is associated with a BLR structure change, but the
issue of the Pop.A/B boundary deserves further scrutiny. Once the
population A/B dichotomy is taken into account the phenomenology of
the \hbbc\ emitting region, as well as inferences about BLR
structure are quite similar over a 5 dex luminosity range that
includes some of the most luminous known quasars.  This overall
scenario is described in more  detail by the following results of
the present paper:
\begin{enumerate}

\item A minimum requirement for quantifying and interpreting BLR
properties in Type 1 AGN involves the Pop. A--B dichotomy.
Spectroscopic averaging without such a discrimination appears to be
equivalent to discussing stellar properties without consideration of
the OBAFGKM spectroscopic sequence

\item Pop. A sources show a minimum FWHM \hb\ that increases with
source luminosity  from $\sim$500-1000 \kms\ to 3000-3500 \kms. The
best boundary to this lower envelope favors an exponent a
$\la$ 0.67 in the Kaspi relation.  Pop. A sources span a FWHM \hb\ range of $\sim$ 4000
\kms\ driven by source orientation and virial motions in an
accretion disk. The virial assumption implies a \mbh\ range of 4dex
($\log$ \mbh = 6--10) for this high accreting population. {  In
the ISAAC sample studied in this paper Pop. A sources show $-0.7 \la \log $ \lledd $\la 0.18$.}

\item \hb\ emission in Pop. B sources involves both unshifted
broad (\hbbc) and redshifted very broad line (\hbvbc) components.
Pop. B shows a ``super'' Baldwin effect with $W$(\hbbc) decreasing
by a factor 4 with increasing $L_\mathrm{bol}$. At the same time
$W$(\hbvbc) remains almost constant implying that it arises in a
photo-ionized optically thick medium. FWHM(\hbbc) component shows no
luminosity dependence after correction for \hbvbc\ broadening. This
yields an \mbh\ range similar to that for Pop. A. {Pop. B is a
lower accreting population with $-1.1  \la  \log$ \lledd $\la -0.2$\
for the objects of this paper. One should  bear in mind the caveats of
\S \ref{bhmass} in analyzing the \lledd\ limits, and that 80\%\ of
Pop. B sources of the ISAAC sample show $\log$ \lledd$\la -0.4$. Considering the
blurring by errors in \lledd\ estimates, and results of previous
papers, intrinsic \lledd\ ranges could be $\approx$ 0.01 -- 0.2
(Pop. B) and $\approx$ 0.2 -- 1 (Pop. A), with a separation at
\lledd $\approx 0.2 \pm 0.1$.  Pop. B sources in the SDSS sample seem to remain a population of lower Eddington ratio  even if the extreme correction to FWHM  implied by  removal of the \hbvbc\ is applied (\S \ref{disc:mbh}). }

\item Most Type 1 AGN do not show broad line profiles consistent with
simple accretion disk models. Perhaps Lorentzian Pop. A profiles can
be accommodated with   thin/slim disk emission models (if disk
emission is very extended) but Pop. B cannot. The redshifted \hbvbc\ in Pop. B sources can not be considered part of a double
peak signature since the asymmetry and the centroid shift at
$\frac{1}{4}$\ intensity (near the profile base is usually too large
to be attributed to a gravitational redshift. Arp 102B -- the
prototypical ``double-peaked'' Balmer line  emitter -- shows a redshift
at line base that is modest, and consistent with the value expected
for gravitational redshift at the inner emitting radius of the
accretion disk according to the model of \citet{chenetal89}. This is
however not the case for many  Pop. B objects that have been
included in so-called double-peaked source compilations   \citep[i.e. as accretion disk candidates: e.g., ][]{stratevaetal03}.

\item \hbvbc\ is the dominant broad line component at high  luminosity
(from $\sim$ 25\% to 60\% of the  \hb\ flux over our
$L_\mathrm{bol}$ range) for Pop. B sources. The geometry and kinematics of the VBLR are
yet unclear. The \hbvbc\ redshift $\Delta v \approx$ 1--2000 \kms\ is
unlikely to be gravitational in nature. The increase in dominance
with redshift favors a connection to the hypothesized infall
involved with the \mbh\ growth. 
Another possibility involves photon downshifting via some form of
scattering \citep[e.g.][and references
therein]{laor06}. 

\end{enumerate}

\acknowledgements

Funding for the SDSS and SDSS-II has been provided by the Alfred P. Sloan Foundation, the
Participating Institutions, the National Science Foundation, the U.S. Department of Energy,
the National Aeronautics and Space Administration, the Japanese Monbukagakusho, the Max
Planck Society, and the Higher Education Funding Council for England. The SDSS Web Site is http://www.sdss.org/.
The SDSS is managed by the Astrophysical Research Consortium for the Participating Institutions. The Participating Institutions are the American Museum of Natural History, Astrophysical Institute Potsdam, University of Basel, University of Cambridge, Case Western Reserve University, University of Chicago, Drexel University, Fermilab, the Institute for Advanced Study, the Japan Participation Group, Johns Hopkins University, the Joint Institute for Nuclear Astrophysics, the Kavli Institute for Particle Astrophysics and Cosmology, the Korean Scientist Group, the Chinese Academy of Sciences (LAMOST), Los Alamos National Laboratory, the Max-Planck-Institute for Astronomy (MPIA), the Max-Planck-Institute for Astrophysics (MPA), New Mexico State University, Ohio State University, University of Pittsburgh, University of Portsmouth, Princeton University, the United States Naval Observatory, and the University of Washington.

\pagebreak
\newpage

%\addtocounter{figure}{-11}
\begin{figure*}
\includegraphics[width=19.6cm, height=21.6cm, angle=0]{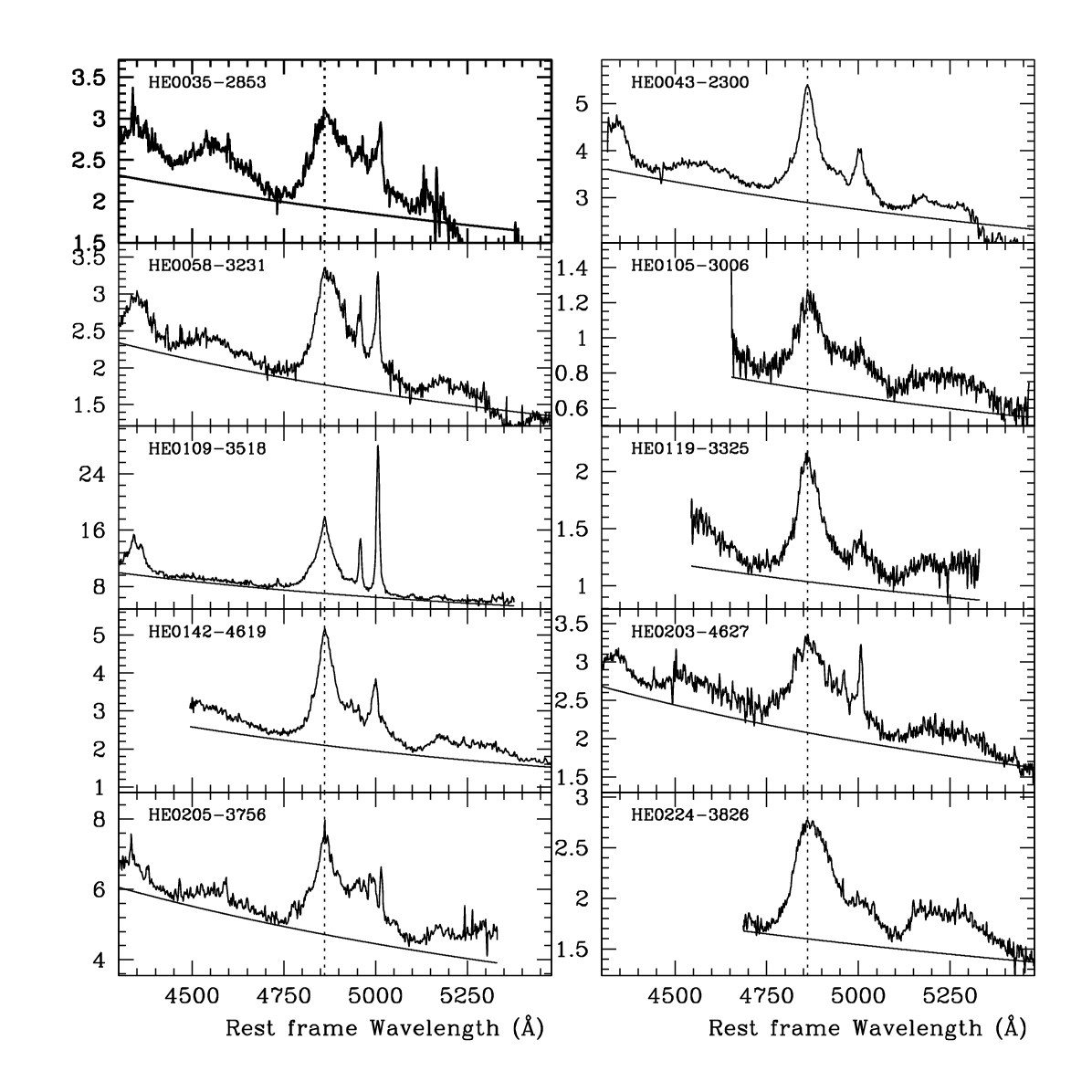}
%\vspace{22cm}
\caption[]{Calibrated VLT-ISAAC spectra for 30 new
intermediate-redshift quasars. Absciss\ae\ are rest-frame
wavelength in \AA, ordinates are rest-frame specific flux in units
of 10$^{-15}$ \ergss\ cm$^{-1}$ \AA$^{-1}$.  The spectrum behavior of HE 0035$-$2853, HE0043$-$2300, HE 0058$-$3231 is fully unreliable beyond $\approx$ 5250 \AA\
 due to heavy atmospheric absorption also affecting the calibration standard. } \label{fig:spectra}
\end{figure*}
\addtocounter{figure}{-1}

\begin{figure*}
%\figurenumber{1}
\includegraphics[width=19.6cm, height=21.6cm, angle=0]{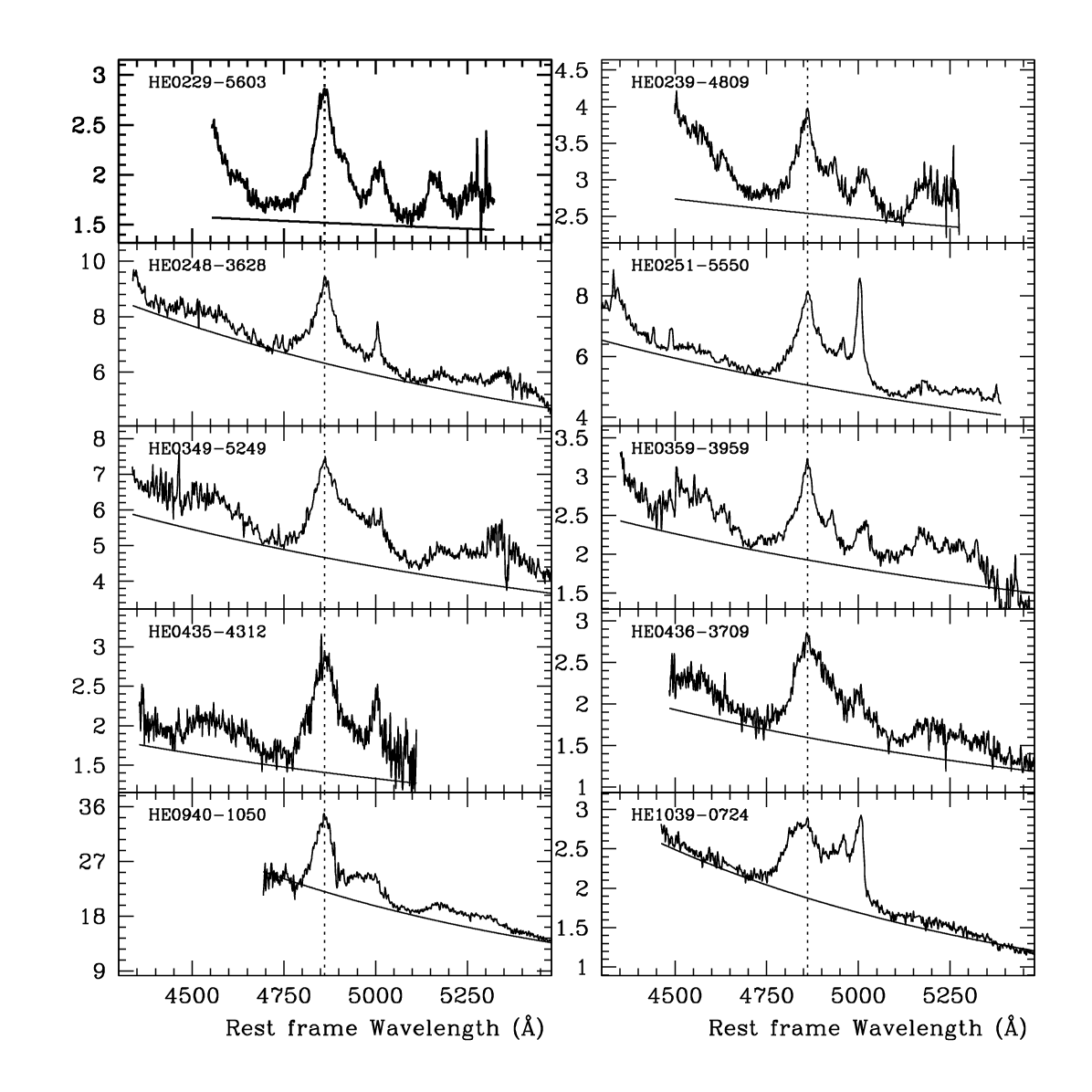}
%\vspace{22cm}
\caption[]{Cont.}
\end{figure*}

\addtocounter{figure}{-1}

\begin{figure*}
%\figurenumber{1}
\includegraphics[width=19.6cm, height=21.6cm, angle=0]{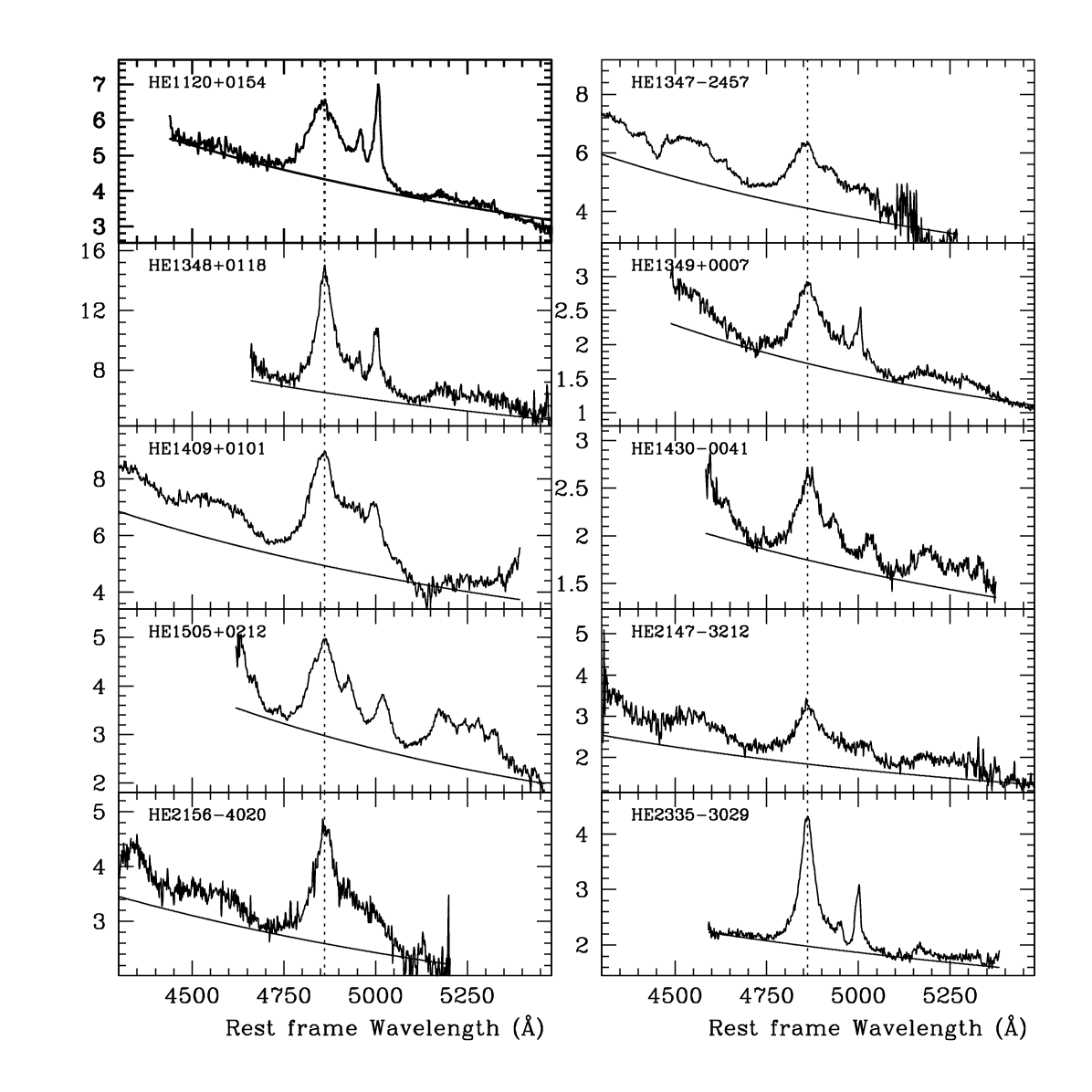}
%\vspace{22cm}
\caption[]{Cont.}
\end{figure*}

\begin{figure*}
\includegraphics[width=19.6cm, height=21.6cm, angle=0]{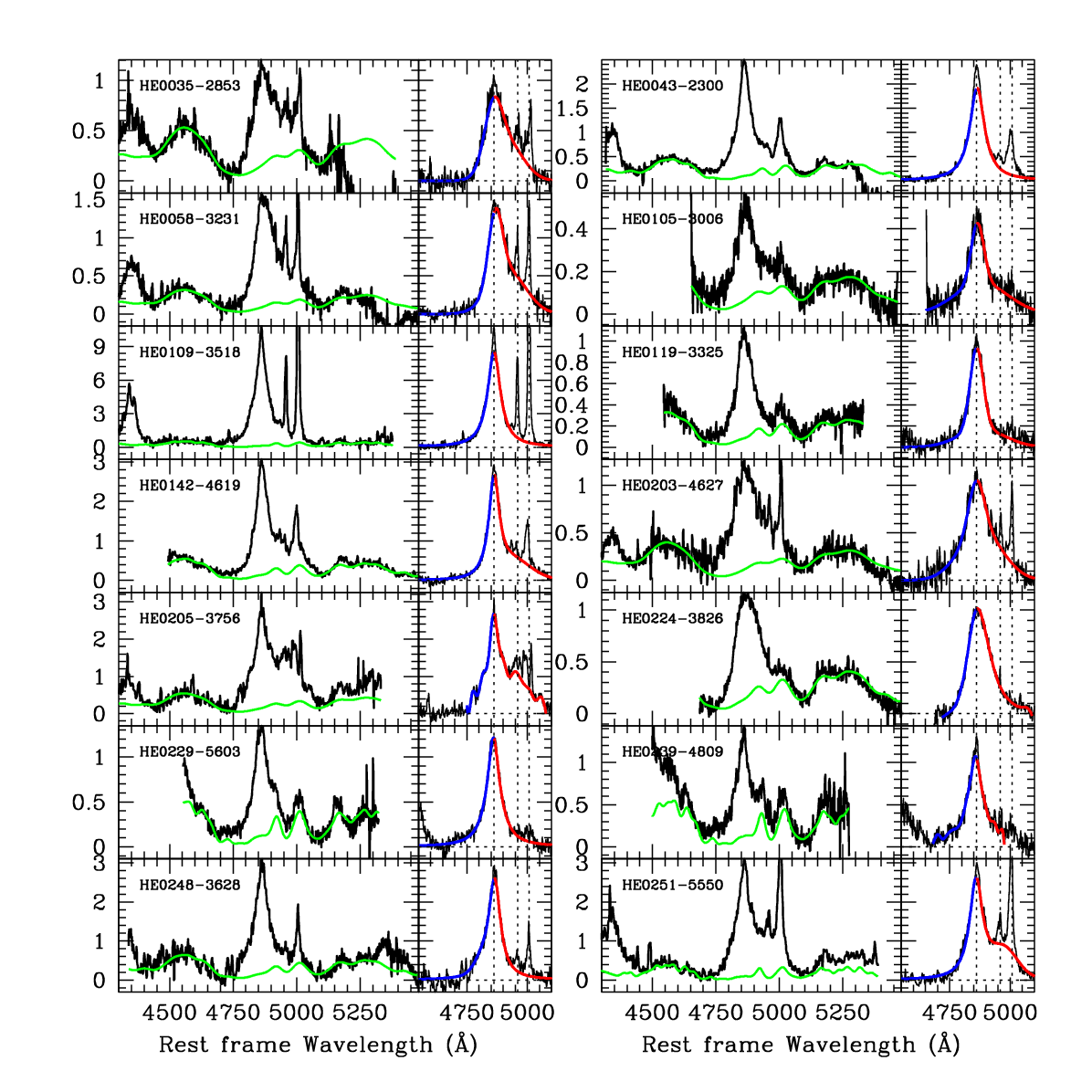}
%\vspace{22cm}
\caption[]{Continuum-subtracted spectra. Left-hand panels: the \hb\ spectral regions. The thin green lines show the \feii\ emission. Right-hand panels: spectra after continuum and \feii\ subtraction. The  \hb\ profile (broad + very broad) is marked by the thick line, blue on the short-wavelength side with respect to the rest wavelength, red on the long-wavelength side. Abscissa is rest frame wavelength in \AA, ordinate is specific flux in units of $10^{-15}$ erg s$^{-1}$ cm$^{-2}$ \AA$^{-1}$.  } \label{fig:hbeta}
\end{figure*}
\addtocounter{figure}{-1}

\begin{figure*}
%\figurenumber{1}
\includegraphics[width=19.6cm, height=21.6cm, angle=0]{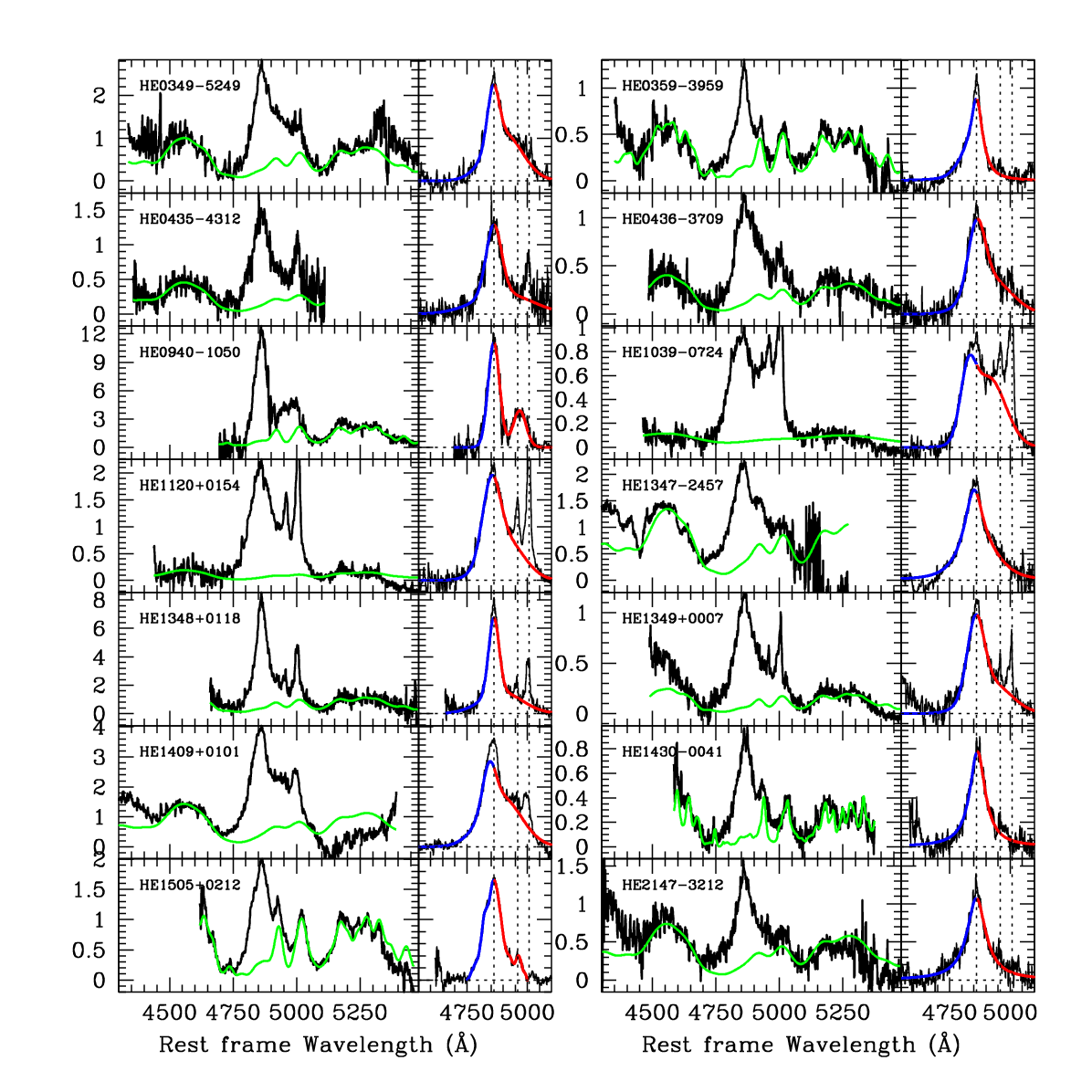}
%\vspace{22cm}
\caption[]{Cont.}
\end{figure*}

\addtocounter{figure}{-1}

\begin{figure*}
%\figurenumber{1}
\includegraphics[width=19.6cm, height=21.6cm, angle=0]{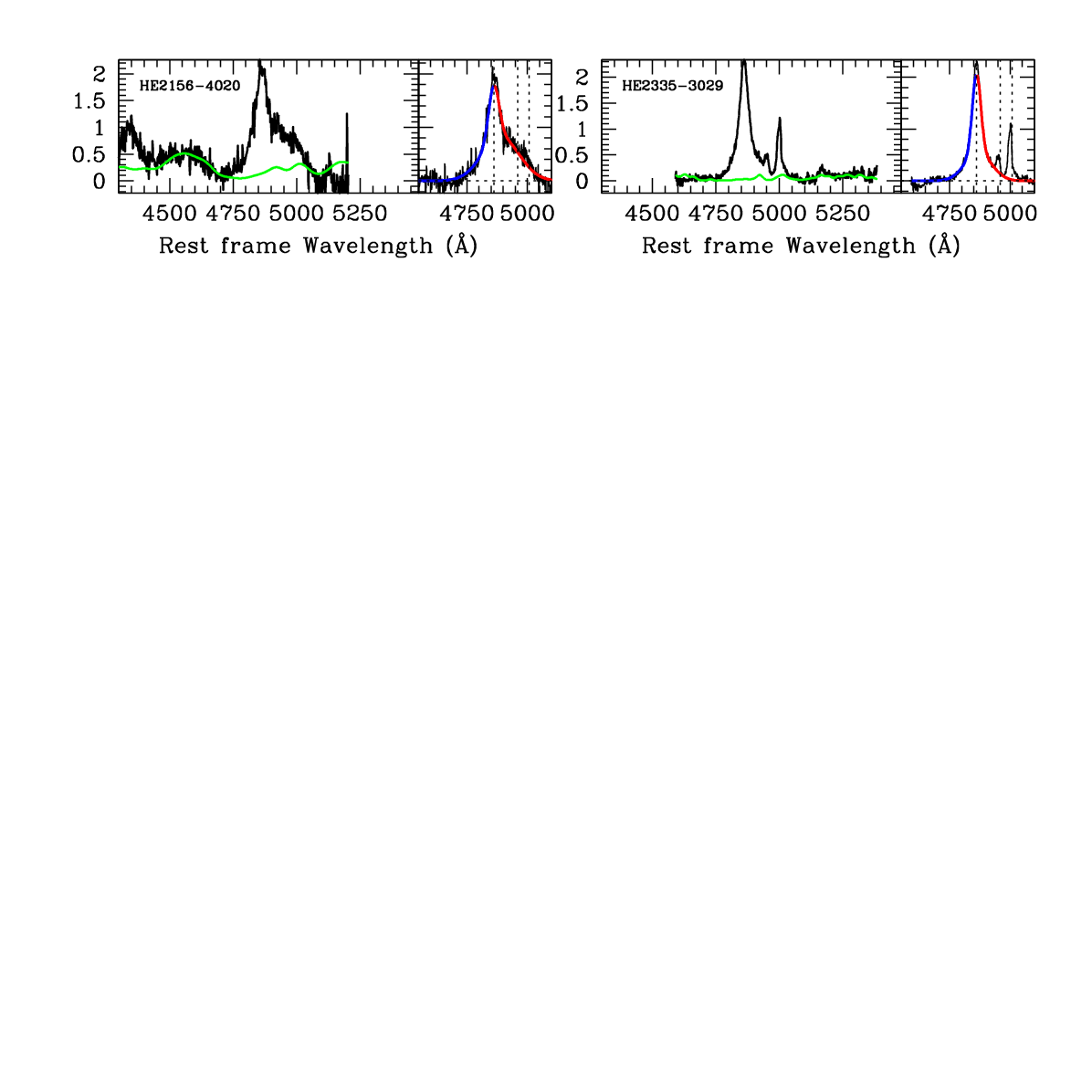}
%\vspace{22cm}
\caption[]{Cont.}
\end{figure*}

\break
\newpage
\vfill
\break
\newpage

%\addtocounter{figure}{2}
\begin{figure*}

\includegraphics[width=9.0cm,height=9.0cm, angle=0]{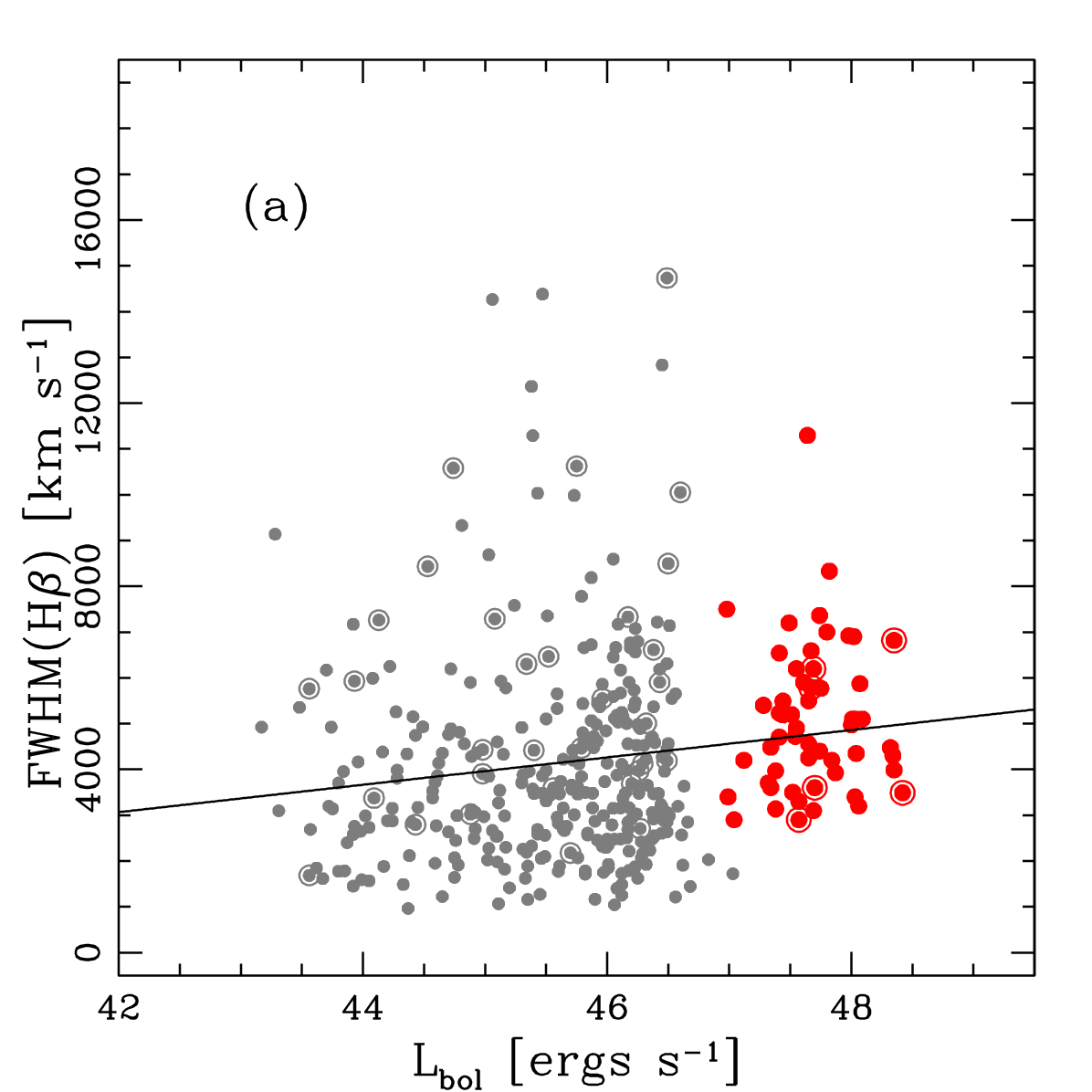}
\includegraphics[width=9.0cm,height=9.0cm, angle=0]{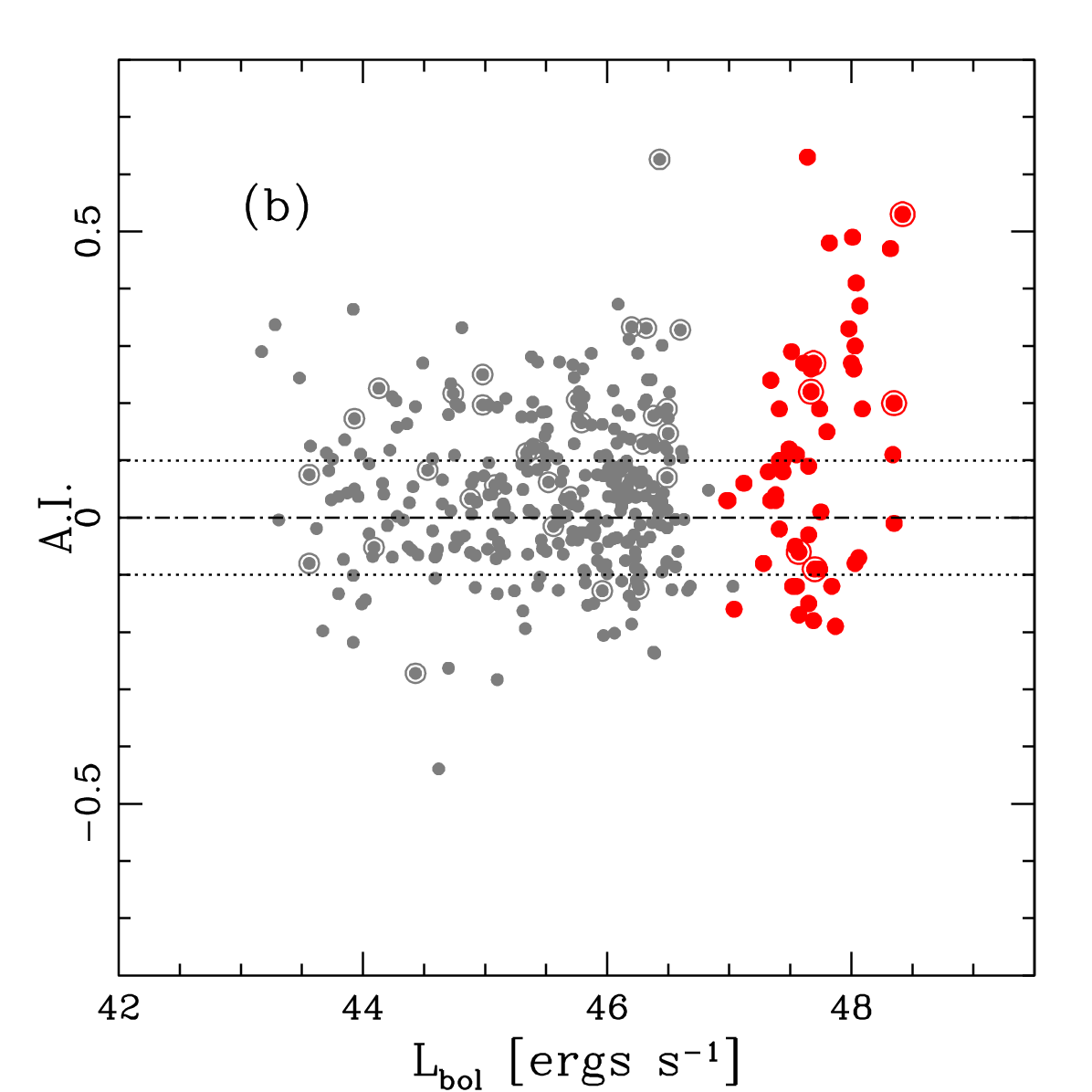}
\includegraphics[width=9.0cm,height=9.0cm, angle=0]{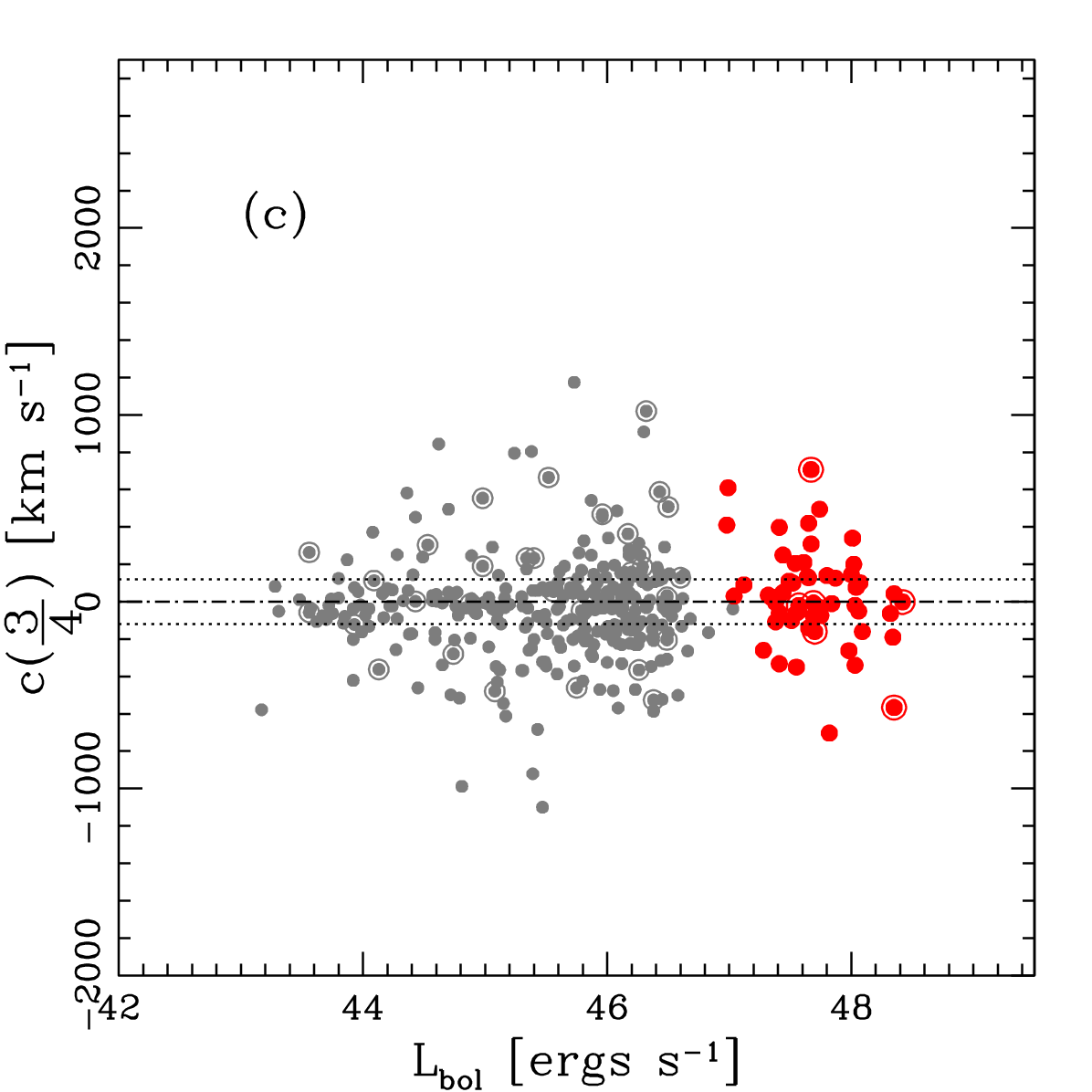}
\includegraphics[width=9.0cm,height=9.0cm, angle=0]{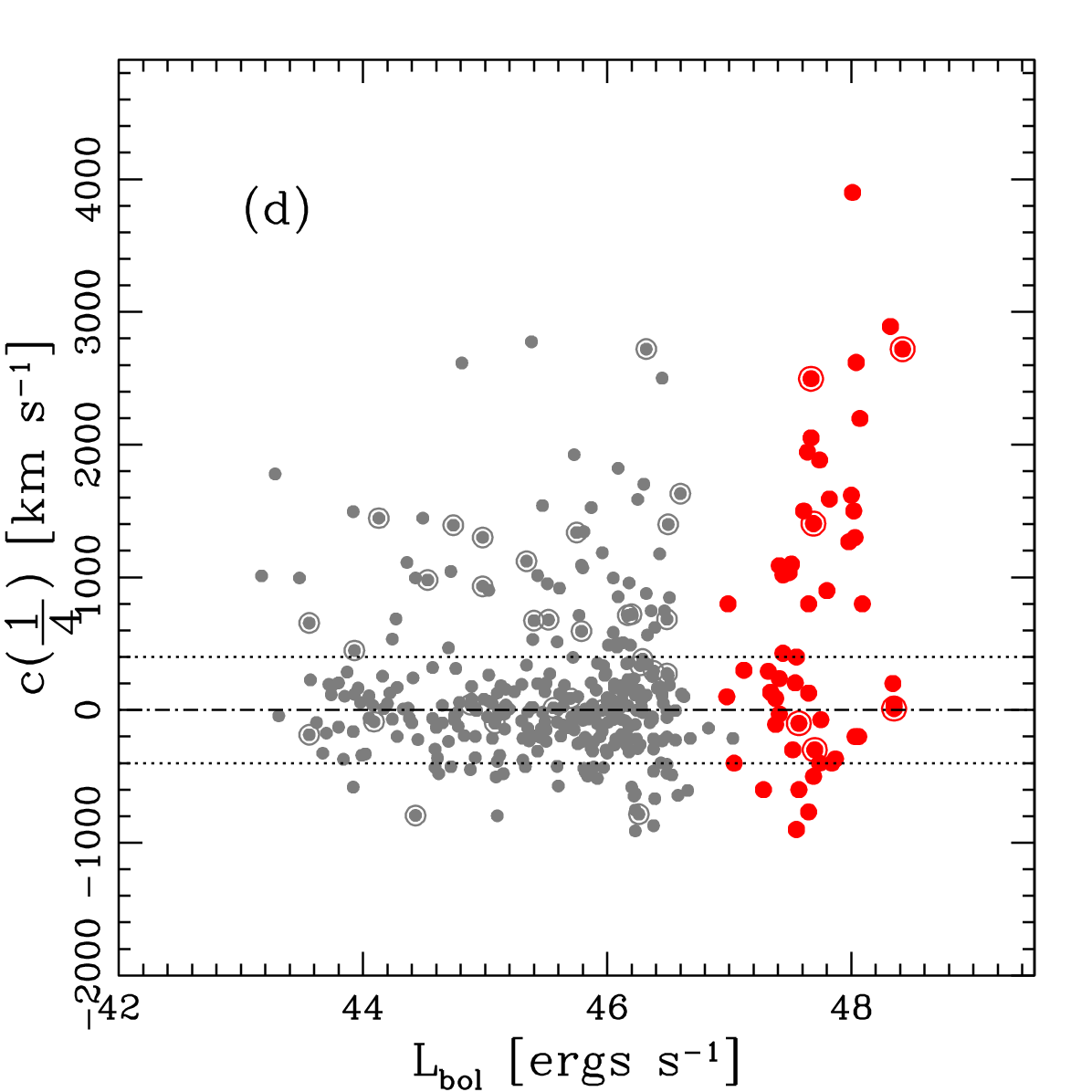}
%\vspace{22cm}
\caption[]{FWHM(\hb) in \kms\ (a), Asymmetry Index (b), centroid at 3/4 fractional heights, in \kms\ (c), and centroid at 1/4 fractional height, in \kms\ (d) as a function of bolometric luminosity in units of \ergss. Small gray circles: data from the SDSS sample of \citet{zamfiretal08}; larger filled circles: ISAAC sample. Circled data points identify RL sources. In the upper left panel, the continuous line shows an unweighted lsq best fit. The dotted lines identify the $\pm 2 \sigma$ confidence limit for A.I., and centroids equal to zero.} \label{fig:lumeff}
\end{figure*}

\begin{figure*}
\begin{center}
\includegraphics[width=9cm,height=9cm, angle=0]{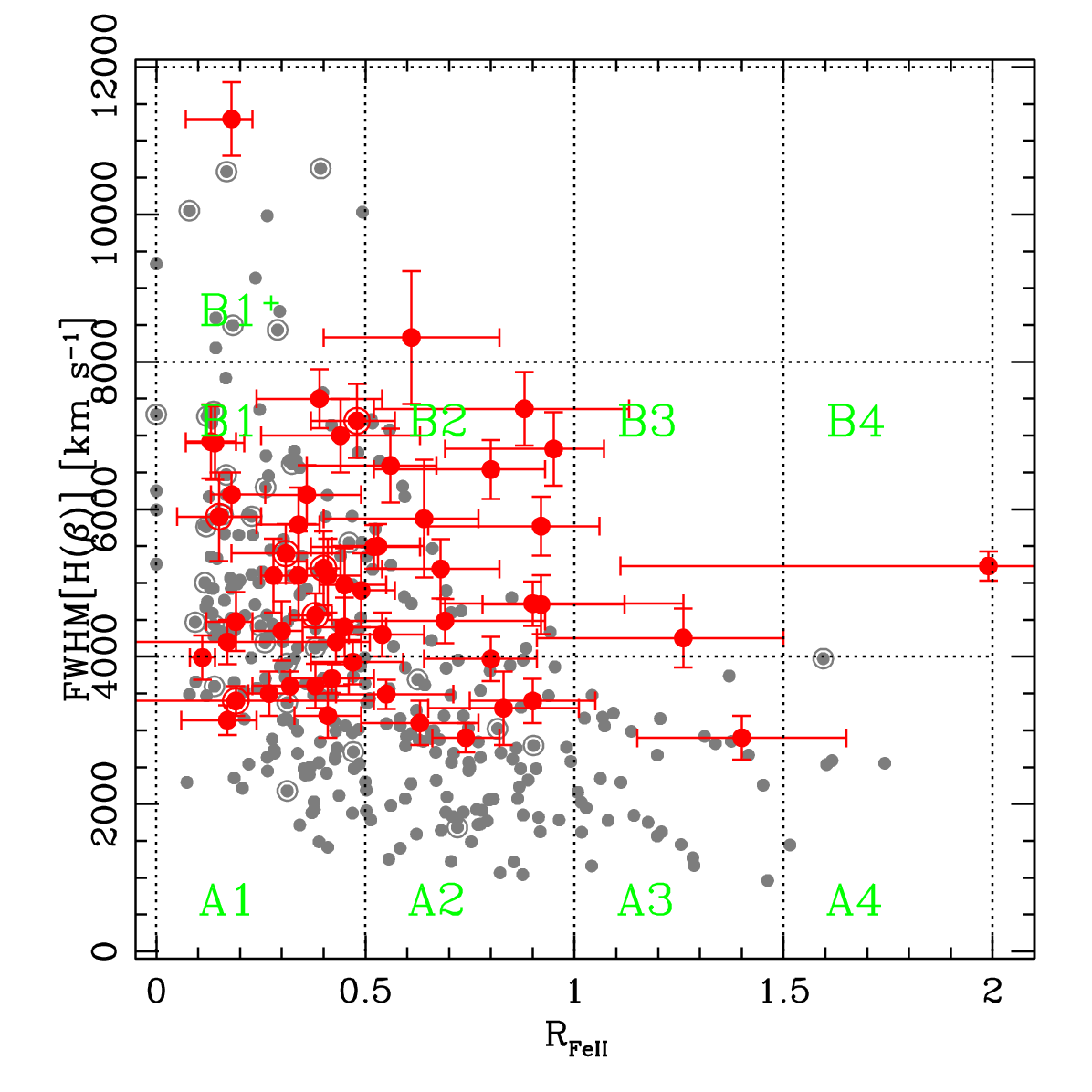}
\end{center}
%\vspace{22cm}
\caption[]{The optical plane of Eigenvector 1 with the VLT sources presented in this paper and in Paper I and II (large red filled circles), and the SDSS sample of \citet[][small grey circles]{zamfiretal08}). Circled data points identify RL sources. Abscissa is \feiiq\ prominence \rfe, ordinate is FWHM(\hb) in \kms. The plane is subdivided in spectral types following \citet{sulenticetal02}. } \label{fig:4de1}
\end{figure*}
\break
\newpage

%\begin{figure*}
%\includegraphics[width=9.0cm,height=9.0cm, angle=0]{fwhmai.eps}
%\includegraphics[width=9.0cm,height=9.0cm, angle=0]{fwhmc14.eps}
%\vspace{22cm}
%\caption[]{Asymmetry index (upper panel) and centroid at 1/4 intensity in \kms\ as a function of FWHM in \kms. The vertical dot-dashed line defines the limit at FWHM~$\approx 4000$~\kms\ that separates Pop.~A and B at low $z$. Filled squares indicate sources belonging to Modified Pop.~A (see text for details). } \label{fig:fwhm}
%\end{figure*}

\begin{figure*}
\includegraphics[angle=0,scale=0.5]{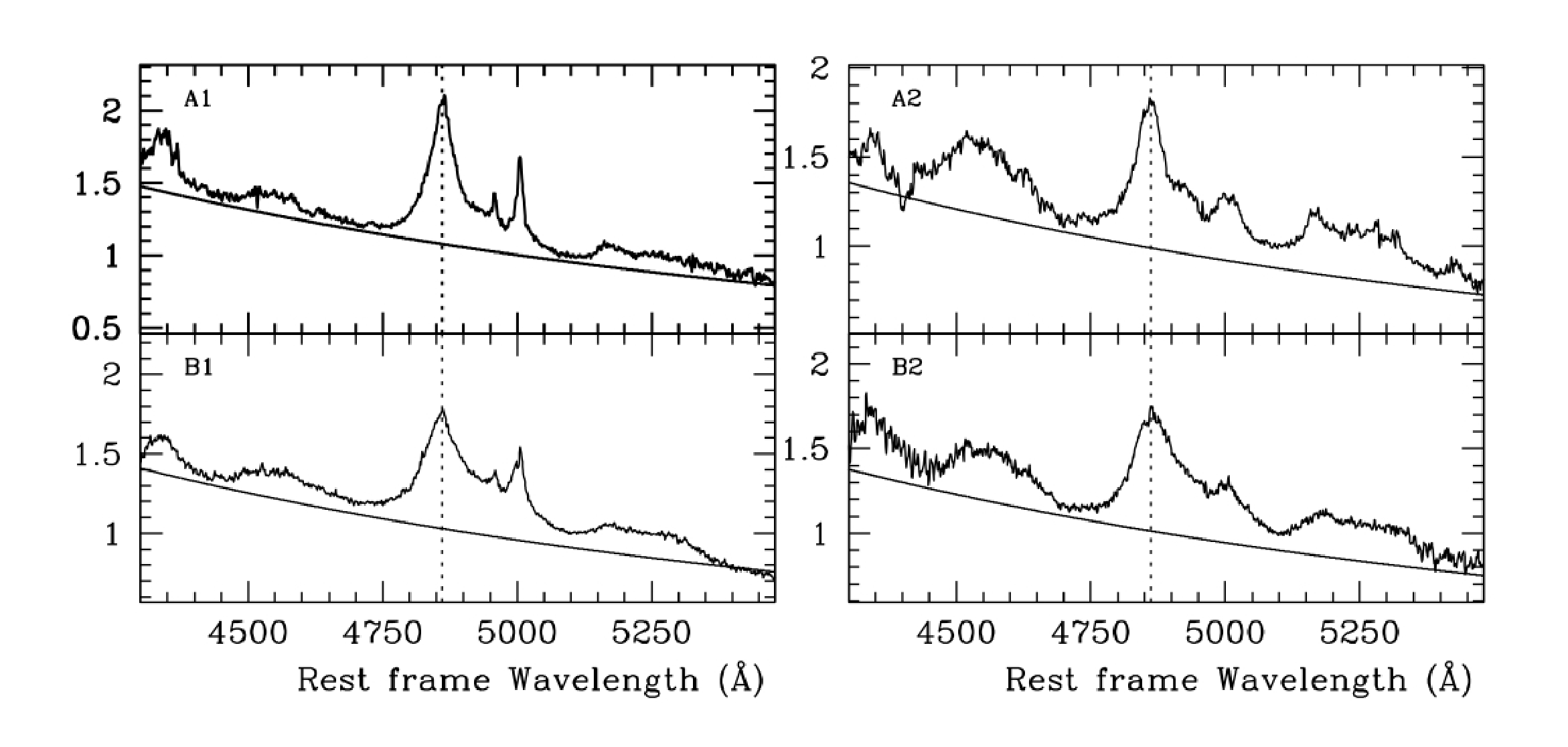}
%\includegraphics[scale=.45,angle=0]{a1s.eps}
%\includegraphics[width=7,height=.7, angle=0]{m.eps}
%\includegraphics[width=5.9cm,height=9cm, angle=0]{bm.eps}
%\vspace{22cm}
\caption[]{Median spectra for bins A1, B1, A2, B2, as defined in Fig. \ref{fig:4de1}. Original spectral have been normalized to unity at $\lambda = $ 5100 \AA. The adopted continuum is shown as a solid line. } \label{fig:4bin}
\end{figure*}

\begin{figure*}
\includegraphics[width=9.0cm,height=9.0cm, angle=0]{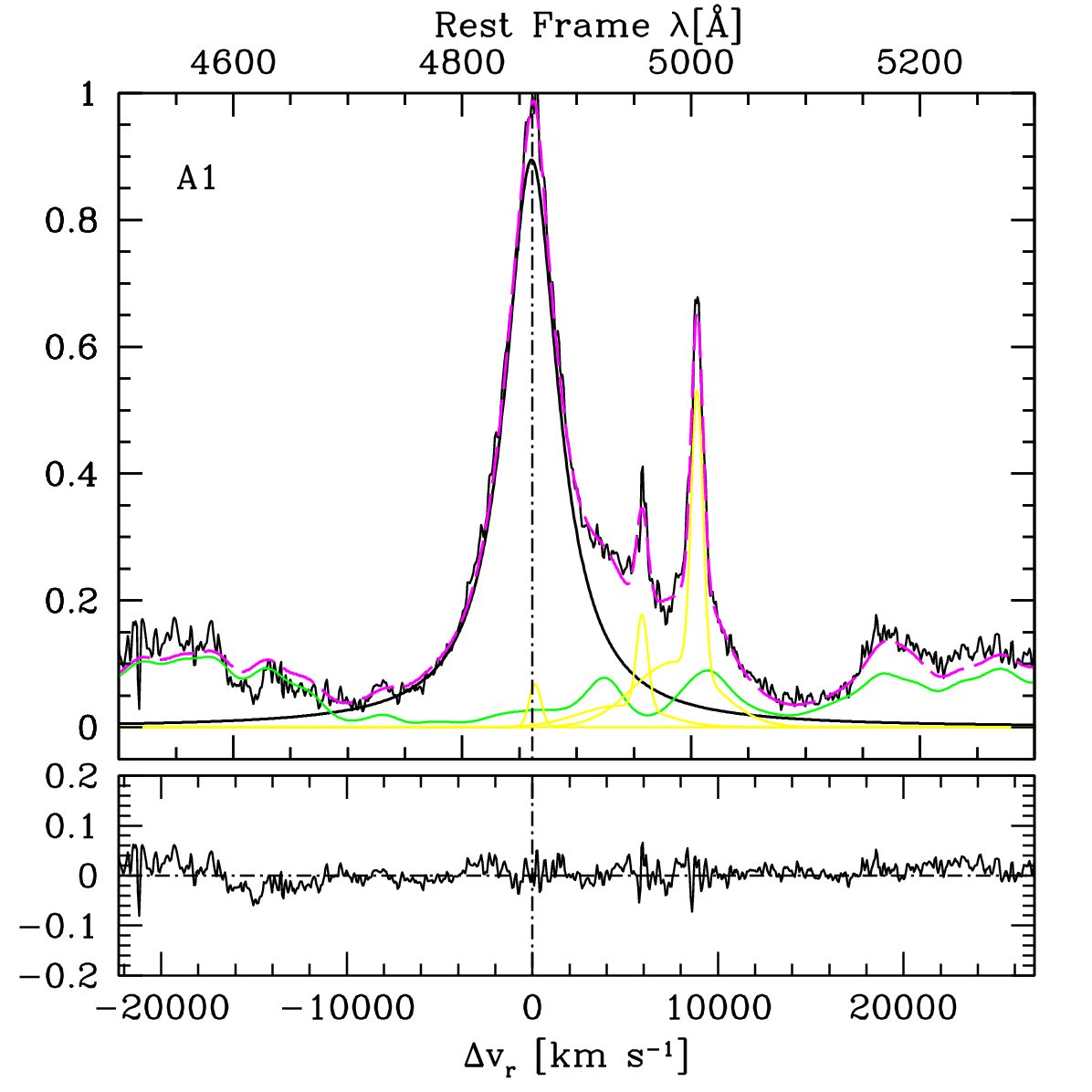}
\includegraphics[width=9.0cm,height=9.0cm, angle=0]{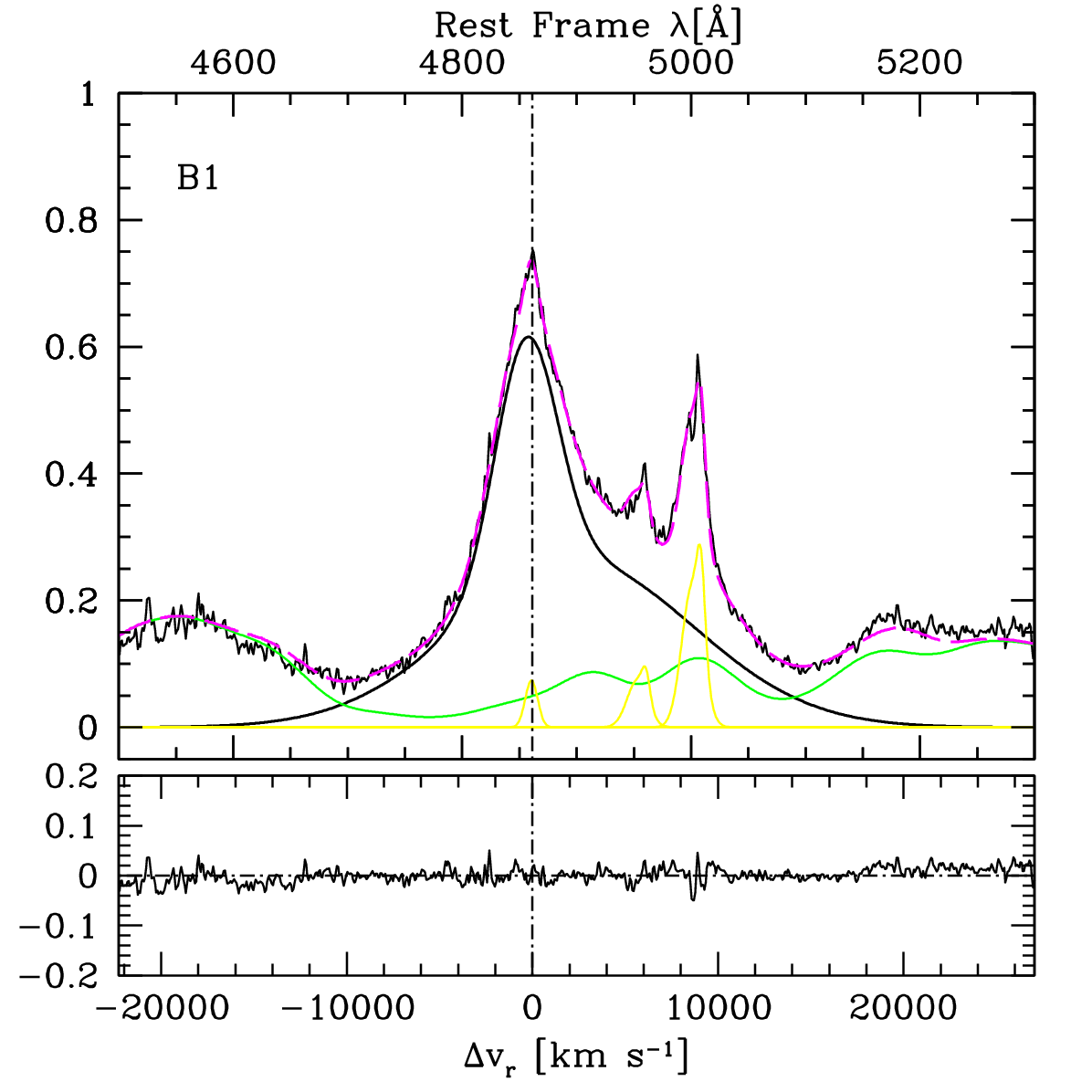}
\includegraphics[width=9.0cm,height=9.0cm, angle=0]{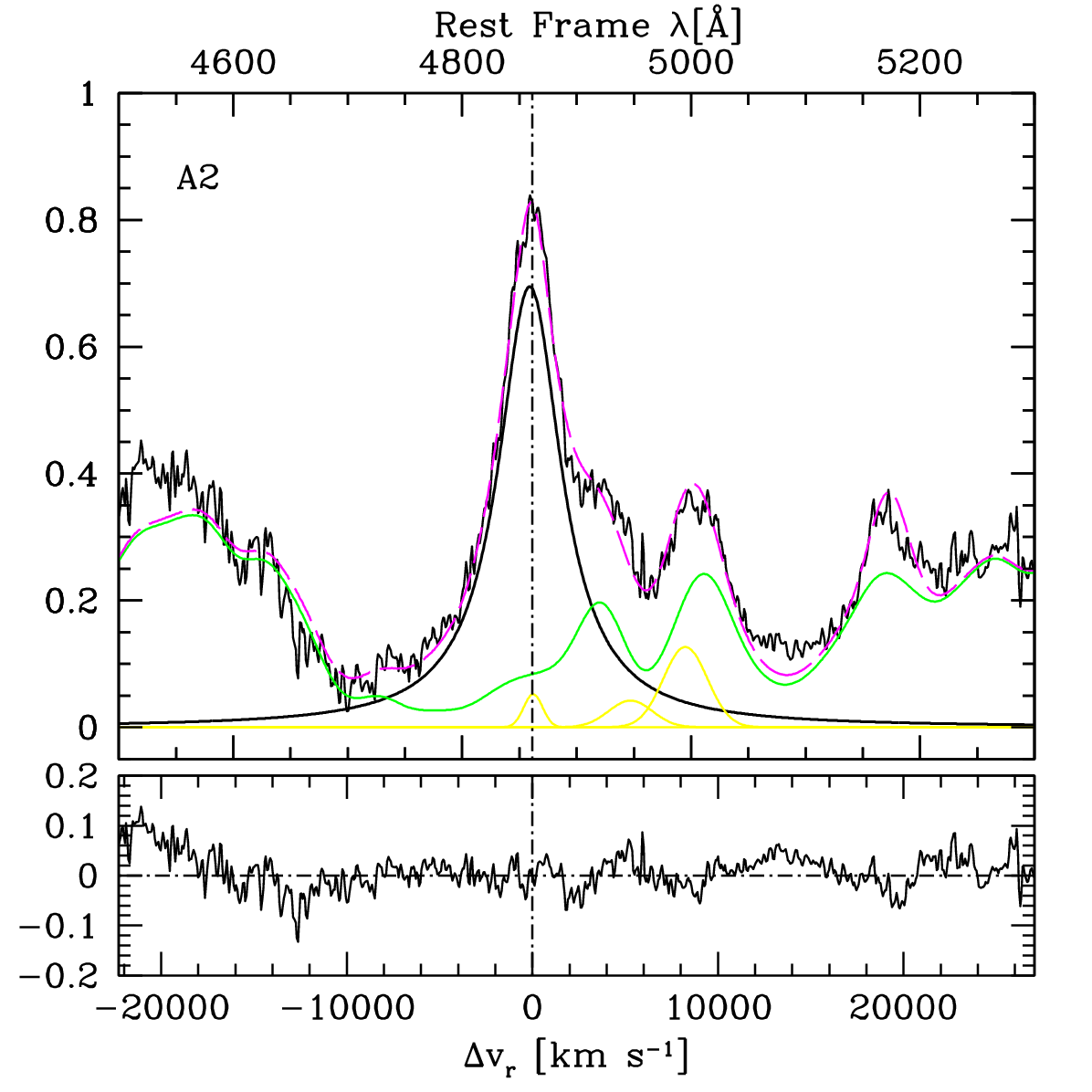}
\includegraphics[width=9.0cm,height=9.0cm, angle=0]{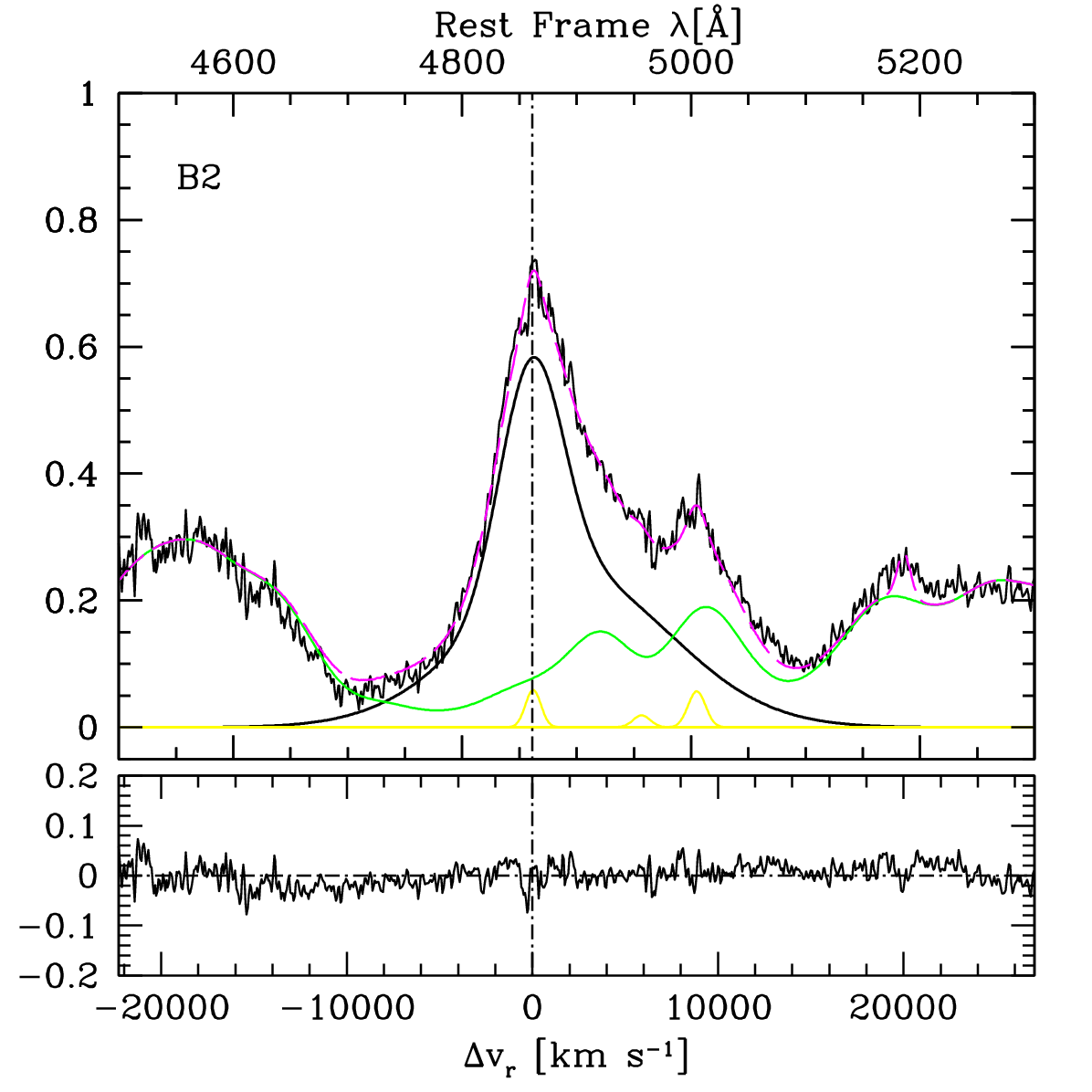}
%\vspace{22cm}
\caption[]{Analysis of the \hb\ profiles for spectral types A1, A2, B1, B2. Median spectra are shown after continuum subtraction (thin line); the model fit obtained with {\tt specfit} is shown as a thick dashed magenta line. The \hb\ profile is represented either by a Lorentzian function or by the sum of two Gaussians (not shown individually; thick lines). The thin green line shows the \feiiopt\ contributions, and the yellow lines the \oiiiopt\ and \hbnc\ emission. Residuals of the fitting procedure are displayed in the window immediately below the one showing the spectral components. Abscissa scale  is rest frame wavelength in \AA\ (top), and radial velocity in \kms\ from rest wavelength of \hb\ (bottom). } \label{fig:4binan}
\end{figure*}

\begin{figure*}
\begin{center}\includegraphics[scale=0.8,angle=0]{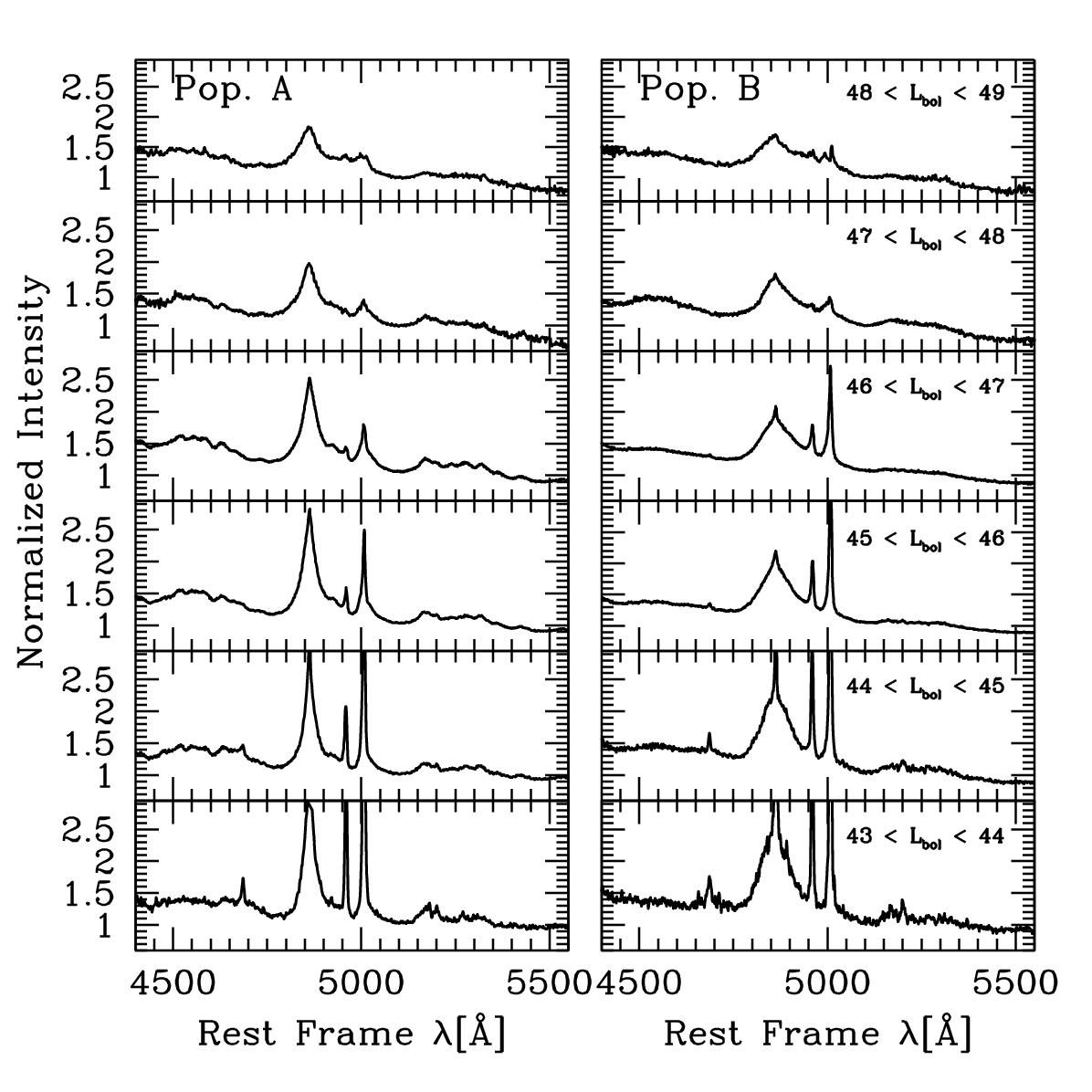}
\end{center}
%\vspace{22cm}
\caption[]{Luminosity effects in the median spectra of Pop.~A and Pop.~B sources, binned in decades of luminosity in the  luminosity ranges 43 $< \log L_{\mathrm bol} < 49$.  Abscissa is rest frame wavelength in \AA, ordinate is normalized intensity (set to unity at $\lambda = 5100$ \AA).  } \label{fig:lbin}
\end{figure*}
\vfill\clearpage

%\vfill\clearpage

\begin{figure*}
\includegraphics[width=9.0cm,height=8.0cm, angle=0]{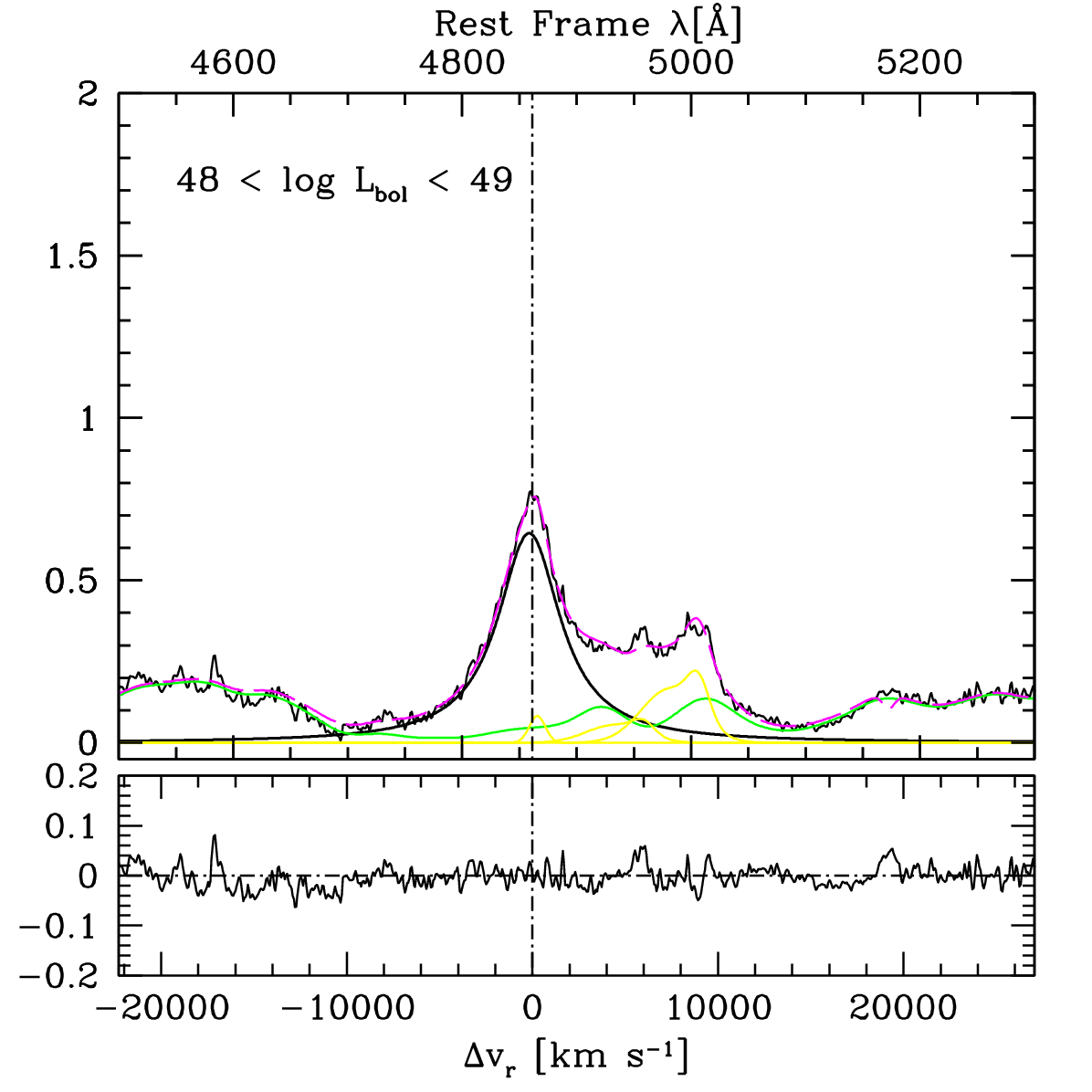}
\includegraphics[width=9.0cm,height=8.0cm, angle=0]{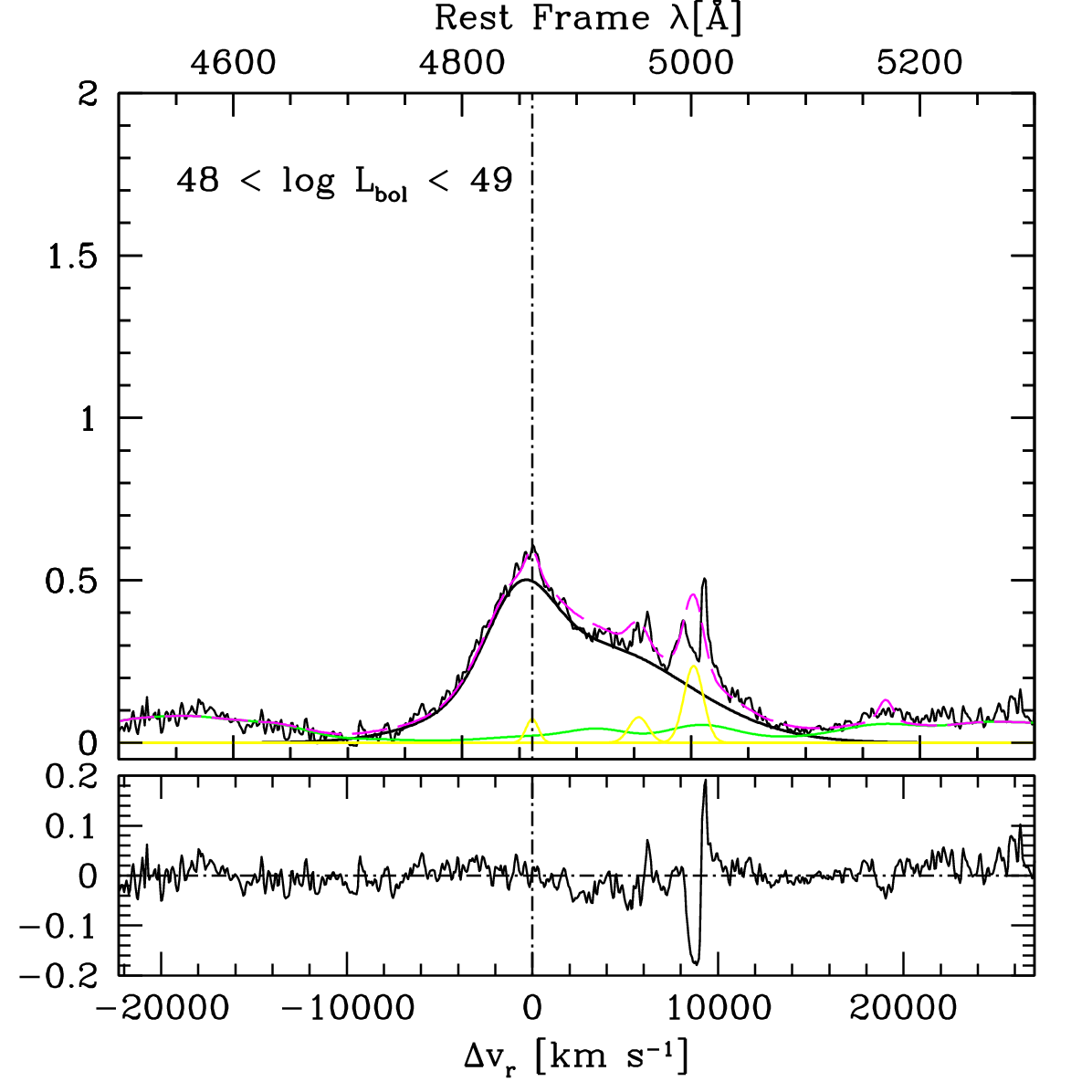}
\includegraphics[width=9.0cm,height=8.0cm, angle=0]{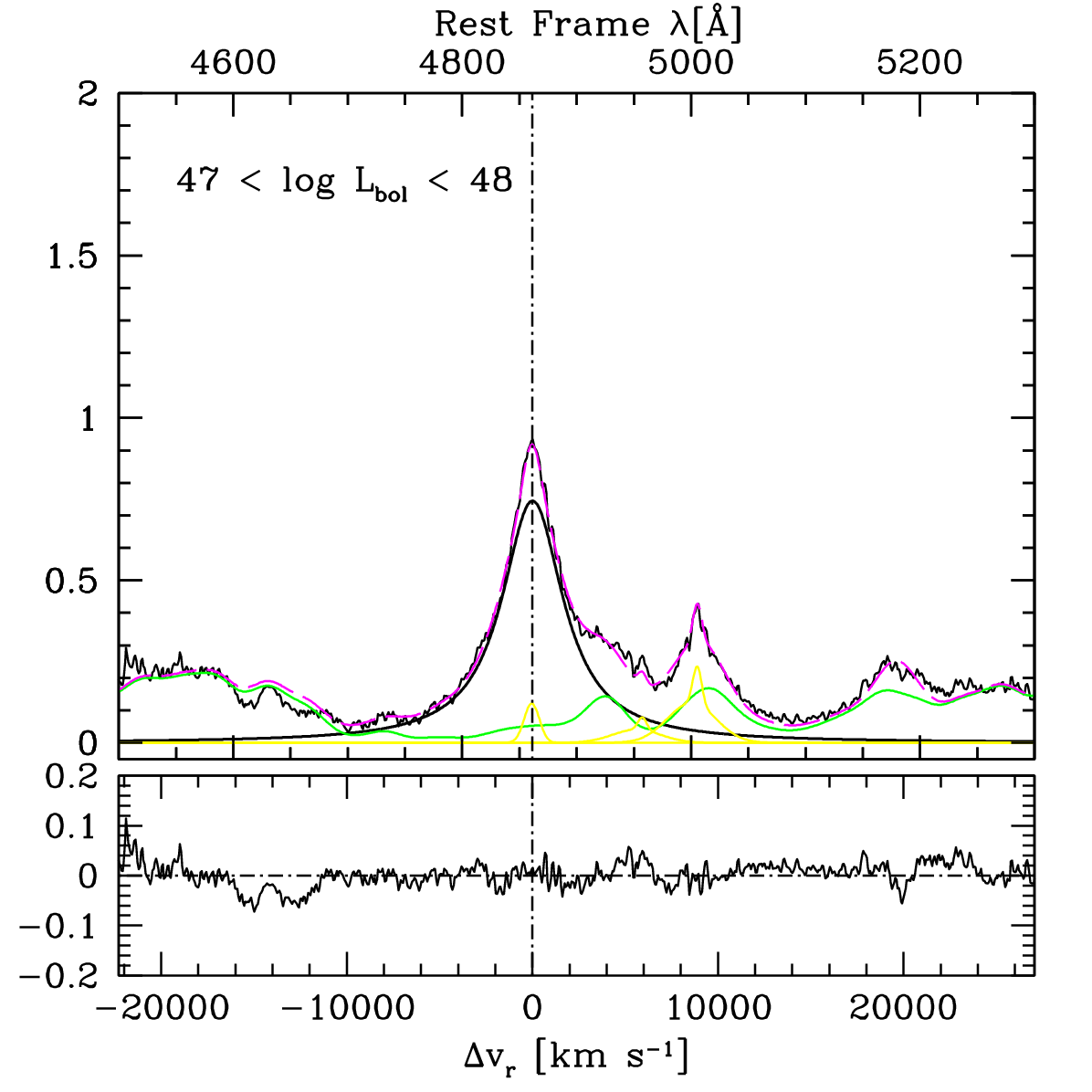}
\includegraphics[width=9.0cm,height=8.0cm, angle=0]{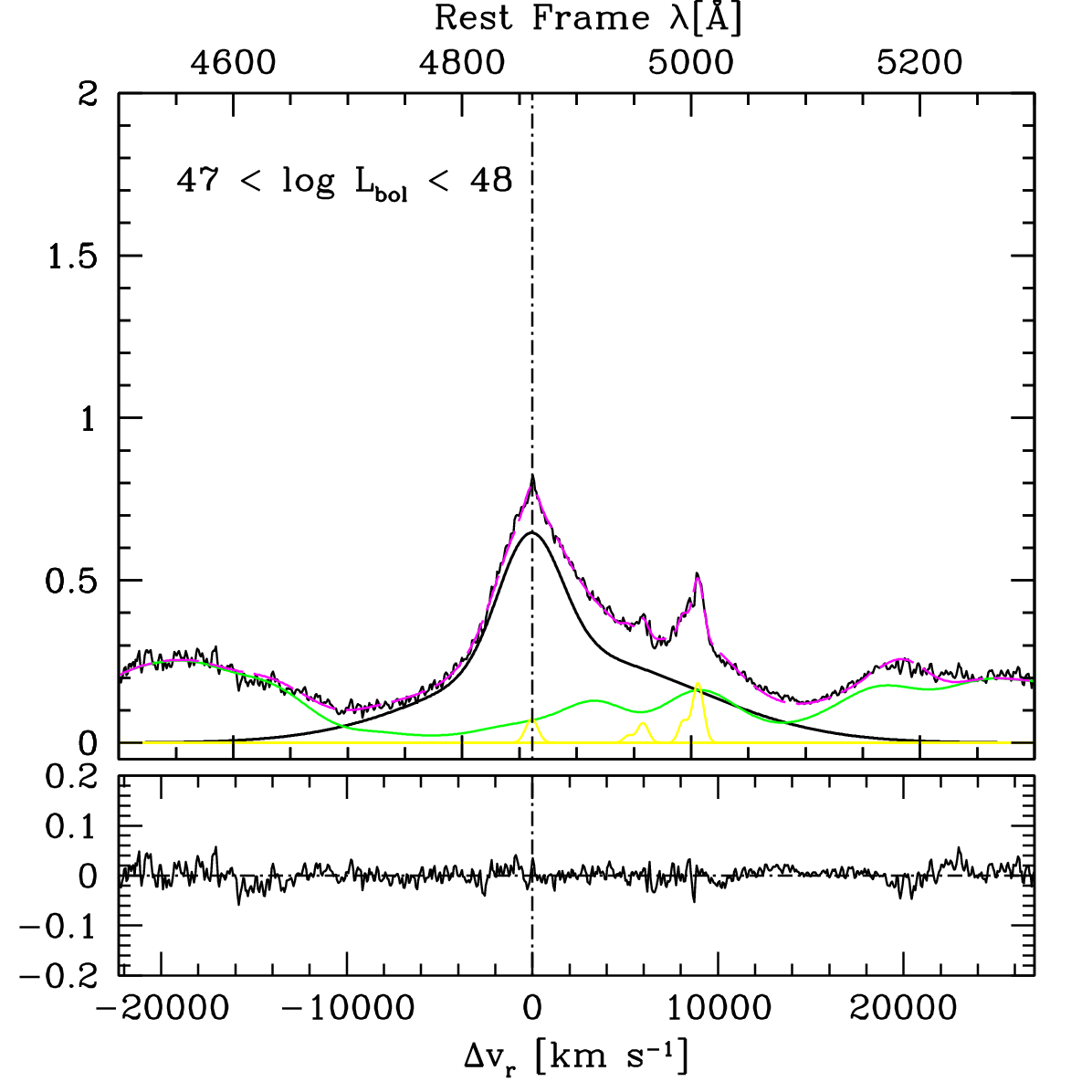}
\includegraphics[width=9.0cm,height=8.0cm, angle=0]{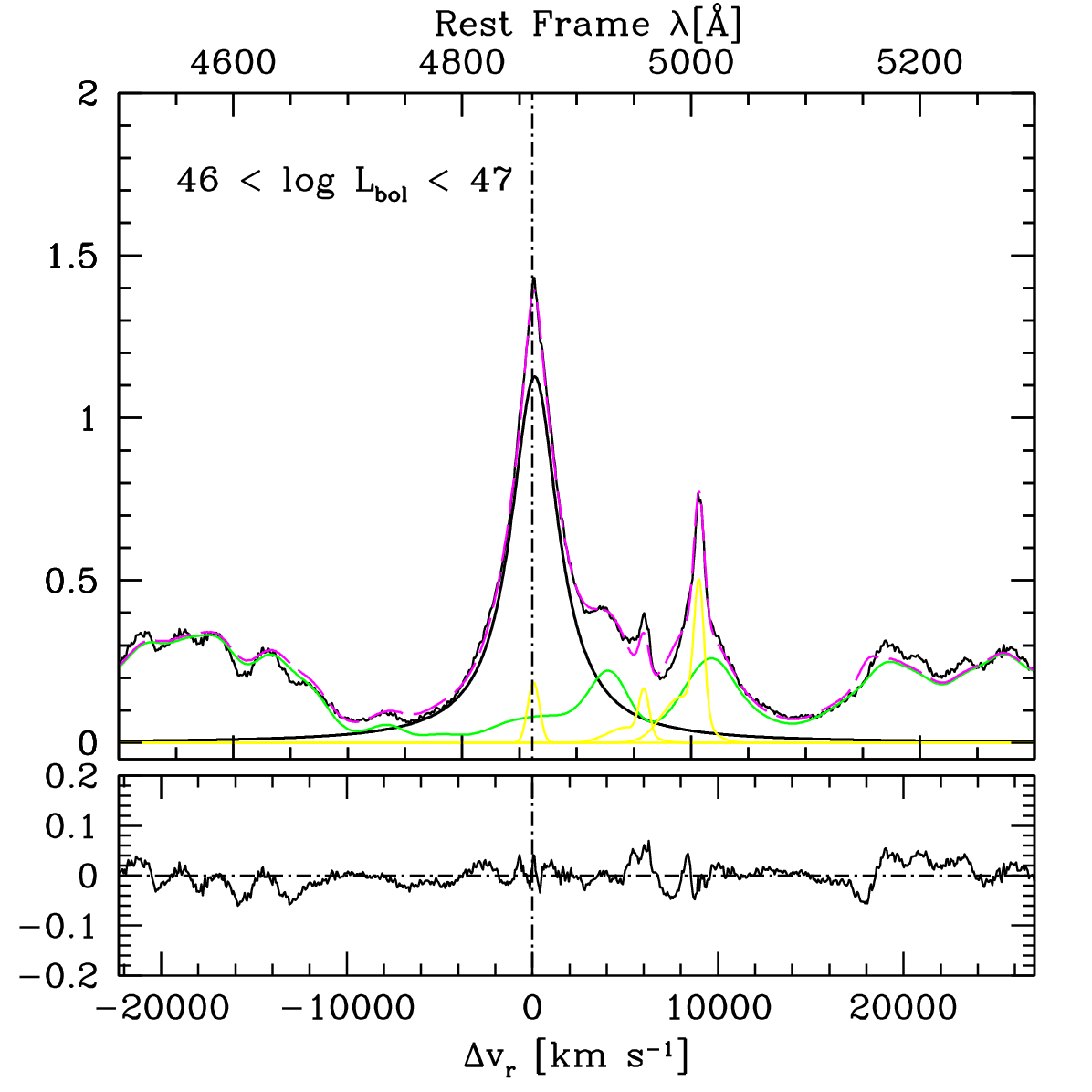}
\includegraphics[width=9.0cm,height=8.0cm, angle=0]{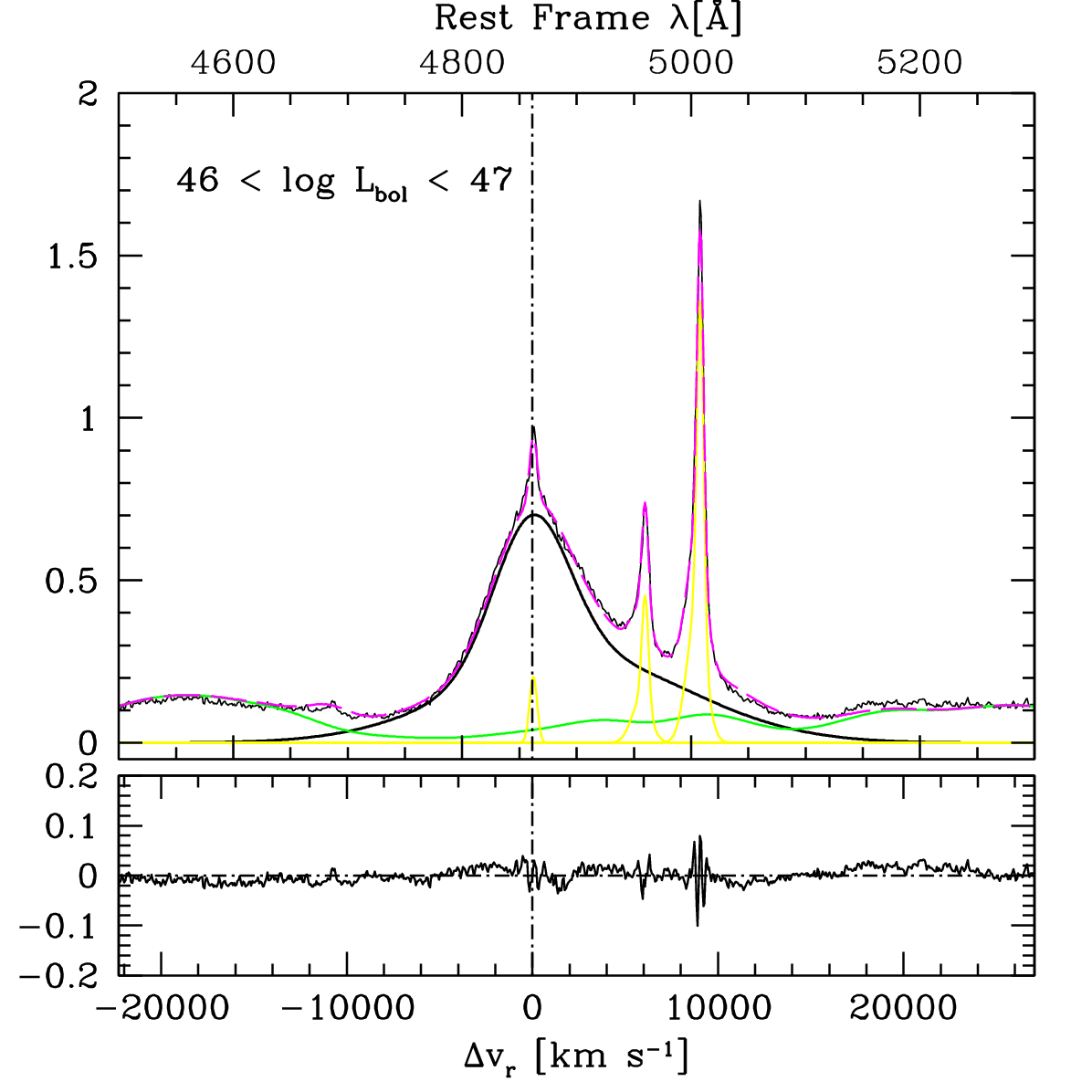}
%\vspace{22cm}
\caption[]{Median spectra analysis of the \hb\ profiles as a function of luminosity, for Pop. A and B (right panels).  The  median spectra have been continuum-subtracted. Lines follow the same coding of Fig. \ref{fig:4binan}. Residuals of the fitting procedure are displayed in the window immediately below the one showing the spectral components. } \label{fig:dcom}
\end{figure*}
\vfill\clearpage

\addtocounter{figure}{-1}
\begin{figure*}
\includegraphics[width=9.0cm,height=8.0cm, angle=0]{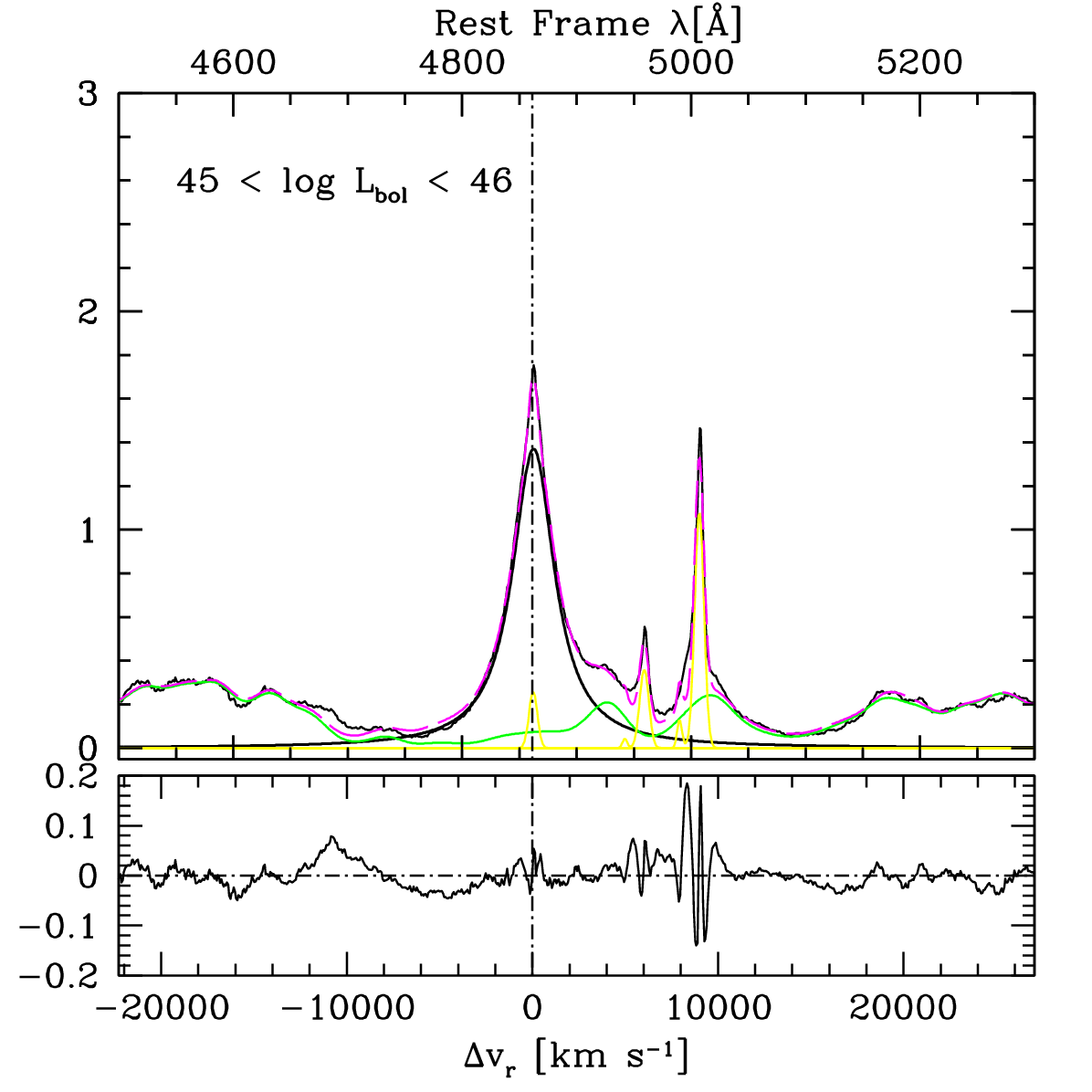}
\includegraphics[width=9.0cm,height=8.0cm, angle=0]{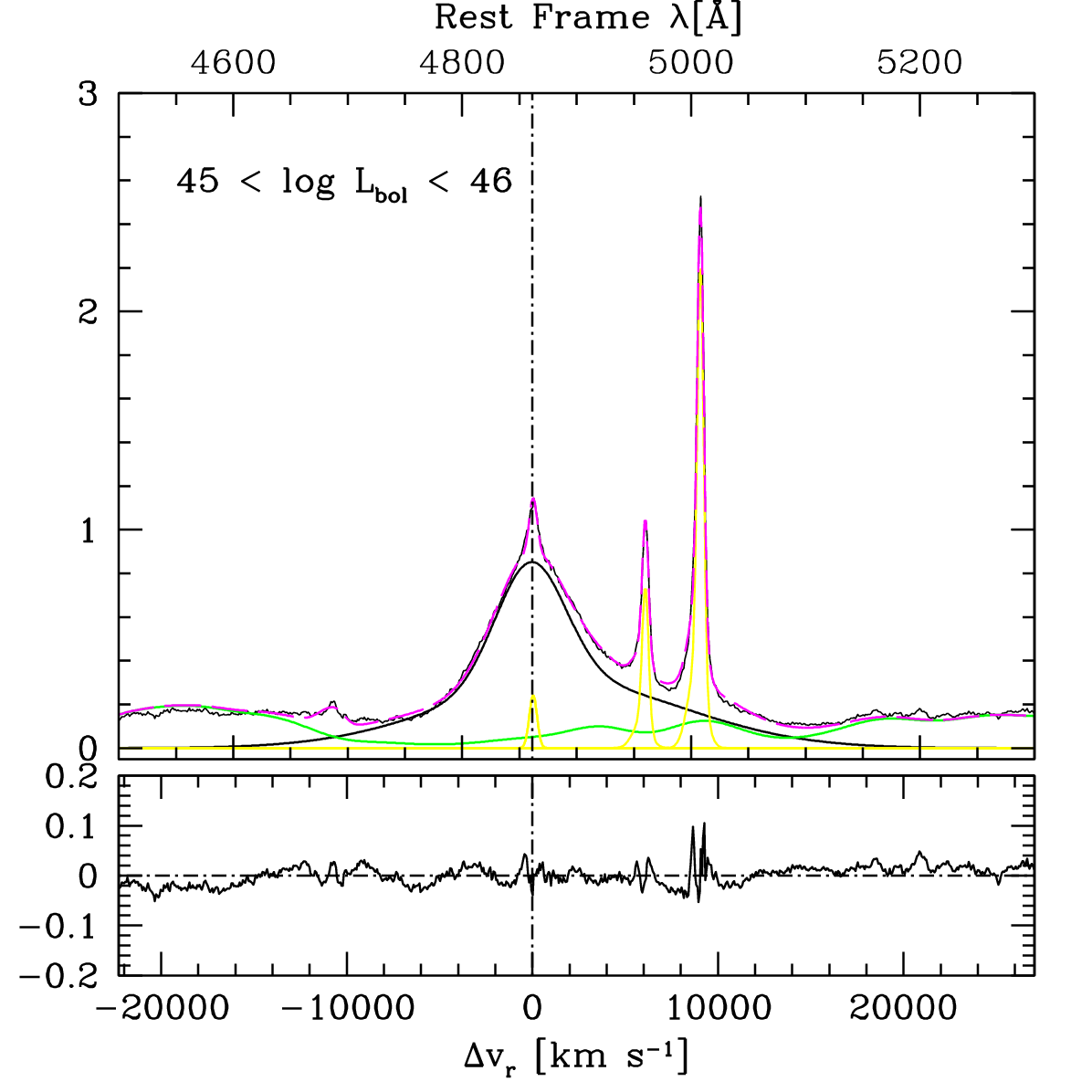}
\includegraphics[width=9.0cm,height=8.0cm, angle=0]{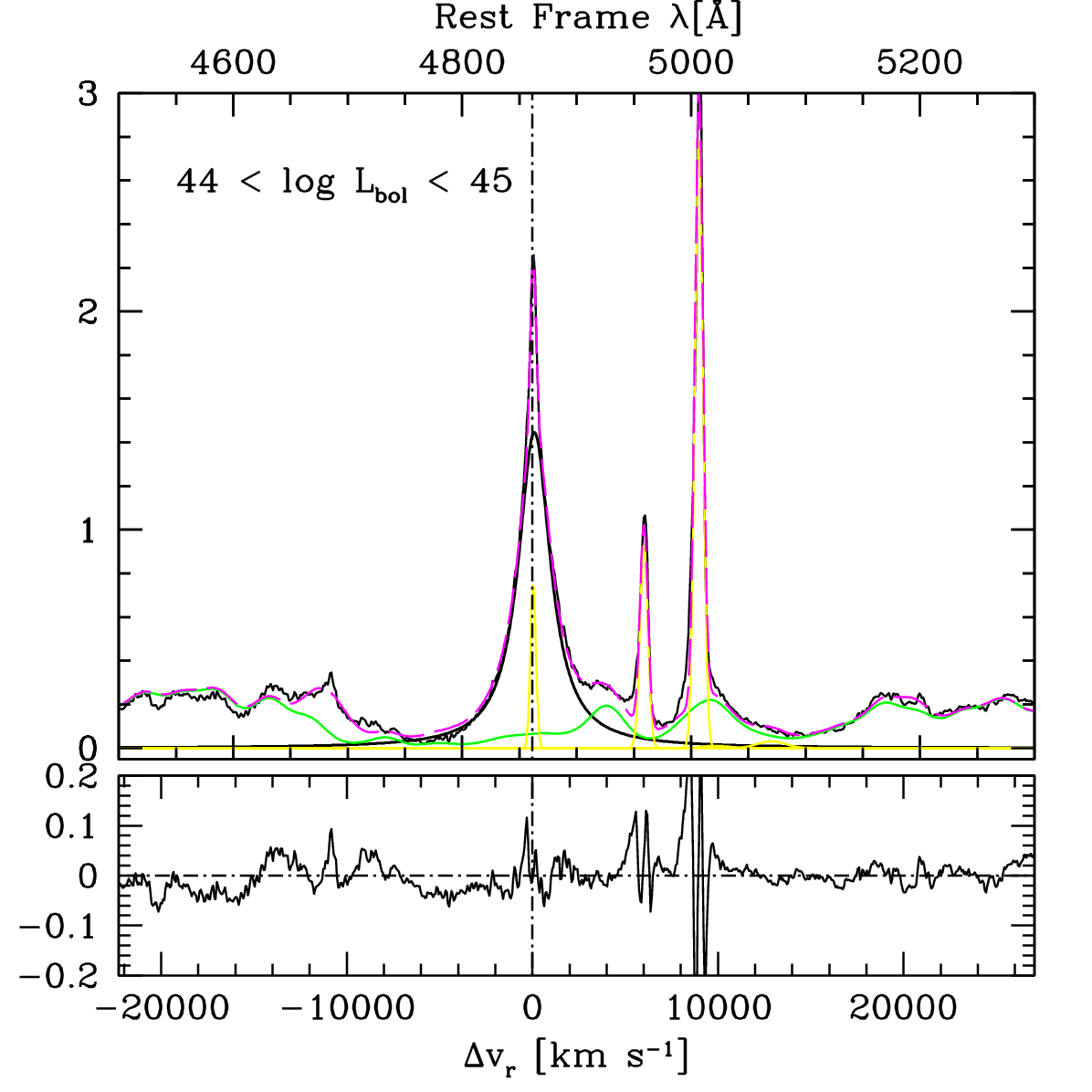}
\includegraphics[width=9.0cm,height=8.0cm, angle=0]{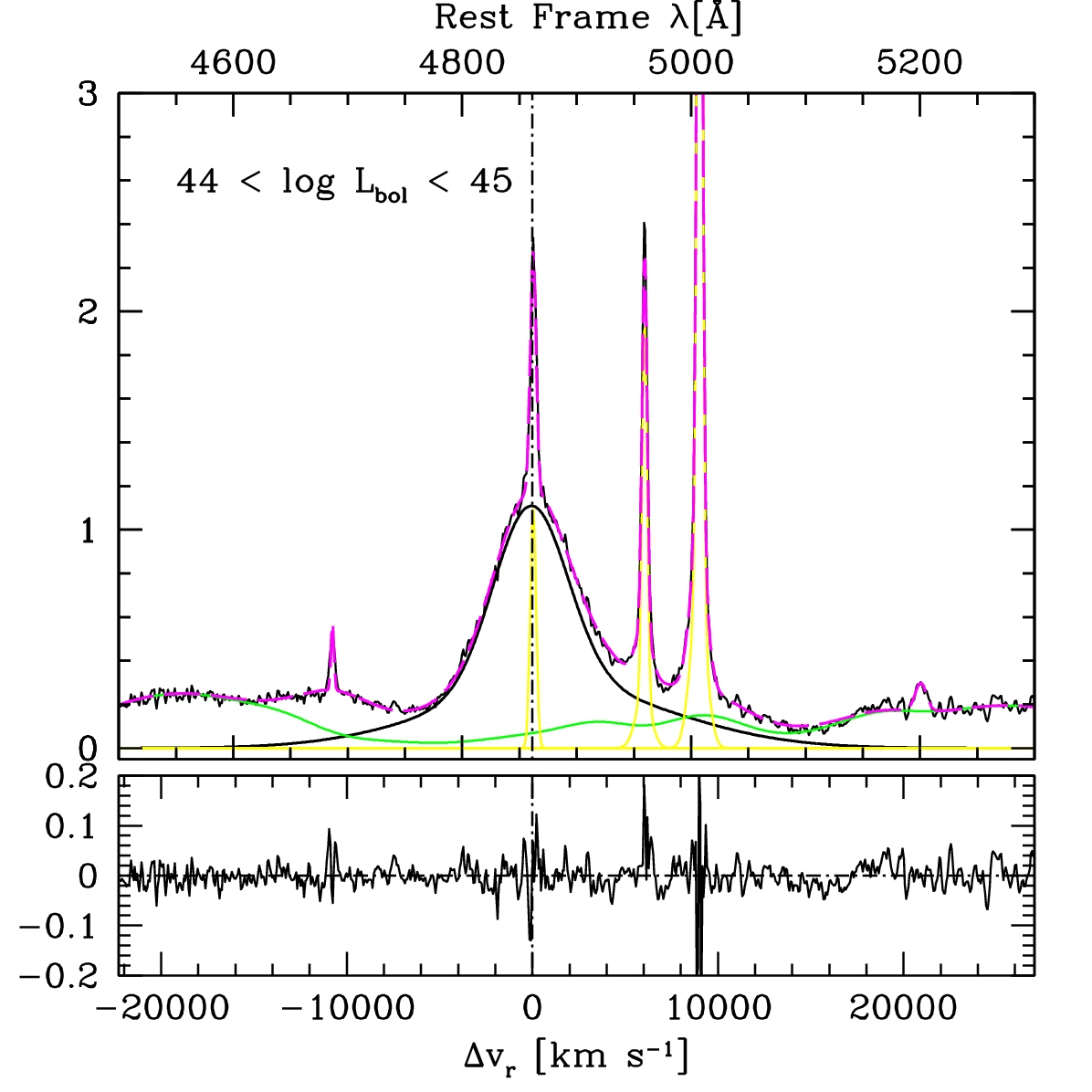}
\includegraphics[width=9.0cm,height=8.0cm, angle=0]{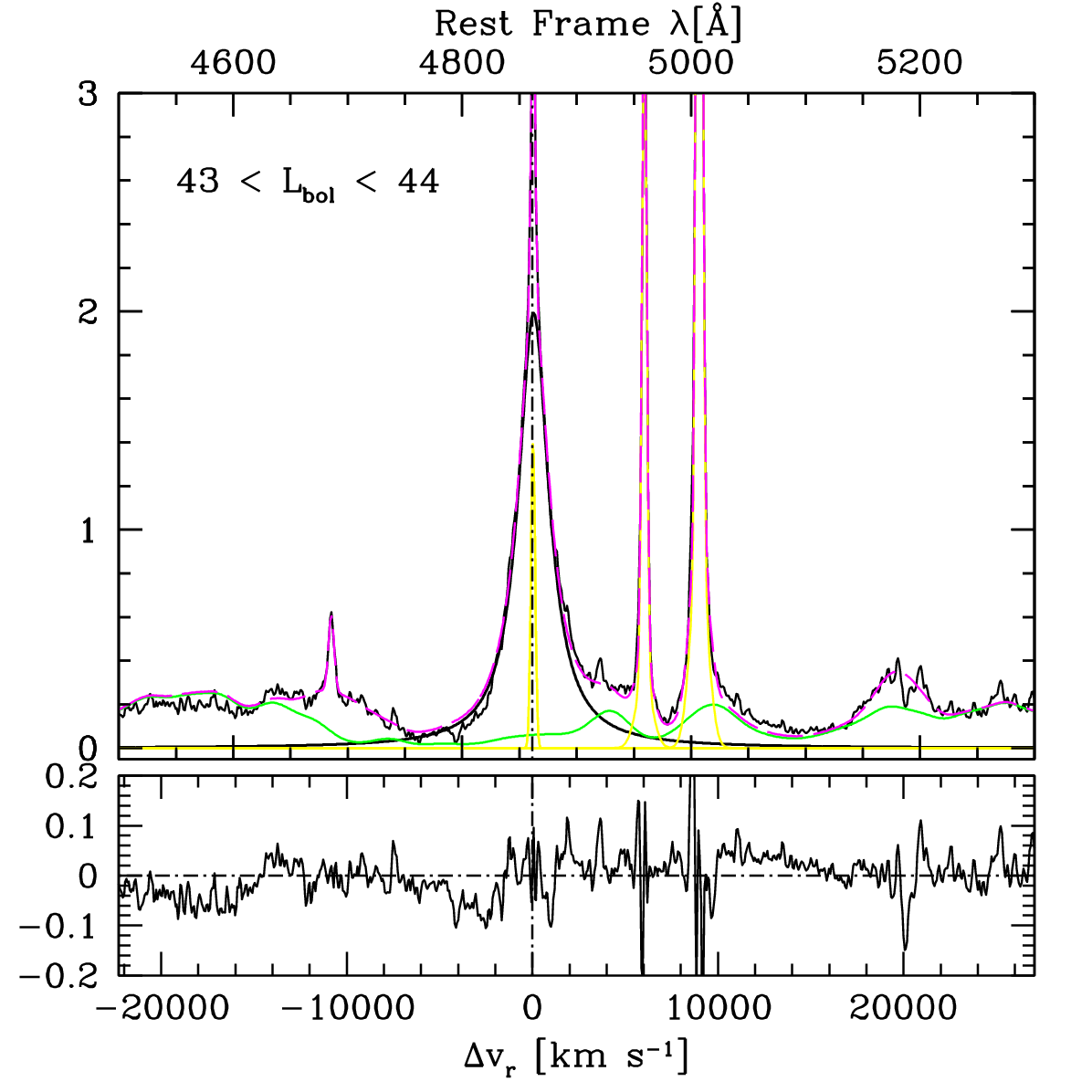}
\includegraphics[width=9.0cm,height=8.0cm, angle=0]{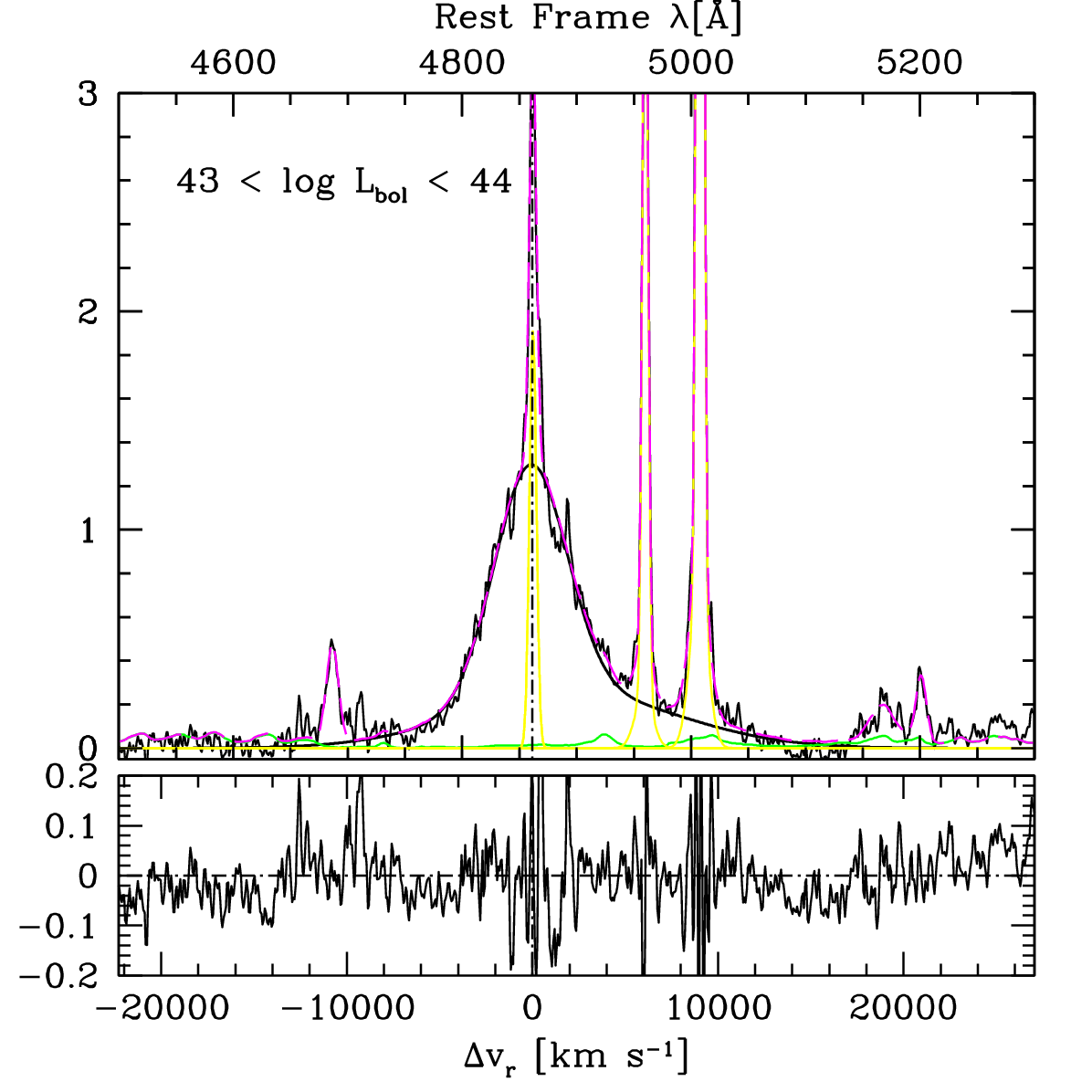}
%\vspace{22cm}
\caption[]{Cont. }
\end{figure*}

\begin{figure*}

\includegraphics[scale=0.3, angle=0]{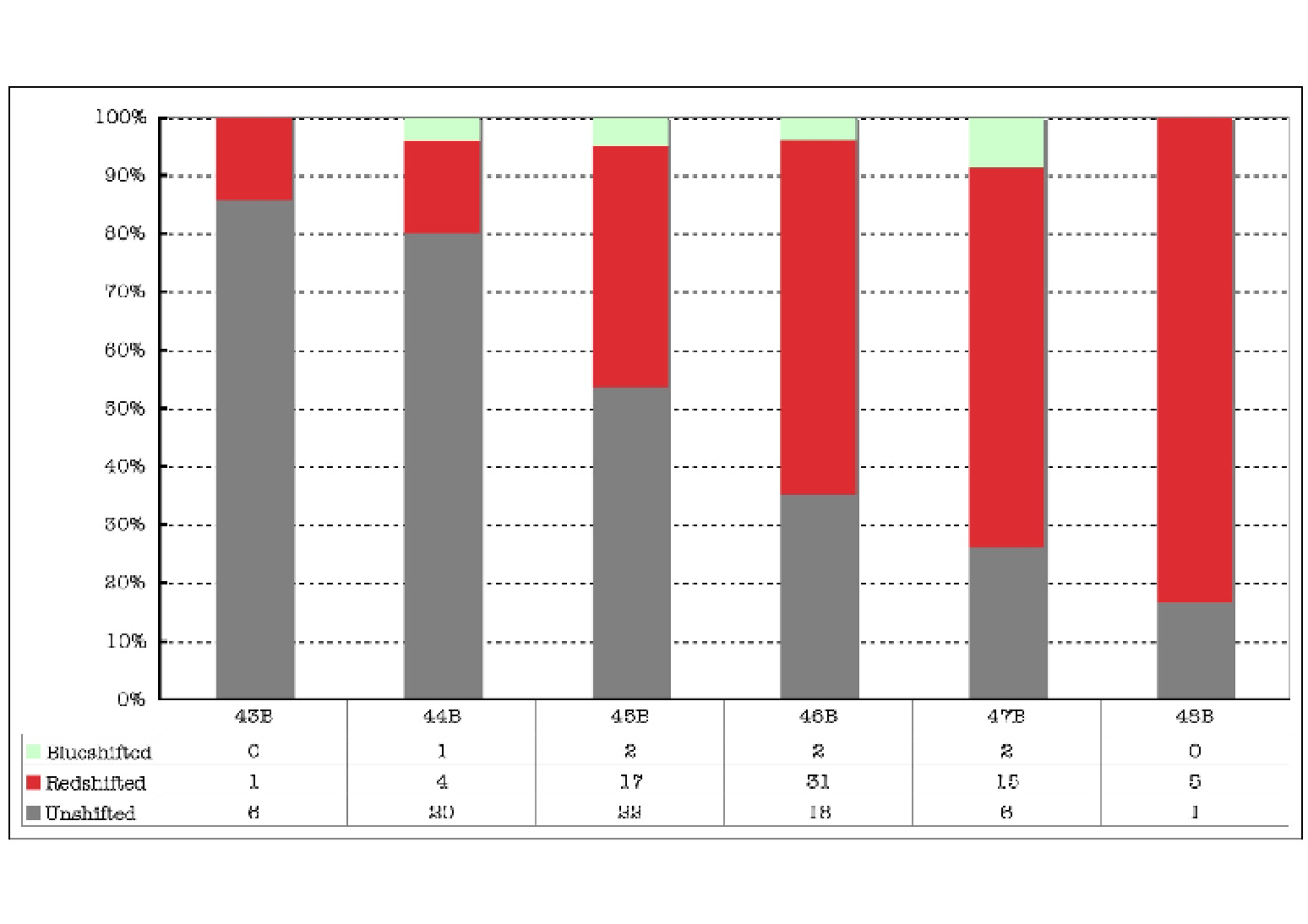}
%\vspace{22cm}
\caption[]{Distribution of unshifted and shifted \hb\ line centroids at 1/4 fractional intensity as a function of the bolometric luminosity ($\log L_\mathrm{bol}$, binned over $\Delta \log L_\mathrm{bol} = 1$, with the naming convention of Tables \ref{tab:decn}, \ref{tab:pro2}, \ref{tab:cen2}) for the Pop. B sources of the merged SDSS ISAAC sample.   Dark grey (bottom): unshifted; grey (middle): redshifted; pale grey (top): blueshifted. ``Unshifted'' means that the centroid measure is 0 within a $2\sigma$ confidence range of $\pm 400$ \kms.} \label{fig:hist}
\end{figure*}

\begin{figure*}
\includegraphics[width=9cm,height=9cm, angle=0]{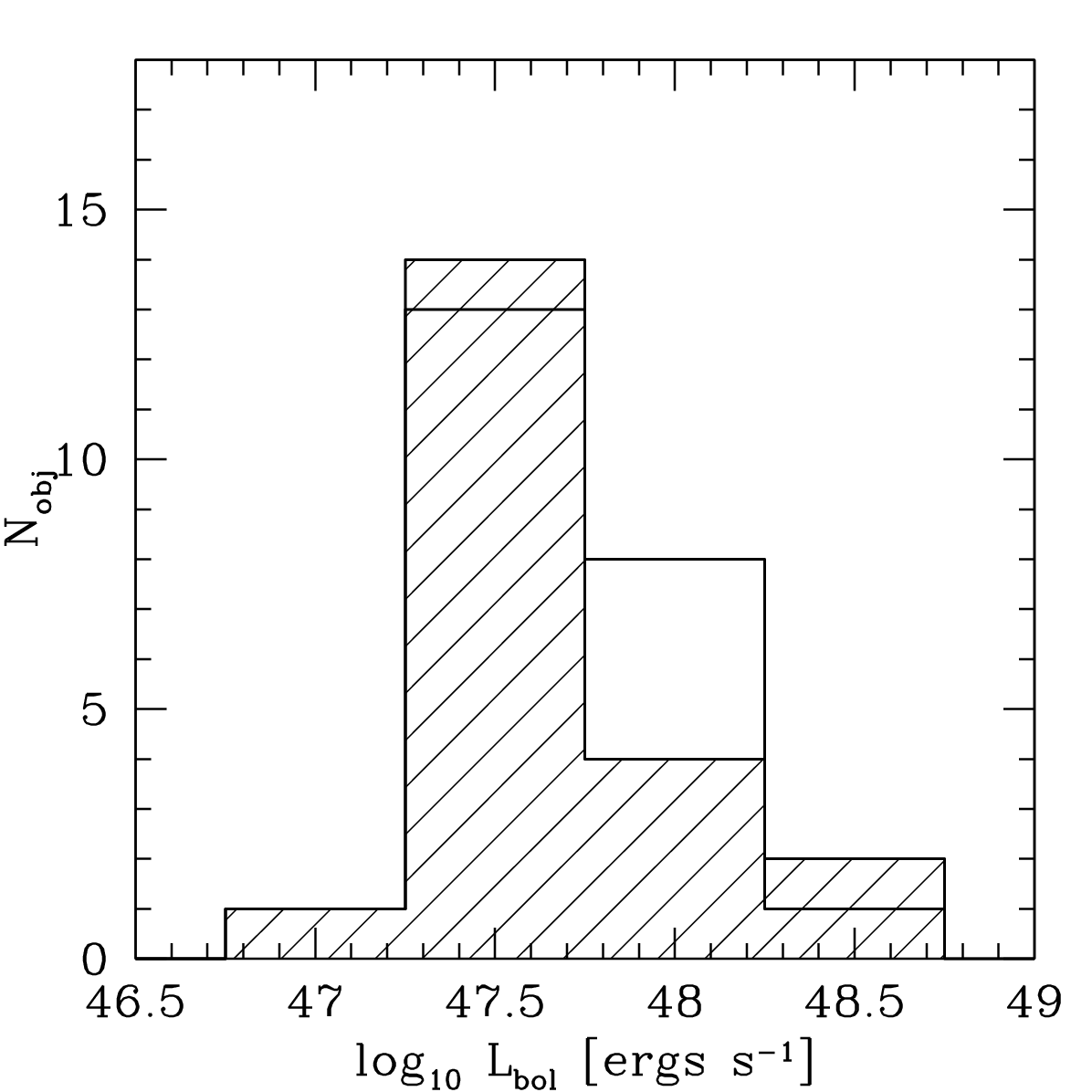}
\includegraphics[width=9cm,height=9cm, angle=0]{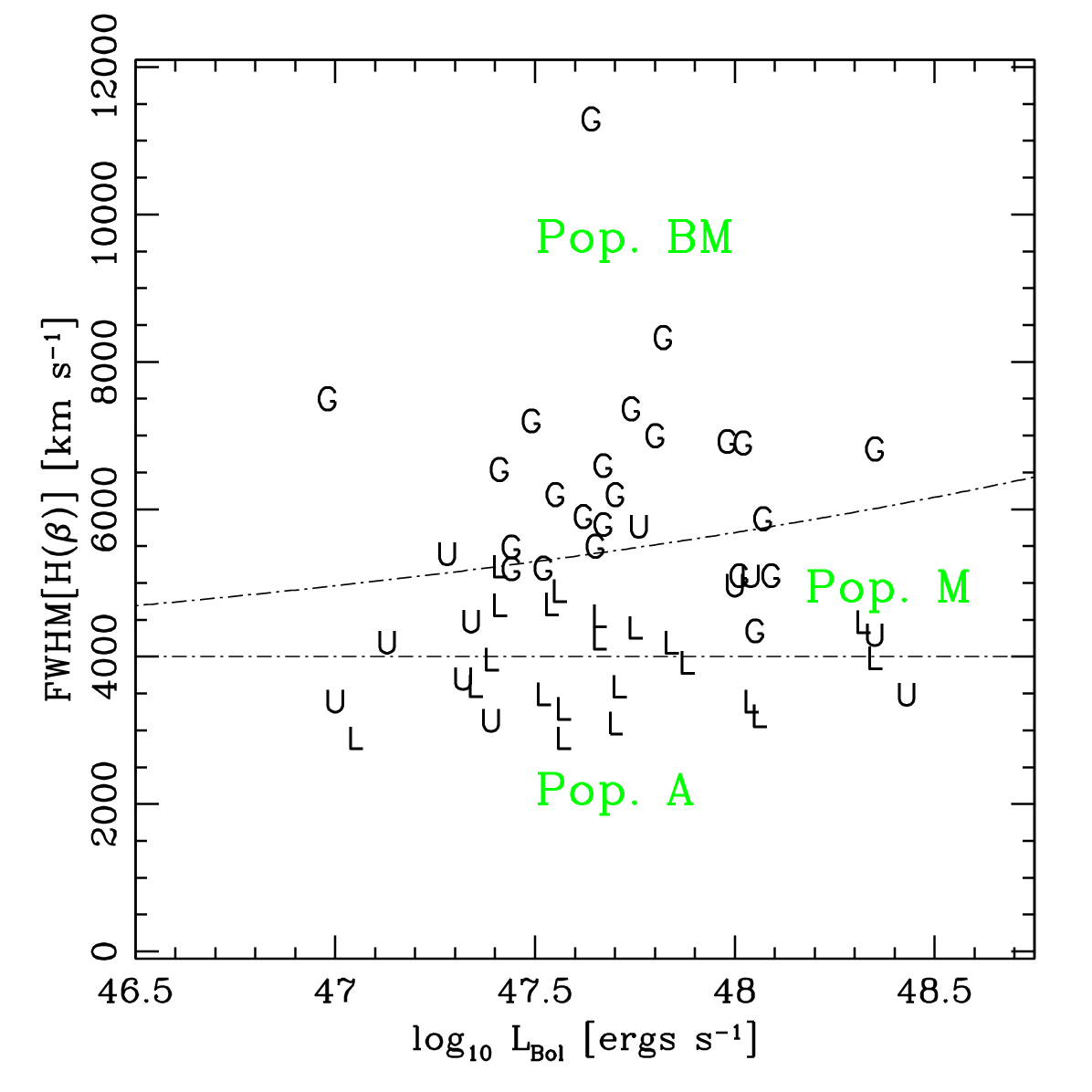}
%\vspace{22cm}
\caption[]{Left: distribution of the ISAAC sources whose \hbbc\  can be fit by a Lorentzian (dashed histogram) and by a Gaussian  as a function of bolometric luminosity in ergs s$^{-1}$.  Right: the same ISAAC sources  in the FWHM(\hb) vs. bolometric luminosity plane. Each source is identified according to the best-fitting function for the core of  \hb: Lorentzian (L), Gaussian (G), or uncertain (U, undecided).  The lower dot-dashed line sets the limit of Pop.~A at 4000~\kms, the upper line traces the luminosity-dependent limit that may be more appropriate. In between the two lines, an intermediate population of sources (called Pop. M) show predominantly Lorentzian profiles. } \label{fig:lb}
\end{figure*}

\begin{figure*}

\includegraphics[width=18.0cm,height=18.0cm, angle=0]{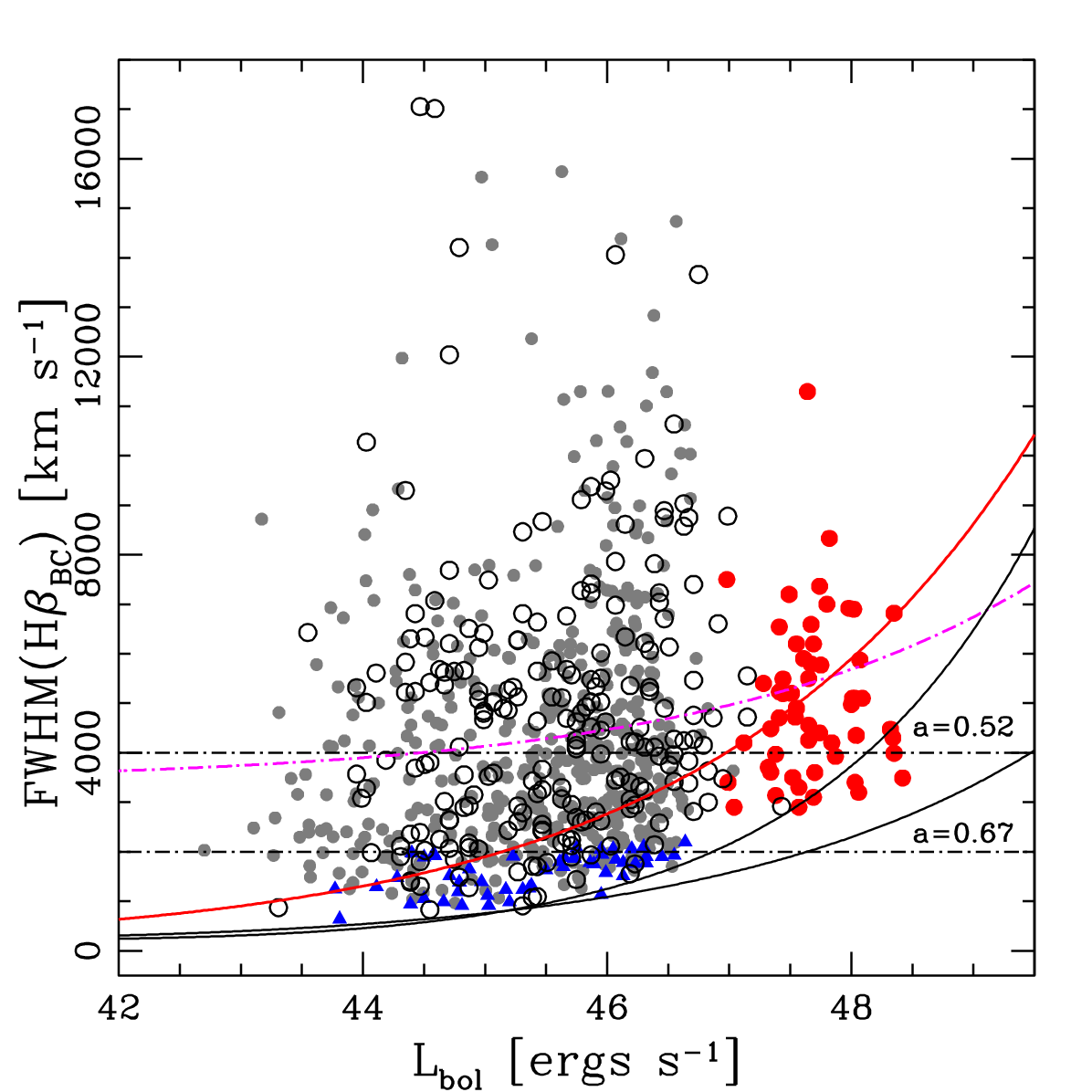}
%\vspace{22cm}
\caption[]{FWHM(\hb) in \kms\  as a function of bolometric luminosity. Small gray circles: data from the SDSS sample of \citet{zamfiretal08} with the ``deep limit" $g < 17.5$; larger filled circles: ISAAC sample; blue filled triangles: data for NLSy1s from \citet{zhouetal06}. The open circles are the data points from the ATLAS of \citet{marzianietal03a}. The horizontal straight, dot dashed lines show the limit of classical NLSy1s sources (2000 \kms) and the limit derived for Population A of \citet{sulenticetal00a} at low $z$\ i.e., 4000 \kms.   The continuous line show the minimum FWHM expected in case of virial motion, for two values of the power-law index in the relationship between BLR size and luminosity: a = 0.52 \citep{bentzetal06} and a = 0.67 \citep[][Paper II]{kaspietal05}. The thick dot-dashed curve at FWHM $\approx$ 4000 \kms\ traces the luminosity-depedent boundary between Pop. A and B partly adopted in this paper. The red thick lines show the FWHM increase in luminosity expected for \lledd = 0.15 (assuming a = 0.67).} \label{fig:kasp}
\end{figure*}

\vfill\clearpage

\Online

\begin{table*}
%\begin{landscape}
\begin{center}
\caption{Basic Properties of Sources and Log of Observations \label{tab:obs}}
    \begin{tabular}{lllllrlrlclllr}
    \hline  \hline
    \noalign{\smallskip}
     Survey name &  Other name & \multicolumn{1}{c}{$\rm m_B^{\mathrm{a}}$}
     &\multicolumn{1}{c}{$z^{\mathrm{b}}$} & \multicolumn{1}{c}{$\Delta z^{\mathrm{b}}$} &
     \multicolumn{1}{c}{Line$^{\mathrm{c}}$} &\multicolumn{1}{c}{$M_{\mathrm B}^
     {\mathrm{d}}$}  & $\log \rm R_K^{\mathrm{e}}$ & \multicolumn{1}{c}{Date$^{\mathrm{f}}$} &
\multicolumn{1}{c}{Band$^{\mathrm{g}}$} & \multicolumn{1}{c}{DIT$^{\mathrm{h}}$}
     & \multicolumn{1}{c}{N$_{\rm exp}^{\mathrm{i}}$} & \multicolumn{1}{c}{Airmass$^{\mathrm{j}}$} &
     \multicolumn{1}{c}{S/N$^{\mathrm{k}}$} \\
%     \multicolumn{1}{c}{} & \multicolumn{1}{c}{} & \multicolumn{1}{c}{\AA} & \multicolumn{1}{c}{\AA}
%     &
\multicolumn{1}{c}{(1)}      & \multicolumn{1}{c}{(2)}         & \multicolumn{1}{c}{(3)}
     & \multicolumn{1}{c}{(4)}    & \multicolumn{1}{c}{(5)} & \multicolumn{1}{c}{(6)}  & \multicolumn{1}{c}{(7)}  & \multicolumn{1}{c}{(8)} & \multicolumn{1}{c}{(9)} & \multicolumn{1}{c}{(10)} & \multicolumn{1}{c}{(11)} & \multicolumn{1}{c}{(12)}   & \multicolumn{1}{c}{(13)} & \multicolumn{1}{c}{(14)}\\
     \hline
HE0035-2853 &       &   17.03   &   1.6377  &   0.0012  &   1,2,3   &   -28.10  &   $<$ 0.21    &   2004/07/13  &   J   &   180 &   20  &   1.12-1.03   &   20  \\
HE0043-2300 &       &   17.06   &   1.5402  &   0.0014  &   1   &   -27.89  &       2.03    &   2004/08/08  &   J   &   180 &   20  &   1.01-1.08   &   40  \\
HE0058-3231 &   Q 0058-3231 &   17.14   &   1.5821  &   0.0002  &   1,2,3   &   -27.94  &   $<$ 0.24    &   2004/07/11  &   J   &   180 &   20  &   1.02-1.01   &   20  \\
HE0105-3006 &   Q 0105-301  &   16.79   &   1.0900  &   0.0004  &   1.3 &   -27.36  &   $<$ 0.12    &   2004/07/28  &   sZ  &   180 &   16  &   1.00-1.01   &   10  \\
HE0109-3518 &   Q 0109-3518 &   16.44   &   2.4057  &   0.0003  &   1,2,3   &   -29.63  &   $<$ -0.04   &   2004/08/02  &   sH  &   180 &   20  &   1.49-1.23   &   20  \\
HE0119-3325 &   Q 0119-3325 &   17.17   &   1.1422  &   0.0012  &   1   &   -27.11  &       1.11    &   2004/08/02  &   sZ  &   180 &   20  &   1.11-1.03   &   12  \\
HE0142-4619 &       &   17.20   &   1.1665  &   0.0017  &   1   &   -27.12  &   $<$ 0.55    &   2004/08/02  &   sZ  &   180 &   20  &   1.07-1.08   &   20  \\
HE0142-4619 &       &       &       &       &       &       &           &   2004/08/03  &   J   &   180 &   20  &   1.62-1.34   &   25  \\
HE0203-4627 &       &   17.34   &   1.4381  &   0.0007  &   1,2,3   &   -27.49  &       2.07    &   2004/08/03  &   J   &   180 &   20  &   1.28-1.15   &   10-15   \\
HE0203-4627 &       &       &       &       &       &       &           &   2004/08/08  &   sZ  &   180 &   20  &   1.20-1.11   &   10  \\
HE0205-3756 &   Q 0205-379  &   17.17   &   2.4335  &   0.0057  &   1   &   -28.87  &       1.06    &   2004/08/03  &   sH  &   180 &   32  &   1.05-1.03   &   25  \\
HE0224-3826 &       &   16.83   &   1.0763  &   0.0012  &   1   &   -27.29  &   $<$ 0.14    &   2004/08/04  &   sZ  &   180 &   16  &   1.06-1.03   &   15  \\
HE0229-5603 &       &   17.07   &   1.1409  &   0.0015  &   1   &   -27.22  &   $<$ -0.29   &   2004/08/18  &   sZ  &   180 &   16  &   1.52-1.36   &   20  \\
HE0239-4809 &       &   17.12   &   1.1660  &   0.0009  &   1   &   -27.27  &       0.84    &   2004/08/22  &   sZ  &   180 &   16  &   1.29-1.18   &   20  \\
HE0248-3628 &       &   16.58   &   1.5355  &   0.0007  &   1   &   -28.45  &       0.55    &   2004/08/22  &   J   &   180 &   12  &   1.08-1.04   &   25  \\
HE0251-5550 &   H25.01  &   16.59   &   2.3505  &   0.0011  &   1   &   -29.56  &   $<$ 0.23    &   2004/08/18  &   sH  &   180 &   20  &   1.31-1.21   &   35  \\
HE0349-5249 &   H27.03  &   16.13   &   1.5409  &   \ldots  &   1   &   -28.93  &       0.76    &   2004/08/22  &   J   &   180 &   12  &   1.17-1.14   &   17.5    \\
HE0359-3959 &       &   17.09   &   1.5209  &   \ldots  &   1   &   -27.89  &       0.22    &   2004/08/28  &   J   &   180 &   20  &   1.14-1.06   &   20  \\
HE0435-4312 &       &   17.10   &   1.2321  &   0.0014  &   1   &   -27.35  &   $<$ 0.51    &   2004/09/17  &   sZ  &   180 &   16  &   1.47-1.28   &   10-5    \\
HE0436-3709 &       &   16.84   &   1.4447  &   0.0023  &   1   &   -27.93  &   $<$ 0.38    &   2004/09/17  &   J   &   180 &   16  &   1.10-1.04   &   10  \\
HE0940-1050 &       &   16.96   &   3.0932  &   \ldots  &   1   &   -29.82  &   $<$     0.10    &   2005/06/10  &   sK  &   180 &   24  &   1.18-1.50   &   10-45   \\
HE1039-0724 &       &   17.16   &   1.4584  &   0.0002  &   1,2,3   &   -27.86  &   $<$     0.20    &   2005/06/23  &   J   &   180 &   20  &   1.19-1.44   &   20  \\
HE1120+0154 &   UM 425  &   16.31   &   1.4720  &   0.0004  &   1,2,3   &   -28.70  &       -0.57   &   2005/06/24  &   J   &   180 &   16  &   1.42-1.75   &   35  \\
HE1347-2457 &       &   16.83   &   2.5986  &   \ldots  &   1   &   -29.64  &       -0.68   &   2005/04/15  &   sH  &   180 &   28  &   1.16-1.54   &   65-10   \\
HE1348+0118 &   Q 1348+0118 &   17.08   &   1.0873  &   0.0012  &   1   &   -27.05  &   $<$     0.24    &   2005/06/13  &   sZ  &   180 &   12  &   1.20-1.29   &   15  \\
HE1349+0007 &   UM 617  &   16.83   &   1.4442  &   0.0010  &   1,2,3   &   -28.00  &       -0.18   &   2005/06/24  &   J   &   180 &   20  &   1.25-1.52   &   15  \\
HE1409+0101 &       &   16.92   &   1.6497  &   0.0039  &   1   &   -28.30  &       0.40    &   2005/07/21  &   J   &   180 &   20  &   1.18-1.38   &   55-25   \\
HE1430-0041 &   Q 1430-0041 &   16.72   &   1.1216  &   \ldots  &   1   &   -27.60  &       -0.39   &   2005/07/21  &   sZ  &   180 &   16  &   1.43-1.78   &   20  \\
HE1505+0212 &       &   17.06   &   1.0943  &   \ldots  &   1   &   -27.28  &       0.51    &   2004/07/18  &   sZ  &   180 &   16  &   1.26-1.53   &   50  \\
HE1505+0212 &       &       &   1.0930  &   0.0240  &   1   &       &           &   2005/05/21  &   sZ  &   180 &   20  &   1.28-1.57   &   30  \\
HE2147-3212 &       &   16.84   &   1.5432  &   \ldots  &   1   &   -28.15  &   $<$ 0.14    &   2004/07/05  &   J   &   180 &   20  &   1.06-1.19   &   15  \\
HE2156-4020 &       &   17.39   &   2.5431  &   0.0010  &   1   &   -28.75  &       -0.09   &   2004/07/10  &   sH  &   180 &   32  &   1.05-1.20   &   15-5    \\
HE2335-3029 &       &   16.96   &   1.1194  &   0.0016  &   1   &   -27.21  &   $<$     0.20    &   2004/07/11  &   sZ  &   180 &   16  &   1.58-1.31   &   25  \\
 \noalign{\smallskip}
\noalign{\smallskip} \hline \hline
\end{tabular}
\end{center}
\begin{list}{}{}
\item[$^{\mathrm{a}}$] Apparent B Johnson magnitude corrected because of Galactic absorption. The value of HE 0436$-$3709 referes to the b Str\"omgren magnitude. \item[$^{\mathrm{b}}$] Redshift, with uncertainty in parentheses.
\item[$^{\mathrm{c}}$] Lines used for redshift calculations: 1: \hb, 2: \oiii. 3: [\ion{O}{iii}]\l4959. \item[$^{\mathrm{d}}$] Absolute B
magnitude, computed for $H_0=70$~\kms~Mpc$^{-1}$,  $\Omega_{\mathrm M} = 0.3$,  $\Omega_{\Lambda} = 0.7$, and $k$-correction spectral index $a=0.6$.
\item[$^{\mathrm{e}}$] Decimal logarithm of the specific flux ratio at 6cm and 4400 \AA\ (effective wavelength of  the
B band). Upper limits are from the NVSS ($\approx$ 2.5 mJy), and would place all undetected sources in the RQ domain.
\item[$^{\mathrm{f}}$] Date refers to time at start of exposure. \item[$^{\mathrm{g}}$] Photometric band.
\item[$^{\mathrm{h}}$] Detector Integration Time (DIT) of ISAAC, in seconds. \item[$^{\mathrm{i}}$] Number of
exposures with integration time equal to DIT. \item[$^{\mathrm{j}}$] Airmass at start and end of exposure.
\item[$^{\mathrm{k}}$] $S/N$\ at continuum level in the proximity of \hb. Two values are reported in case of different
$S/N$\  on the blue and red side of \hb\ (blue side first). The $S/N$\ value is with N estimated at a $2 \sigma$\
confidence level i.e., 2 times the rms.
\end{list}
%\end{landscape}
\end{table*}

\begin{table*}
\begin{center}
\caption{Measurements of Fluxes, Equivalent Widths and FWHM of
Strongest Lines} \label{tab:broad}
    %\begin{tabular}{lllllllll}
    \begin{tabular}{lcccc}
    \hline  \hline     \noalign{\smallskip}
     Object name
     & \multicolumn{1}{c}{F(\hb)$^{\mathrm{a}}$}
     & \multicolumn{1}{c}{W(\hb)$^{\mathrm{b}}$}
     &\multicolumn{1}{c}{F(\feiiq)$^{\mathrm{c}}$}
     & \multicolumn{1}{c}{FWHM(\feii)$^{\mathrm{d}}$}
    \\
\multicolumn{1}{c}{(1)}      & \multicolumn{1}{c}{(2)}         & \multicolumn{1}{c}{(3)}
     & \multicolumn{1}{c}{(4)}    & \multicolumn{1}{c}{(5)} %& \multicolumn{1}{c}{(6)} & \multicolumn{1}{c}{(7)}
       \\
     \hline
     \noalign{\smallskip}
     \baselineskip=24pt
HE0035$-$2853   &   115 $   _{- 10  }^{+    30  }$  &   60  $_{-    8   }^{+    17  }$  &   101 $_{-    40  }^{+    10  }$      &   6400    $_{ -1700   }^{+    700 }$  \\
HE0043$-$2300   &   214 $   _{- 20  }^{+    20  }$  &   74  $_{-    10  }^{+    10  }$  &   82  $_{-    10  }^{+    5   }$      &   3900    $_{ -1300   }^{+    700 }$  \\
HE0058$-$3231   &   173 $   _{- 20  }^{+    20  }$  &   100 $_{-    15  }^{+    15  }$  &   60  $_{-    15  }^{+    5   }$      &   5800    $_{ -2200   }^{+    900 }$  \\
HE0105$-$3006   &   58  $   _{- 8   }^{+    8   }$  &   83  $_{-    14  }^{+    14  }$  &   39  $_{-    6   }^{+    6   }$      &   6000    $_{ -2100   }^{+    1200    }$  \\
HE0109$-$3518   &   831 $   _{- 100 }^{+    100 }$  &   119 $_{-    19  }^{+    19  }$  &   95  $_{-    20  }^{+    20  }$      &   \ldots  $_{     }^{     }$  \\
HE0119$-$3325   &   93  $   _{- 15  }^{+    15  }$  &   90  $_{-    17  }^{+    17  }$  &   64  $_{-    20  }^{+    20  }$      &   4500    $_{ -2100   }^{+    2000    }$  \\
HE0142$-$4619   &   274 $   _{- 30  }^{+    30  }$  &   133 $_{-    20  }^{+    20  }$  &   104 $_{-    40  }^{+    20  }$      &   4100    $_{ -1400   }^{+    1400    }$  \\
HE0203$-$4627   &   159 $   _{- 30  }^{+    20  }$  &   77  $_{-    16  }^{+    12  }$  &   76  $_{-    10  }^{+    10  }$      &   6600    $_{ -1300   }^{+    2000    }$  \\
HE0205$-$3756   &   343 $   _{- 40  }^{+    40  }$  &   74  $_{-    11  }^{+    11  }$  &   102 $_{-    20  }^{+    10  }$      &   4900    $_{ -1800   }^{+    1100    }$  \\
HE0224$-$3826   &   123 $   _{- 35  }^{+    15  }$  &   77  $_{-    23  }^{+    12  }$  &   98  $_{-    20  }^{+    10  }$  $^{\mathrm{e}}$ &   5100    $_{ -1700   }^{ 700 }$  \\
HE0229$-$5603   &   118 $   _{- 20  }^{+    10  }$  &   77  $_{-    15  }^{+    10  }$  &   94  $_{-    10  }^{+    10  }$      &   3000    $_{ -900    }^{+    1600    }$  \\
HE0239$-$4809   &   111 $   _{- 10  }^{+    10  }$  &   44  $_{-    6   }^{+    6   }$  &   102 $_{-    12  }^{+    20  }$      &   2500    $_{ -1200   }^{+    1200    }$  \\
HE0248$-$3628   &   250 $   _{- 30  }^{+    60  }$  &   40  $_{-    6   }^{+    10  }$  &   119 $_{-    20  }^{+    10  }$      &   4300    $_{ -1800   }^{+    1100    }$  \\
HE0251$-$5550   &   353 $   _{- 40  }^{+    40  }$  &   71  $_{-    11  }^{+    11  }$  &   67  $_{-    25  }^{+    10  }$      &   2200    $_{ -600    }^{+    2300    }$  \\
HE0349$-$5249   &   289 $   _{- 50  }^{+    50  }$  &   63  $_{-    13  }^{+    13  }$  &   186 $_{-    60  }^{+    20  }$      &   4700    $_{ -2200   }^{+    600 }$  \\
HE0359$-$3959   &   88  $   _{- 15  }^{+    15  }$  &   45  $_{-    9   }^{+    9   }$  &   111 $_{-    24  }^{+    10  }$      &   2600    $_{ -1200   }^{+    800 }$  \\
HE0435$-$4312   &   165 $   _{- 30  }^{+    30  }$  &   118 $_{-    24  }^{+    24  }$  &   86  $_{-    20  }^{+    10  }$      &   5900    $_{ -1800   }^{+    1500    }$  \\
HE0436$-$3709   &   139 $   _{- 20  }^{+    20  }$  &   88  $_{-    15  }^{+    15  }$  &   78  $_{-    25  }^{+    10  }$      &   5200    $_{ -2200   }^{+    1900    }$  \\
HE0940$-$1050   &   922 $   _{- 100 }^{+    200 }$  &   43  $_{-    6   }^{+    10  }$  &   505 $_{-    100 }^{+    100 }$  $^{\mathrm{e}}$ &   3000    $_{ -900    }^{+    700 }$  \\
HE1039$-$0724   &   138 $   _{- 20  }^{+    20  }$  &   75  $_{-    13  }^{+    13  }$  &   26  $_{-    15  }^{+    5   }$      &   \ldots  $_{     }^{     }$  \\
HE1120+0154 &   270 $   _{- 60  }^{+    60  }$  &   63  $_{-    15  }^{+    15  }$  &   35  $_{-    15  }^{+    15  }$      &   \ldots  $_{     }^{     }$  \\
HE1347$-$2457   &   264 $   _{- 35  }^{+    20  }$  &   64  $_{-    11  }^{+    8   }$  &   252 $_{-    60  }^{+    25  }$      &   5100    $_{ -1700   }^{+    800 }$  \\
HE1348+0118 &   611 $   _{- 60  }^{+    60  }$  &   95  $_{-    13  }^{+    13  }$  &   255 $_{-    55  }^{+    55  }$      &   4300    $_{ -2400   }^{+    600 }$  \\
HE1349+0007 &   129 $   _{- 20  }^{+    20  }$  &   76  $_{-    14  }^{+    14  }$  &   47  $_{-    10  }^{+    15  }$      &   4000    $_{ -500    }^{+    1500    }$  \\
HE1409+0101 &   443 $   _{- 100 }^{+    150 }$  &   91  $_{-    22  }^{+    32  }$  &   272 $_{-    70  }^{+    20  }$      &   6300    $_{ -2100   }^{+    2000    }$  \\
HE1430$-$0041   &   87  $   _{- 18  }^{+    9   }$  &   50  $_{-    12  }^{+    7   }$  &   78  $_{-    10  }^{+    30  }$      &   1000    $_{ -200    }^{+    1000    }$  \\
HE1505+0212 &   176 $   _{- 20  }^{+    20  }$  &   59  $_{-    9   }^{+    9   }$  &   350 $_{-    150 }^{+    50  }$      &   2700    $_{ -900    }^{+    700 }$  \\
HE2147$-$3212   &   151 $   _{- 10  }^{+    20  }$  &   82  $_{-    10  }^{+    14  }$  &   139 $_{-    40  }^{+    10  }$      &   5800    $_{ -1300   }^{+    700 }$  \\
HE2156$-$4020   &   212 $   _{- 20  }^{+    35  }$  &   82  $_{-    11  }^{+    16  }$  &   95  $_{-    10  }^{+    15  }$      &   5200    $_{ -3000   }^{+    1900    }$  \\
HE2335$-$3029   &   142 $   _{- 10  }^{+    20  }$  &   72  $_{-    9   }^{+    12  }$  &   25  $_{-    15  }^{+    10  }$      &   2000    $_{ -800    }^{+    2000    }$  \\\hline
\end{tabular}
\end{center}
\begin{list}{}{}
\item[$^{\mathrm{a}}$] Rest frame flux of \hb\ in \AA\ $\pm 2\sigma$ confidence level uncertainty.
\item[$^{\mathrm{b}}$] Rest frame equivalent width of \hb\ in \AA\ $\pm 2\sigma$ confidence level uncertainty.
\item[$^{\mathrm{c}}$] Rest frame flux of the \feiiq\ blend in units of \ergss\ cm$^{-2}$.
\item[$^{\mathrm{d}}$] FWHM of lines in the blend. See text for details.
\item[$^{\mathrm{e}}$]  Rest frame flux of the  \feiiq\ blend estimated from the \feii$\lambda$5270 blend.  See text for details.
\end{list}
\end{table*}

\begin{table*}
\begin{center}
\caption{Measurements of Fluxes and Equivalent Widths of Narrow
Lines} \label{tab:narrow}
    %\begin{tabular}{lllllllll}
    \begin{tabular}{lcccc}
    \hline  \hline
    \noalign{\smallskip}
     Object name
     %& \multicolumn{1}{c}{F(\hbbc)$^{\mathrm{a}}$}
     %& \multicolumn{1}{c}{W(\hbbc)$^{\mathrm{b}}$}
     %&\multicolumn{1}{c}{W(\feiiq)$^{\mathrm{c}}$}
     %& \multicolumn{1}{c}{FWHM(\feiiq)$^{\mathrm{d}}$}
     &\multicolumn{1}{c}{F(\hbnc)$^{\mathrm{a}}$}
     &\multicolumn{1}{c}{W(\hbnc)$^{\mathrm{b}}$}
      & \multicolumn{1}{c}{F(\oiii)$^{\mathrm{c}}$}
       & \multicolumn{1}{c}{W(\oiii)$^{\mathrm{d}}$}

    \\
\multicolumn{1}{c}{(1)}  &
     \multicolumn{1}{c}{(2)}    & \multicolumn{1}{c}{(3)} &
     \multicolumn{1}{c}{(4)} & \multicolumn{1}{c}{(5)}
       \\
     \hline
     \noalign{\smallskip}
HE0035$-$2853   &       4.0 $_{-    1.0 }^{+    1.0 }$  &       2.1 $_{-    0.6 }^{+    0.6 }$  &       12.4        $_{-    2.0 }^{+    2.0 }$&     6.7 $_{-    1.2 }^{+    1.2 }$  \\
HE0043$-$2300   &       12.0    $_{-    12.0    }^{+    1.0 }$  &       4.1 $_{-    4.2 }^{+    0.5 }$  &       27.8        $_{-    3.0 }^{+    5.0 }$&     10.1    $_{-    1.5 }^{+    2.0 }$  \\
HE0058$-$3231   &       1.8 $_{-    1.0 }^{+    3.0 }$  &       1.0 $_{-    0.6 }^{+    1.7 }$  &       13.1        $_{-    1.5 }^{+    4.0 }$&     7.9 $_{-    1.2 }^{+    2.4 }$  \\
HE0105$-$3006   &       0.3 $_{-    0.3 }^{+    2.0 }$  &       0.4 $_{-    0.4 }^{+    2.9 }$  &       1.2     $_{-    1.2 }^{+    1.5 }$&     1.8 $_{-    1.7 }^{+    2.2 }$  \\
HE0109$-$3518   &       24.4    $_{-    20.0    }^{+    10.0    }$  &       3.5 $_{-    2.9 }^{ 1.5 }$  &       247.5       $_{-    25.0    }^{+    25.0    }$&     38.5    $_{-    5.3 }^{+    5.3 }$  \\
HE0119$-$3325   &       0.8 $_{-    0.8 }^{+    2.0 }$  &       0.7 $_{-    0.8 }^{+    1.9 }$  &       1.6     $_{-    1.0 }^{+    1.5 }$&     1.6 $_{-    1.0 }^{+    1.5 }$  \\
HE0142$-$4619   &       3.9 $_{-    2.0 }^{+    7.0 }$  &       1.9 $_{-    1.0 }^{+    3.4 }$  &       20.0        $_{-    5.0 }^{+    5.0 }$&     10.4    $_{-    2.6 }^{+    2.6 }$  \\
HE0203$-$4627   &       1.0 $_{-    1.0 }^{+    2.0 }$  &       0.5 $_{-    0.5 }^{+    1.0 }$  &       8.3     $_{-    2.0 }^{+    2.0 }$&     4.2 $_{-    1.1 }^{+    1.1 }$  \\
HE0205$-$3756   &   $<$ 7.0                     &   $<$ 1.5                     &       33.0        $_{-    10.0    }^{+    10.0    }$&     7.4 $_{-    2.3 }^{+    2.3 }$  \\
HE0224$-$3826   &       1.0 $_{-    1.0 }^{+    1.0 }$  &       0.0 $_{-    0.6 }^{+    0.6 }$  &   $<$ 3.0                     &   $<$ 1.9                     \\
HE0229$-$5603   &   $<$ 1.0 $_{-        }^{+        }$  &   $<$ 0.7                     &       4.0     $_{-    4.0 }^{+    3.0 }$&     2.7 $_{-    2.6 }^{+    2.0 }$  \\
HE0239$-$4809   &       2.8 $_{-    1.0 }^{+    0.3 }$  &       1.1 $_{-    0.4 }^{+    0.2 }$  &       7.2     $_{-    2.0 }^{+    2.0 }$&     2.9 $_{-    0.8 }^{+    0.8 }$  \\
HE0248$-$3628   &       4.8 $_{-    2.0 }^{+    2.0 }$  &       0.8 $_{-    0.3 }^{+    0.3 }$  &       15.0        $_{-    3.5 }^{+    3.5 }$&     2.6 $_{-    0.6 }^{+    0.6 }$  \\
HE0251$-$5550   &       5.6 $_{-    5.0 }^{+    2.0 }$  &       1.1 $_{-    1.0 }^{+    0.4 }$  &       49.0        $_{-    10.0    }^{+    10.0    }$&     10.3    $_{-    2.2 }^{+    2.2 }$  \\
HE0349$-$5249   &       3.3 $_{-    1.0 }^{+    3.0 }$  &       0.7 $_{-    0.2 }^{+    0.7 }$  &       11.0        $_{-    5.0 }^{+    3.0 }$&     2.5 $_{-    1.1 }^{+    0.7 }$  \\
HE0359$-$3959   &       4.3 $_{-    4.0 }^{+    1.0 }$  &       2.3 $_{-    2.1 }^{+    0.6 }$  &       0.8     $_{-    0.4 }^{+    2.0 }$&     0.4 $_{-    0.2 }^{+    1.0 }$  \\
HE0435$-$4312   &       0.4 $_{-    0.4 }^{+    3.0 }$  &       0.3 $_{-    0.3 }^{+    2.1 }$  &       13.4        $_{-    5.0 }^{+    4.0 }$&     10.1    $_{-    3.7 }^{+    3.0 }$  \\
HE0436$-$3709   &       1.5 $_{-    0.5 }^{+    1.5 }$  &       0.9 $_{-    0.3 }^{+    1.0 }$  &       3.2     $_{-    1.0 }^{+    1.0 }$&     2.1 $_{-    0.7 }^{+    0.7 }$  \\
HE0940$-$1050   &       5.5 $_{-    1.5 }^{+    2.5 }$  &       0.3 $_{-    0.1 }^{+    0.1 }$  &       11.1        $_{-    11.0    }^{+    100.0   }$&     0.6 $_{-    0.5 }^{+    4.7 }$  \\
HE1039$-$0724   &       5.0 $_{-    2.5 }^{+    2.5 }$  &       2.7 $_{-    1.4 }^{ 1.4 }$  &       26.2        $_{-    7.0 }^{+    4.0 }$&     15.5    $_{-    4.1 }^{+    2.7 }$  \\
HE1120+0154 &       2.2 $_{-    0.2 }^{+    4.0 }$  &       0.5 $_{-    0.1 }^{ 0.9 }$  &       52.6        $_{-    8.0 }^{+    8.0 }$&     13.1    $_{-    2.3 }^{+    2.3 }$  \\
HE1347$-$2457   &       5.0 $_{-    3.0 }^{+    1.0 }$  &       1.2 $_{-    0.7 }^{+    0.3 }$  &   $<$ 3.0     $_{-        }^{+        }$&     5.8 $_{-    0.6 }^{+    0.6 }$  \\
HE1348+0118 &       18.7    $_{-    7.0 }^{+    2.0 }$  &       2.9 $_{-    1.1 }^{+    0.4 }$  &       60.7        $_{-    20.0    }^{+    10.0    }$&     10.1    $_{-    3.3 }^{+    1.9 }$  \\
HE1349+0007 &       2.6 $_{-    0.5 }^{+    0.5 }$  &       1.5 $_{-    0.3 }^{+    0.3 }$  &       9.4     $_{-    2.0 }^{+    2.0 }$&     6.1 $_{-    1.3 }^{+    1.3 }$  \\
HE1409+0101 &       34.8    $_{-    15.0    }^{+    3.0 }$  &       7.1 $_{-    3.2 }^{+    0.9 }$  &       35.7        $_{-    10.0    }^{+    5.0 }$&     7.8 $_{-    2.2 }^{+    1.3 }$  \\
HE1430$-$0041   &   $<$ 2.0                     &   $<$ 1.2           &   $<$ 2.0                     &   $<$ 1.2                     \\
HE1505+0212 &   $<$ 1.0                     &   $<$ 0.3                  &   $<$ 2.0                     &   $<$ 0.7                     \\
HE2147$-$3212   &       2.0 $_{-    1.0 }^{+    1.0 }$  &       1.1 $_{-    0.6 }^{+    0.6 }$  &       1.8     $_{-    1.0 }^{+    1.0 }$&     1.1 $_{-    0.6 }^{+    0.6 }$  \\
HE2156$-$4020   &       5.0 $_{-    5.0 }^{+    2.0 }$  &       1.9 $_{-    2.0 }^{+    0.8 }$  &       3.8     $_{-    3.8 }^{+    1.0 }$&     1.5 $_{-    1.5 }^{+    0.4 }$  \\
HE2335$-$3029   &       4.0 $_{-    2.0 }^{+    2.0 }$  &       2.0 $_{-    1.0 }^{+    1.0 }$  &       19.0        $_{-    2.0 }^{+    4.0 }$&     10.2    $_{-    1.4 }^{+    2.3 }$  \\
\noalign{\smallskip} \hline \hline
\end{tabular}
\end{center}
\begin{list}{}{}
\item[$^{\mathrm{a}}$] Rest frame flux of \hbnc\ in units of 10$^{-15}$\ \ergss\ cm$^{-2}$.
 \item[$^{\mathrm{b}}$] Rest frame equivalent width of \hbnc\ in \AA.
 \item[$^{\mathrm{c}}$] Rest frame flux of \oiii\ in units of 10$^{-15}$\ \ergss\ cm$^{-2}$.
\item[$^{\mathrm{d}}$] Rest frame equivalent width of the \oiii\ line in \AA.
 % \item[$^{\mathrm{e}}$] Contribution $R$\ with respect to the total \oiii\ flux of a second semi-broad or blushifted \oiii\ component. See text for details.
\end{list}
\end{table*}

\begin{table*}
\begin{center}
\caption{\hb\ Line Profile Measurements} \label{tab:profs}
    \begin{tabular}{lllccrrrr}
    \hline  \hline
    \noalign{\smallskip}
     \multicolumn{1}{c}{Source}
     & \multicolumn{1}{c}{FWZI$^{\mathrm{a}}$}
     & \multicolumn{1}{c}{$\Delta^{\mathrm{a,b}}$}
     & \multicolumn{1}{c}{FWHM$^{\mathrm{a}}$}
     & \multicolumn{1}{c}{$\Delta^{\mathrm{a,b}}$}
     %& \multicolumn{1}{c}{$\Delta^{\mathrm{a,b}}$}
     & \multicolumn{1}{c}{A.I.$^{\mathrm{c}}$}
     & \multicolumn{1}{c}{$\Delta^{\mathrm{b}}$}
     & \multicolumn{1}{c}{Kurt.$^{\mathrm{d}}$}
     & \multicolumn{1}{c}{$\Delta^{\mathrm{b}}$}
     \\
\multicolumn{1}{c}{(1)}      & \multicolumn{1}{c}{(2)}         & \multicolumn{1}{c}{(3)}
     & \multicolumn{1}{c}{(4)}    & \multicolumn{1}{c}{(5)} & \multicolumn{1}{c}{(6)} & \multicolumn{1}{c}{(7)}
     & \multicolumn{1}{c}{(8)}   & \multicolumn{1}{c}{(9)} \\    \hline
     \noalign{\smallskip}
HE0035-2853 &   23300   &   3600    &   7400    &   500 &   0.19    &   0.07    &   0.37    &   0.04    \\
HE0043-2300 &   25000   &   4500    &   4600    &   300 &   -0.03   &   0.10    &   0.33    &   0.06    \\
HE0058-3231 &   25200   &   4400    &   5800    &   500 &   0.22    &   0.08    &   0.30    &   0.04    \\
HE0105-3006 &   20700   &   5000    &   5200    &   400 &   0.10    &   0.17    &   0.28    &   0.08    \\
HE0109-3518 &   24400   &   5900    &   4000    &   300 &   -0.01   &   0.10    &   0.33    &   0.06    \\
HE0119-3325 &   25300   &   6800    &   4500    &   300 &   0.03    &   0.09    &   0.41    &   0.06    \\
HE0142-4619 &   26900   &   4200    &   3600    &   300 &   0.24    &   0.17    &   0.26    &   0.08    \\
HE0203-4627 &   27800   &   5100    &   7200    &   500 &   0.12    &   0.09    &   0.32    &   0.05    \\
HE0205-3756 &   20700   &   4400    &   4400    &   400 &   0.41    &   0.06    &   0.16    &   0.03    \\
HE0224-3826 &   20800   &   2900    &   6500    &   400 &   0.10    &   0.07    &   0.40    &   0.05    \\
HE0229-5603 &   26500   &   6300    &   4000    &   300 &   0.03    &   0.10    &   0.31    &   0.06    \\
HE0239-4809 &   19700   &   3900    &   4700    &   400 &   0.19    &   0.14    &   0.27    &   0.06    \\
HE0248-3628 &   22200   &   4400    &   3900    &   300 &   -0.19   &   0.10    &   0.30    &   0.05    \\
HE0251-5550 &   25700   &   4400    &   4500    &   400 &   0.47    &   0.07    &   0.19    &   0.03    \\
HE0349-5249 &   26500   &   4000    &   5900    &   800 &   0.37    &   0.07    &   0.27    &   0.04    \\
HE0359-3959 &   18500   &   4800    &   4300    &   400 &   -0.15   &   0.09    &   0.30    &   0.05    \\
HE0435-4312 &   25300   &   6100    &   5500    &   300 &   0.08    &   0.14    &   0.38    &   0.08    \\
HE0436-3709 &   26700   &   4300    &   6600    &   500 &   0.26    &   0.09    &   0.32    &   0.05    \\
HE0940-1050 &   15400   &   1400    &   3500    &   200 &   0.53    &   0.04    &   0.22    &   0.02    \\
HE1039-0724 &   22900   &   3300    &   11300   &   500 &   0.63    &   0.04    &   0.48    &   0.03    \\
HE1120+0154 &   25800   &   4300    &   6900    &   500 &   0.33    &   0.07    &   0.35    &   0.04    \\
HE1347-2457 &   24700   &   3600    &   6800    &   500 &   0.20    &   0.09    &   0.31    &   0.05    \\
HE1348+0118 &   23600   &   4400    &   3700    &   200 &   0.08    &   0.15    &   0.38    &   0.08    \\
HE1349+0007 &   23400   &   3500    &   6200    &   400 &   0.27    &   0.10    &   0.34    &   0.06    \\
HE1409+0101 &   21600   &   4600    &   8300    &   900 &   0.48    &   0.07    &   0.28    &   0.03    \\
HE1430-0041 &   19100   &   3600    &   4700    &   300 &   -0.05   &   0.10    &   0.33    &   0.06    \\
HE1505+0212 &   18400   &   2300    &   5200    &   200 &   -0.02   &   0.25    &   0.37    &   0.11    \\
HE2147-3212 &   25900   &   3900    &   5800    &   400 &   0.01    &   0.10    &   0.33    &   0.05    \\
HE2156-4020 &   26200   &   4400    &   5000    &   500 &   0.27    &   0.10    &   0.26    &   0.05    \\
HE2335-3029 &   18100   &   4000    &   3100    &   200 &   0.04    &   0.11    &   0.38    &   0.07    \\
\noalign{\smallskip} \hline \hline
\end{tabular}
\end{center}
\begin{list}{}{}
\item[$^{\mathrm{a}}$] In units of \kms. \item[$^{\mathrm{b}}$]
2$\sigma$ confidence level uncertainty. \item[$^{\mathrm{c}}$]
Asymmetry index defined as in Marziani et al.\ (1996).
\item[$^{\mathrm{d}}$] Kurtosis parameter as in Marziani et al.\
(1996).
\end{list}\end{table*}

\begin{table*}
\begin{center}
\caption{\hb\ Line Centroids at Different Fractional Heights}
\label{tab:centroids}
    \begin{tabular}{lrrrrrrrrrr}
    \hline  \hline
    \noalign{\smallskip}
     \multicolumn{1}{c}{Source}
     & \multicolumn{1}{c}{c(0/4)$^{\mathrm{a}}$}
     & \multicolumn{1}{c}{$\Delta^{\mathrm{a,b}}$}
     & \multicolumn{1}{c}{c(1/4)$^{\mathrm{a}}$}
     & \multicolumn{1}{c}{$\Delta^{\mathrm{a,b}}$}
     & \multicolumn{1}{c}{c(1/2)$^{\mathrm{a}}$}
     & \multicolumn{1}{c}{$\Delta^{\mathrm{a,b}}$}
     & \multicolumn{1}{c}{c(3/4)$^{\mathrm{a}}$}
     & \multicolumn{1}{c}{$\Delta^{\mathrm{a,b}}$}
     & \multicolumn{1}{c}{c(0.9)$^{\mathrm{a}}$}
     & \multicolumn{1}{c}{$\Delta^{\mathrm{a,b}}$}\\
       \multicolumn{1}{c}{(1)}
     & \multicolumn{1}{c}{(2)}
     & \multicolumn{1}{c}{(3)}
     & \multicolumn{1}{c}{(4)}
     & \multicolumn{1}{c}{(5)}  & \multicolumn{1}{c}{(6)}
      & \multicolumn{1}{c}{(7)}  & \multicolumn{1}{c}{(8)}  &
      \multicolumn{1}{c}{(9)}
      & \multicolumn{1}{c}{(10)}  & \multicolumn{1}{c}{(11)}
    \\    \hline
     \noalign{\smallskip}
HE0035-2853 &   3100    &   1800    &   1900    &   400 &   930 &   270 &   490 &   200 &   370 &   130 \\
HE0043-2300 &   100 &   2200    &   100 &   400 &   130 &   160 &   130 &   120 &   130 &   80  \\
HE0058-3231 &   3500    &   2200    &   2500    &   500 &   1010    &   240 &   710 &   150 &   640 &   100 \\
HE0105-3006 &   2900    &   2500    &   1000    &   1000    &   290 &   180 &   250 &   140 &   240 &   90  \\
HE0109-3518 &   400 &   3000    &   0   &   300 &   40  &   140 &   40  &   110 &   40  &   70  \\
HE0119-3325 &   800 &   3400    &   100 &   300 &   40  &   130 &   30  &   120 &   20  &   80  \\
HE0142-4619 &   1300    &   3400    &   100 &   300 &   40  &   130 &   30  &   120 &   20  &   80  \\
HE0203-4627 &   -300    &   2100    &   1000    &   800 &   200 &   150 &   110 &   100 &   90  &   60  \\
HE0205-3756 &   2900    &   2200    &   2600    &   500 &   580 &   220 &   80  &   110 &   10  &   60  \\
HE0224-3826 &   3400    &   1500    &   1100    &   400 &   650 &   200 &   400 &   180 &   300 &   120 \\
HE0229-5603 &   1500    &   3200    &   -100    &   300 &   -110    &   150 &   -110    &   110 &   -110    &   60  \\
HE0239-4809 &   -100    &   2000    &   200 &   600 &   -330    &   180 &   -330    &   130 &   -310    &   70  \\
HE0248-3628 &   500 &   2200    &   -400    &   400 &   10  &   160 &   130 &   110 &   150 &   60  \\
HE0251-5550 &   2900    &   2200    &   2900    &   600 &   90  &   200 &   -60 &   120 &   -90 &   70  \\
HE0349-5249 &   2700    &   2000    &   2200    &   500 &   640 &   420 &   100 &   150 &   20  &   90  \\
HE0359-3959 &   -700    &   2400    &   -800    &   400 &   -320    &   180 &   -140    &   120 &   -90 &   70  \\
HE0435-4312 &   2700    &   3100    &   400 &   600 &   100 &   160 &   50  &   150 &   40  &   100 \\
HE0436-3709 &   2600    &   2100    &   2100    &   600 &   570 &   240 &   310 &   180 &   240 &   120 \\
HE0940-1050 &   3200    &   700 &   2700    &   200 &   0   &   90  &   0   &   90  &   0   &   70  \\
HE1039-0724 &   2000    &   1600    &   1900    &   300 &   1270    &   250 &   140 &   830 &   -1380   &   160 \\
HE1120+0154 &   2700    &   2200    &   1300    &   500 &   60  &   260 &   -260    &   190 &   -350    &   120 \\
HE1347-2457 &   2400    &   1800    &   0   &   600 &   -390    &   270 &   -570    &   190 &   -610    &   110 \\
HE1348+0118 &   3400    &   2200    &   300 &   400 &   70  &   110 &   30  &   100 &   30  &   70  \\
HE1349+0007 &   3000    &   1700    &   1400    &   700 &   180 &   200 &   -10 &   170 &   -60 &   110 \\
HE1409+0101 &   800 &   2300    &   1600    &   500 &   510 &   460 &   -700    &   210 &   -860    &   110 \\
HE1430-0041 &   -400    &   1800    &   200 &   400 &   200 &   170 &   200 &   130 &   200 &   80  \\
HE1505+0212 &   -800    &   1200    &   0   &   900 &   -390    &   110 &   -60 &   260 &   30  &   80  \\
HE2147-3212 &   -1300   &   2000    &   -100    &   500 &   -70 &   210 &   -70 &   160 &   -70 &   100 \\
HE2156-4020 &   1700    &   2200    &   1600    &   500 &   390 &   250 &   150 &   130 &   100 &   80  \\
HE2335-3029 &   800 &   2000    &   100 &   300 &   10  &   90  &   -10 &   80  &   -10 &   60  \\
\noalign{\smallskip} \hline \hline
\end{tabular}
\end{center}
\begin{list}{}{}
\item[$^{\mathrm{a}}$] In units of \kms. \item[$^{\mathrm{b}}$] 2$\sigma$ confidence level uncertainty.
\end{list}
\end{table*}
\vfill
\newpage\clearpage
\begin{table*}
\begin{center}
\caption{Measurements on  the
Broad Lines of Median Spectra} \label{tab:decn}
    %\begin{tabular}{llllllllll}
    \begin{tabular}{lcccccccc}
    \hline  \hline     \noalign{\smallskip}
     Object name
     & \multicolumn{1}{c}{W(\hb)$^{\mathrm{a}}$}
     &\multicolumn{1}{c}{W(\feiiq)$^{\mathrm{a}}$}
     & \multicolumn{1}{c}{FWHM(\feii)$^{\mathrm{b}}$}
     & \multicolumn{1}{c}{F(\hbbc)/F(\hb)$^{\mathrm{c}}$}
     & \multicolumn{1}{c}{FWHM(\hbbc)$^{\mathrm{d}}$}
          & \multicolumn{1}{c}{$\log$ \mbh\ \hbbc$^{\mathrm{e}}$}
               & \multicolumn{1}{c}{$\log$ \lledd\ \hbbc$^{\mathrm{e}}$}
    \\
\multicolumn{1}{c}{(1)}      & \multicolumn{1}{c}{(2)}         & \multicolumn{1}{c}{(3)}
     & \multicolumn{1}{c}{(4)}    & \multicolumn{1}{c}{(5)}&  \multicolumn{1}{c}{(6)}
   &\multicolumn{1}{c}{(7)}& \multicolumn{1}{c}{(8)}
       \\
     \hline
     \noalign{\smallskip}
     \baselineskip=24pt
A1  &   72  $   _{- 11  }^{+    11  }$  &   26  $_{-4   }^{+3   }$  &   2700    $_{-1100    }^{+    1100    }$ & 1.00 & \ldots & \ldots & \ldots\\ %&      0.9 $_{-        }^{+        }$  &   15      $_{-    10.0    }^{+        }$& \\
A2  &   65  $   _{- 52  }^{+    10  }$  &   49  $_{-    11  }^{+    13  }$  &   3700    $_{-1400    }^{+    2000    }$& 1.00 & \ldots& \ldots & \ldots\\   %&      1.1 $_{-        }^{+        }$  &   6.1     $_{-        }^{+        }$& \\
B1  &   86  $   _{- 13  }^{+    13  }$  &   26  $_{-    6   }^{+    5   }$  &   5200    $_{-2300    }^{+    2400    }$ &0.27 & 4000& \ldots & \ldots \\  %&      1.0 $_{-        }^{+        }$  &   6.1     $_{-        }^{+        }$& \\
B2  &   70  $   _{- 11  }^{+    11  }$  &   44  $_{-    14  }^{+    8   }$  &   5000    $_{-1700    }^{+800 }$& 0.32 & 4000 & \ldots & \ldots\\  %&      1.0 $_{-        }^{+        }$  &   1.088       $_{-        }^{+        }$& \\
        \\
A   &   61  $   _{-13       }^{+10      }$  &   25  $_{-7   }^{+7       }$  &   2700    $_{ -1200   }^{+1500            }$& 1.00 & \ldots& \ldots & \ldots\\   %&      2.6 $_{-        }^{+        }$  &   9.5     $_{-        }^{+        }$& \\
M   &   67  $   _{-11       }^{+10      }$  &   35  $_{-6       }^{+7       }$  &   3800    $_{-1200        }^{+1450    }$& 1.00 & \ldots& \ldots & \ldots\\   %&      0.5 $_{-        }^{+        }$  &   2.001       $_{-        }^{+        }$& \\
MB  &   86  $   _{-13       }^{+10      }$  &   31  $_{-7       }^{+6       }$  &   5000    $_{-1800        }^{+1600        }$& 0.27 & 4100& \ldots & \ldots\\   %&      1.0 $_{-        }^{+        }$  &   3.37        $_{-        }^{+        }$& \\
\\
43A   &   91  $_{-20      }^{+10      }$  &   36  $_{-7   }^{+7}$ &   3000    $_{ -750    }^{+500 }$& 1.00 & \ldots  & 6.1 & -0.74\\       %&      6.6 $_{-        }^{+        }$  &   65.59       $_{-        }^{+        }$ & \\
44A   &   69  $_{-15      }^{+15      }$  &   38  $_{-10      }^{+10      }$  &   2600    $_{ -750    }^{+500     }$& 1.00 & \ldots   &6.8 & -0.47 \\   %&      4.5 $_{-        }^{+        }$  &   25.55       $_{-        }^{+        }$& \\
45A   &   86  $_{-20      }^{+10      }$  &   43  $_{-10      }^{+10      }$  &   2800    $_{-500     }^{+750     }$& 1.00 & \ldots &7.8 &-0.43\\   %&      2.4 $_{-        }^{+        }$  &   11.67       $_{-        }^{+        }$& \\
46A   &   80  $   _{-10       }^{+10      }$  &   47  $_{-10      }^{+10      }$  &   3000    $_{ -600    }^{+600     }$& 1.00 & \ldots & 8.6 & -0.26\\   %&      2.4 $_{-        }^{+        }$  &   10.58       $_{-        }^{+        }$& \\
47A   &   68  $   _{-11       }^{+10      }$  &   30  $_{-8       }^{+8       }$  &   3000    $_{-1200        }^{+1400    }$& 1.00 & \ldots & 9.6 & -0.20\\   %&      2.0 $_{-        }^{+        }$  &   6.6     $_{-        }^{+        }$& \\
48A   &   60  $   _{-11       }^{+11      }$  &27 $_{-8       }^{+5       }$  &   3800    $_{-1100        }^{+1500        }$& 1.00 &  \ldots & 10.3 & +0.11\\  %&      1.5 $_{-        }^{+        }$  &   13.14       $_{-        }^{+        }$& \\
                                                                    \\
43B   &   130 $_{-20      }^{+20}$    &   8   $_{-7       }^{+10      }$  &   \ldots$^{\mathrm{f}}$ & 0.59 & 4600 & 7.1 & -0.68\\  %&      14.7    $_{-        }^{+        }$  &   84.7        $_{-        }^{+        }$& \\
44B   &   125 $_{-30      }^{+10      }$  &   38  $_{-20      }^{+5       }$  &   5600    $_{-1800        }^{+600     }$& 0.49 & 4700 & 7.7 & -1.37\\   %&      7.0 $_{-        }^{+        }$  &   31.3        $_{-        }^{+        }$& \\
45B   &   111 $   _{-20       }^{+15}$    &   29  $_{-15      }^{+5       }$  &   4900    $_{-800     }^{+500     }$ & 0.35 & 4400  & 8.4 & -0.98\\ %&      2.1 $_{-        }^{+        }$  &   21      $_{-        }^{+        }$& \\
46B   &   93  $   _{-20       }^{+10      }$  &   22  $_{-10      }^{+5       }$  &   5900    $_{-1200        }^{+350     }$  & 0.37 & 4800& 9.1 & -0.73 \\    %&      1.7 $_{-        }^{+        }$  &   15.47       $_{-        }^{+        }$& \\
47B   &   92  $   _{-14       }^{+13      }$  &38 $_{-7       }^{+7       }$& 4900    $_{-2000    }^{+1600}$& 0.27 & 4000 & 9.6& -0.24\\   %&      1.0 $_{-        }^{+        }$  &   3.1     $_{-        }^{+        }$& \\
48B   &   75  $   _{-11       }^{+9       }$  &   12  $_{-3}^{+3  }$  &   4600    $_{-1700    }^{+1200    }$& 0.23 & 4300 &10.3 & +0.03\\   %&      1.0 $_{-        }^{+        }$  &   5.055       $_{-        }^{+        }$& \\

\hline
\end{tabular}
\end{center}
\begin{list}{}{}
\item[$^{\mathrm{a}}$] Equivalent width of \hb\  (\hbbc + \hbvbc)   and \feiiq in \AA\ $\pm 2\sigma$ confidence level uncertainty. Note that those values have been computed on median spectra with flux normalized to unity at $\lambda = 5100$ \AA. Considering that the continuum shape is not flat, but that there is however little dispersion in continuum shape across the median spectra, it is W(\hb) $\approx$ I(\hb)/1.1, and W(\feiiq)$\approx$I(\feiiq)/1.25 \AA.
\item[$^{\mathrm{b}}$] FWHM of lines in the blend in units of \kms\ computed by {\tt specfit} as for the individual sources.  Uncertainty is at $\pm 2\sigma$ confidence level. See text for details.
\item[$^{\mathrm{c}}$]  Intensity ratio of the \hbbc\ to total \hb\ line emission i.e., \hbbc\ and \hbvbc.
\item[$^{\mathrm{d}}$]  FWHM of the \hbbc\ component i.e., after removing \hbvbc.
\item[$^{\mathrm{e}}$]  Logarithm of \mbh, in solar masses, and of \lledd. Values have been computed following Paper II,  using the FWHM(\hbbc) reported in Col. (6), and assuming the average bin luminsosity. Values are therefore only indicative. No \mbh\ or \lledd\ has been computed for median in spectral types since they are normalized median spectra made regardless of their luminosity.
\item[$^{\mathrm{e}}$] \feiiopt\ too faint for FWHM to be meaningfully constrained. 
\end{list}
\end{table*}

\newpage\clearpage

\begin{table*}
\begin{center}
\caption{\hb\ Line Profile Measurements on Spectral Types} \label{tab:pro2}
    \begin{tabular}{lllccrrrr}
    \hline  \hline
    \noalign{\smallskip}
     \multicolumn{1}{c}{Source}
     & \multicolumn{1}{c}{FWZI$^{\mathrm{a}}$}
     & \multicolumn{1}{c}{$\Delta^{\mathrm{a,b}}$}
     & \multicolumn{1}{c}{FWHM$^{\mathrm{a}}$}
     & \multicolumn{1}{c}{$\Delta^{\mathrm{a,b}}$}
     %& \multicolumn{1}{c}{$\Delta^{\mathrm{a,b}}$}
     & \multicolumn{1}{c}{A.I.$^{\mathrm{c}}$}
     & \multicolumn{1}{c}{$\Delta^{\mathrm{b}}$}
     & \multicolumn{1}{c}{Kurt.$^{\mathrm{d}}$}
     & \multicolumn{1}{c}{$\Delta^{\mathrm{b}}$}
     \\
\multicolumn{1}{c}{(1)}      & \multicolumn{1}{c}{(2)}         & \multicolumn{1}{c}{(3)}
     & \multicolumn{1}{c}{(4)}    & \multicolumn{1}{c}{(5)} & \multicolumn{1}{c}{(6)} & \multicolumn{1}{c}{(7)}
     & \multicolumn{1}{c}{(8)}   & \multicolumn{1}{c}{(9)} \\    \hline
     \noalign{\smallskip}
A1  &   35000   &   14000   &   3500    &   300 &   0.01    &   0.10    &   0.33    &   0.06    \\
A2  &   41100   &   16000   &   4100    &   300 &   0.05    &   0.10    &   0.33    &   0.06    \\
B1  &   34700   &   6000    &   5900    &   600 &   0.37    &   0.10    &   0.25    &   0.04    \\
B2  &   29200   &   5000    &   5700    &   500 &   0.26    &   0.09    &   0.29    &   0.05    \\
\\
A   &   37100   &   15000   &   3700    &   300 &   0.02    &   0.10    &   0.33    &   0.06    \\
M   &   43700   &   17000   &   4400    &   300 &   0.01    &   0.10    &   0.33    &   0.06    \\
BM  &   34400   &   6000    &   6300    &   700 &   0.35    &   0.09    &   0.25    &   0.04    \\
\\
43A &   19800   &   8000    &   2000    &   100 &   -0.05   &   0.11    &   0.33    &   0.07    \\
44A &   20700   &   8000    &   2100    &   100 &   -0.07   &   0.11    &   0.33    &   0.07    \\
45A &   27300   &   11000   &   2700    &   200 &   -0.01   &   0.10    &   0.33    &   0.06    \\
46A &   31000   &   12000   &   3100    &   200 &   -0.04   &   0.10    &   0.33    &   0.06    \\
47A &   40000   &   16000   &   4000    &   300 &   0.01    &   0.10    &   0.33    &   0.06    \\
48A &   40900   &   16000   &   4100    &   300 &   0.05    &   0.10    &   0.33    &   0.06    \\
\\
43B &   27900   &   6400    &   5200    &   300 &   0.06    &   0.09    &   0.41    &   0.05    \\
44B &   31600   &   6600    &   5700    &   300 &   0.07    &   0.11    &   0.39    &   0.06    \\
45B &   34800   &   6400    &   6000    &   400 &   0.20    &   0.13    &   0.30    &   0.06    \\
46B &   32900   &   5900    &   6500    &   500 &   0.22    &   0.10    &   0.31    &   0.05    \\
47B &   36900   &   6400    &   5900    &   500 &   0.29    &   0.11    &   0.25    &   0.05    \\
48B &   29400   &   4600    &   9300    &   900 &   0.44    &   0.06    &   0.29    &   0.03    \\

\noalign{\smallskip} \hline \hline
\end{tabular}
\end{center}
\begin{list}{}{}
\item[$^{\mathrm{a}}$] In units of \kms. \item[$^{\mathrm{b}}$]
2$\sigma$ confidence level uncertainty. \item[$^{\mathrm{c}}$]
Asymmetry index defined as in Marziani et al.\ (1996).
\item[$^{\mathrm{d}}$] Kurtosis parameter as in Marziani et al.\
(1996).
\end{list}\end{table*}

\begin{table*}
\begin{center}
\caption{\hb\ Line Centroids at Different Fractional Heights on Spectral Types}
\label{tab:cen2}
    \begin{tabular}{lrrrrrrrrrr}
    \hline  \hline
    \noalign{\smallskip}
     \multicolumn{1}{c}{Source}
     & \multicolumn{1}{c}{c(0/4)$^{\mathrm{a}}$}
     & \multicolumn{1}{c}{$\Delta^{\mathrm{a,b}}$}
     & \multicolumn{1}{c}{c(1/4)$^{\mathrm{a}}$}
     & \multicolumn{1}{c}{$\Delta^{\mathrm{a,b}}$}
     & \multicolumn{1}{c}{c(1/2)$^{\mathrm{a}}$}
     & \multicolumn{1}{c}{$\Delta^{\mathrm{a,b}}$}
     & \multicolumn{1}{c}{c(3/4)$^{\mathrm{a}}$}
     & \multicolumn{1}{c}{$\Delta^{\mathrm{a,b}}$}
     & \multicolumn{1}{c}{c(0.9)$^{\mathrm{a}}$}
     & \multicolumn{1}{c}{$\Delta^{\mathrm{a,b}}$}\\
       \multicolumn{1}{c}{(1)}
     & \multicolumn{1}{c}{(2)}
     & \multicolumn{1}{c}{(3)}
     & \multicolumn{1}{c}{(4)}
     & \multicolumn{1}{c}{(5)}  & \multicolumn{1}{c}{(6)}
      & \multicolumn{1}{c}{(7)}  & \multicolumn{1}{c}{(8)}  &
      \multicolumn{1}{c}{(9)}
      & \multicolumn{1}{c}{(10)}  & \multicolumn{1}{c}{(11)}
    \\    \hline
     \noalign{\smallskip}
A1  &   0   &   6900    &   0   &   300 &   -30 &   130 &   -30 &   100 &   -30 &   60  \\
A2  &   -100    &   8200    &   -200    &   300 &   -170    &   150 &   -170    &   110 &   -170    &   70  \\
B1  &   2400    &   3000    &   2100    &   700 &   130 &   280 &   -130    &   150 &   -170    &   100 \\
B2  &   2200    &   2600    &   1700    &   600 &   380 &   230 &   170 &   150 &   130 &   100 \\
\\
A   &   -100    &   7400    &   -100    &   300 &   -70 &   130 &   -70 &   100 &   -70 &   60  \\
M   &   0   &   8700    &   0   &   400 &   -40 &   160 &   -40 &   120 &   -40 &   70  \\
BM  &   2400    &   2900    &   2200    &   700 &   270 &   330 &   -60 &   160 &   -120    &   100 \\

\\
43A &   100 &   3900    &   100 &   200 &   90  &   70  &   90  &   50  &   90  &   30  \\
44A &   200 &   4100    &   100 &   200 &   130 &   70  &   130 &   60  &   130 &   40  \\
45A &   0   &   5400    &   0   &   200 &   30  &   100 &   30  &   80  &   30  &   50  \\
46A &   100 &   6200    &   100 &   300 &   120 &   110 &   120 &   90  &   110 &   50  \\
47A &   0   &   7900    &   0   &   300 &   -20 &   140 &   -20 &   110 &   -20 &   70  \\
48A &   -100    &   8100    &   -200    &   300 &   -170    &   150 &   -170    &   110 &   -170    &   70  \\
43B &   2100    &   3200    &   200 &   300 &   40  &   150 &   0   &   140 &   -10 &   100 \\
44B &   1200    &   3300    &   200 &   500 &   10  &   170 &   -30 &   150 &   -40 &   100 \\
45B &   1700    &   3200    &   1200    &   800 &   140 &   210 &   40  &   160 &   20  &   100 \\
46B &   2400    &   3000    &   1600    &   700 &   370 &   230 &   190 &   170 &   150 &   110 \\
47B &   2100    &   3200    &   2000    &   800 &   220 &   260 &   30  &   160 &   0   &   100 \\
48B &   3100    &   2300    &   2700    &   400 &   1550    &   440 &   -20 &   240 &   -230    &   120 \\
\noalign{\smallskip} \hline \hline
\end{tabular}
\end{center}
\begin{list}{}{}
\item[$^{\mathrm{a}}$] In units of \kms. \item[$^{\mathrm{b}}$] 2$\sigma$ confidence level uncertainty.
\end{list}
\end{table*}
\vfill

\begin{figure*}
\includegraphics[angle=0,scale=0.5]{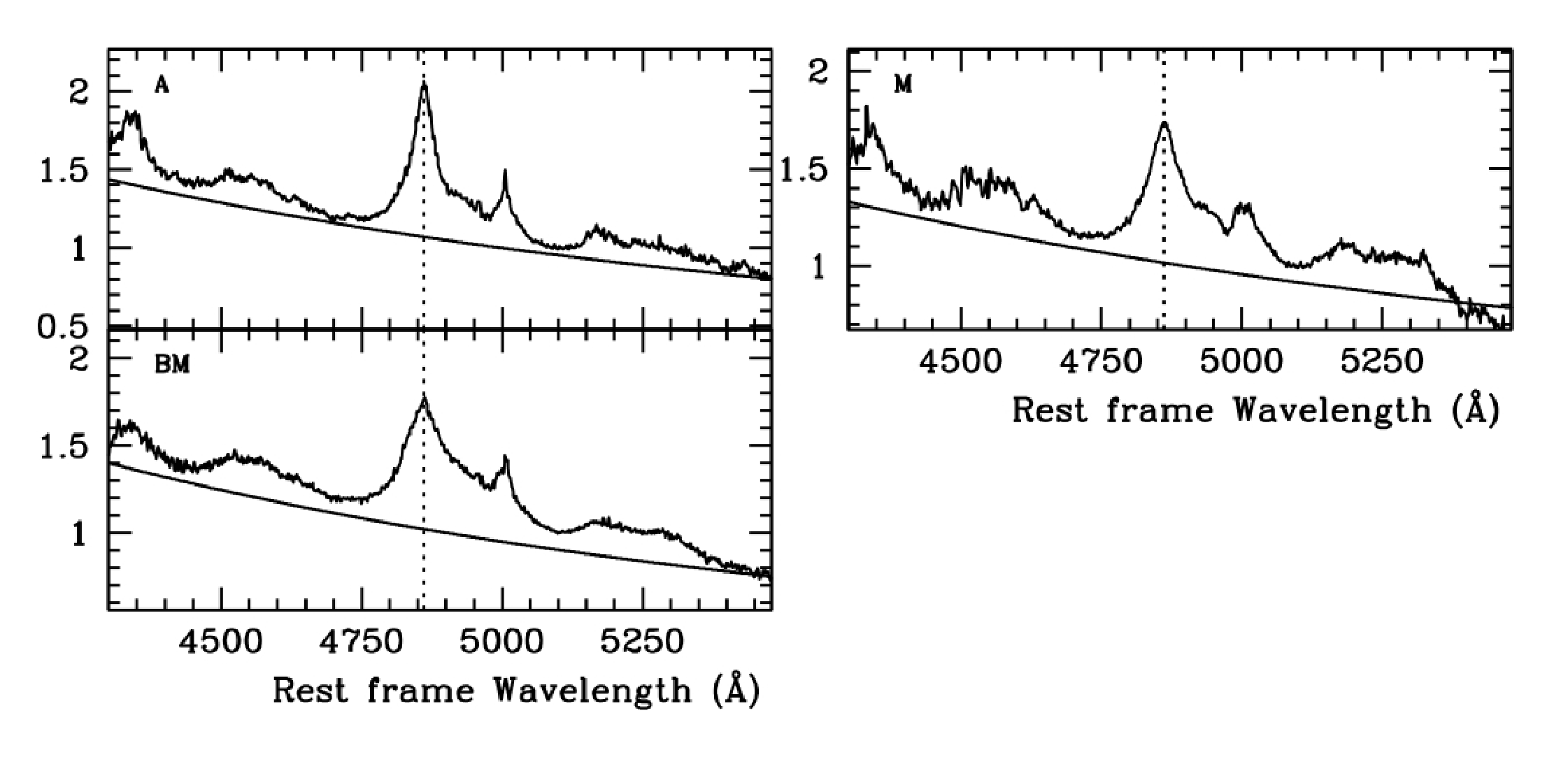}
%\includegraphics[scale=.45,angle=0]{a1s.eps}
%\includegraphics[width=7,height=.7, angle=0]{m.eps}
%\includegraphics[width=5.9cm,height=9cm, angle=0]{bm.eps}
%\vspace{22cm}
\caption[]{Median spectra for Population A, intermediate spectra in region M, and modified Population B (BM), as defined in Fig. \ref{fig:lb}.  } \label{fig:medianmod}
\end{figure*}

\begin{figure*}
\includegraphics[width=9cm,height=9cm, angle=0]{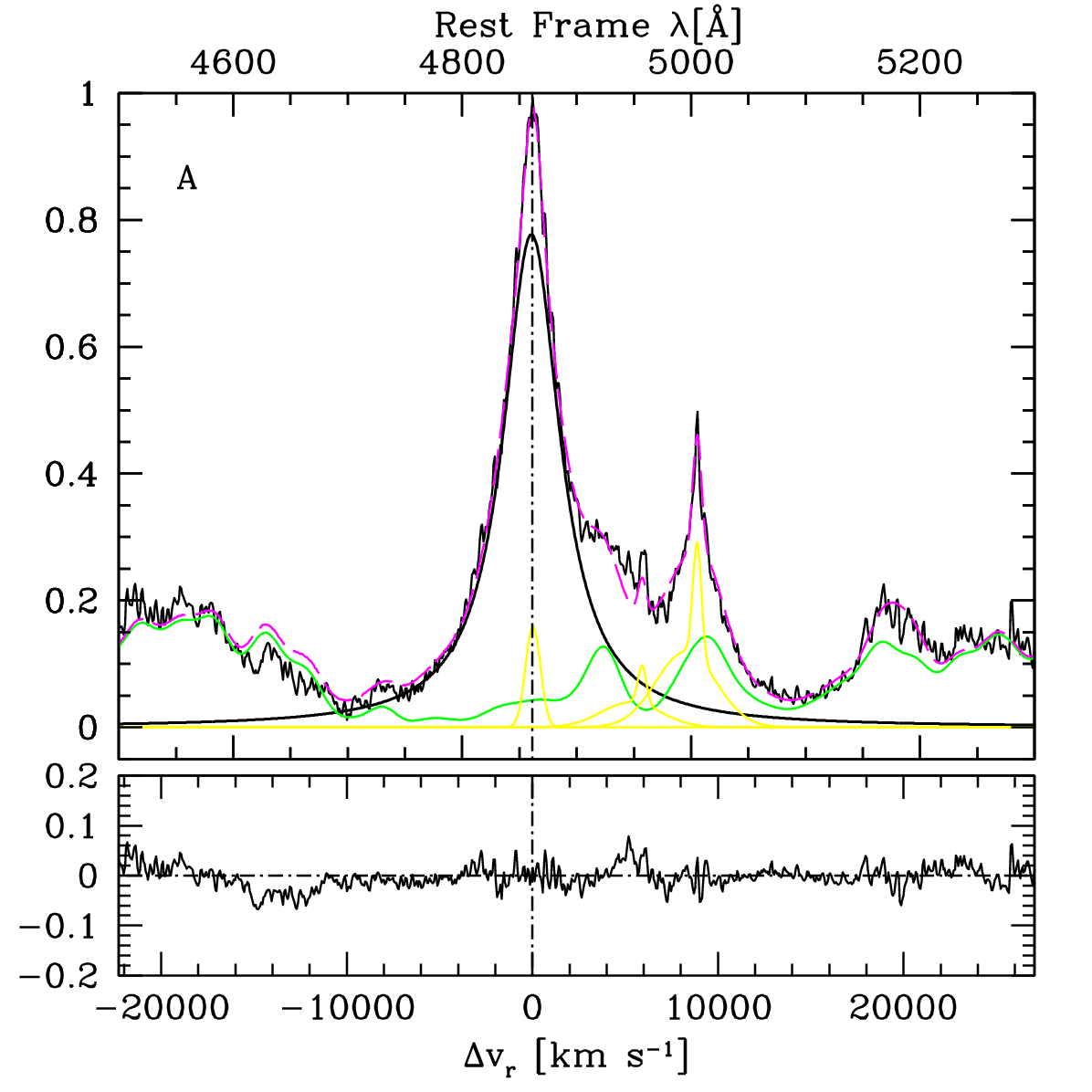}
\includegraphics[width=9cm,height=9cm, angle=0]{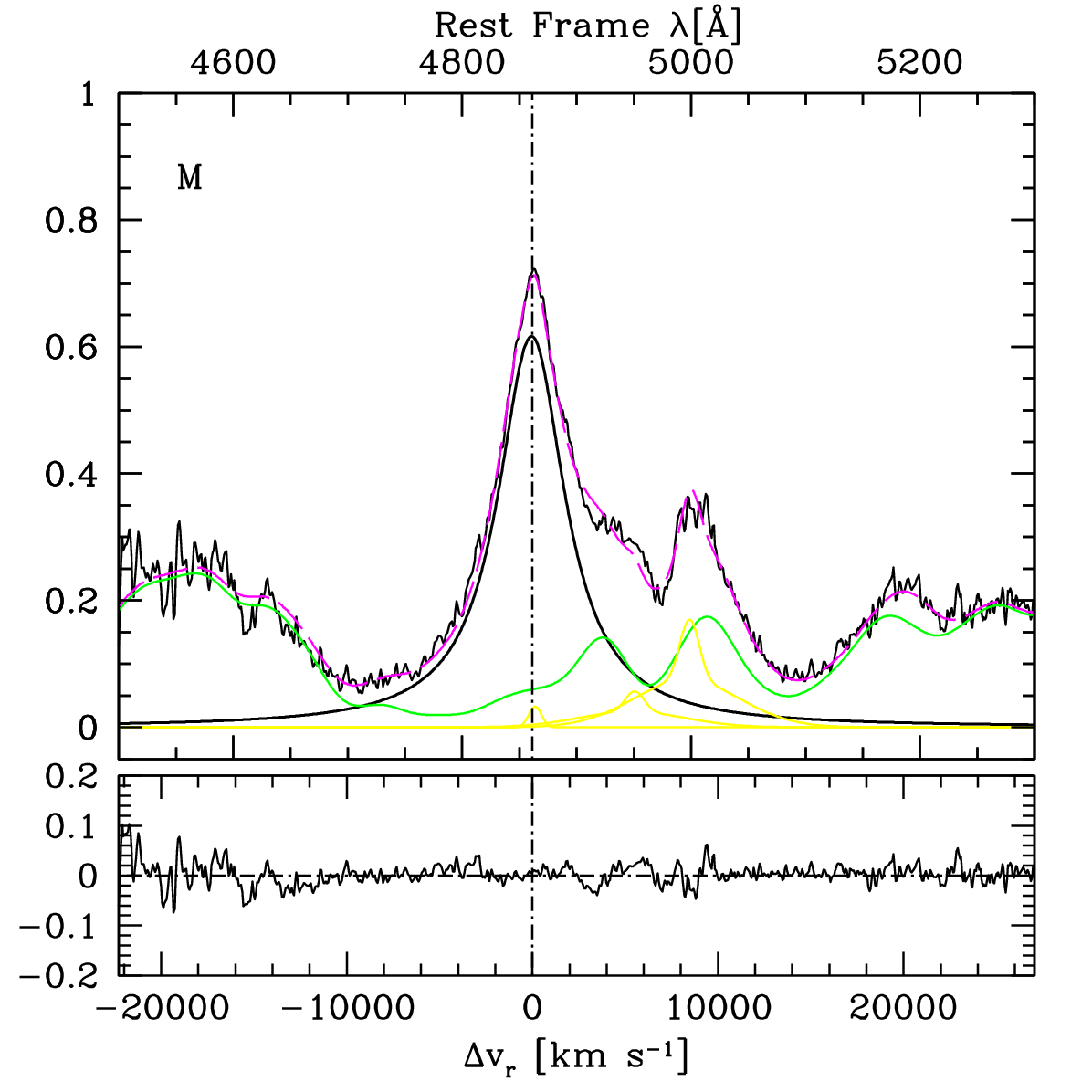}
\includegraphics[width=9cm,height=9cm, angle=0]{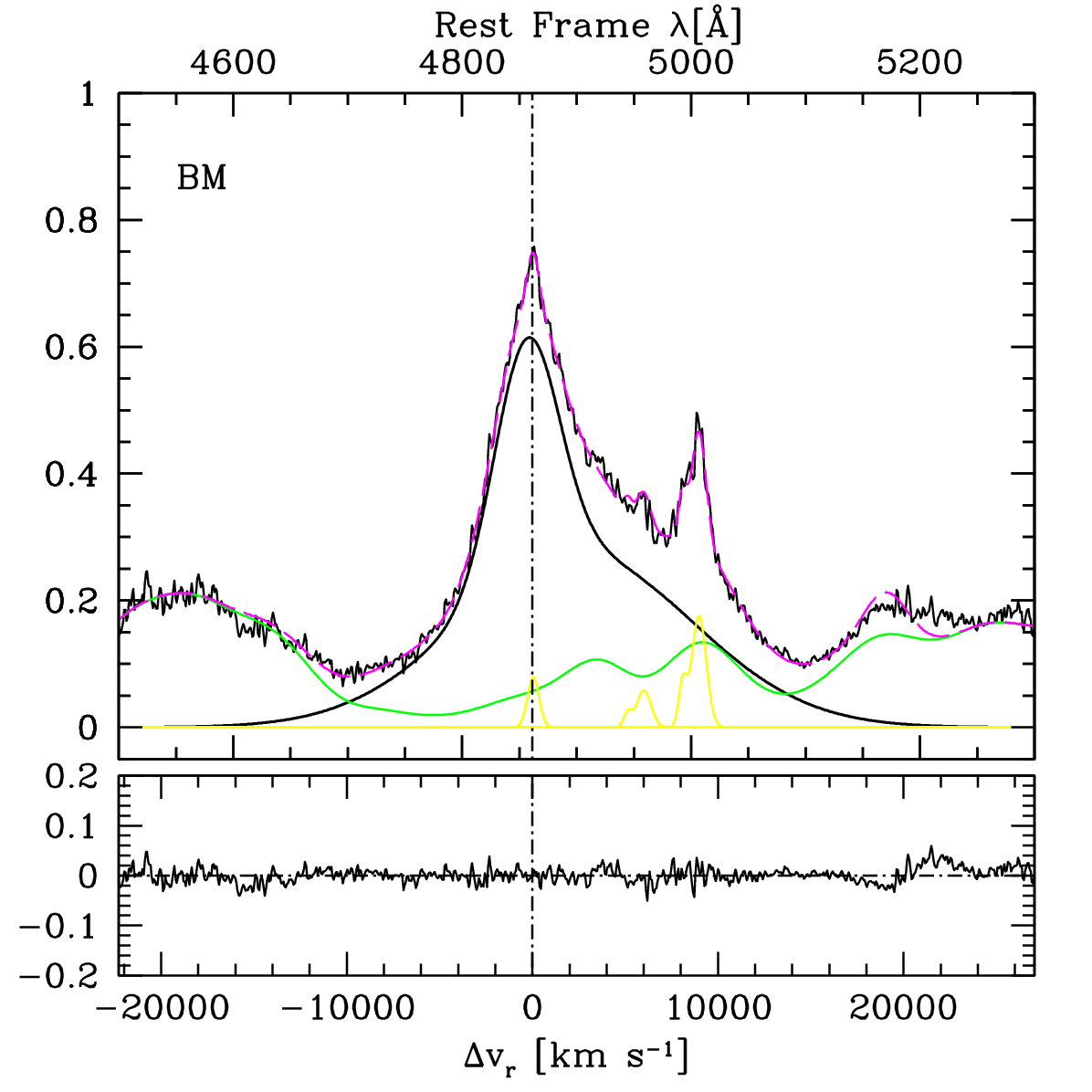}
%\vspace{22cm}
\caption[]{Median spectra analysis of the \hb\ profiles for Population A, ``middle" Population M, and Population MB. Units and line formats are the same as for Fig. \ref{fig:dcom}. } \label{fig:medianmodan}
\end{figure*}

\begin{figure*}
\includegraphics[width=9.0cm,height=9cm, angle=0]{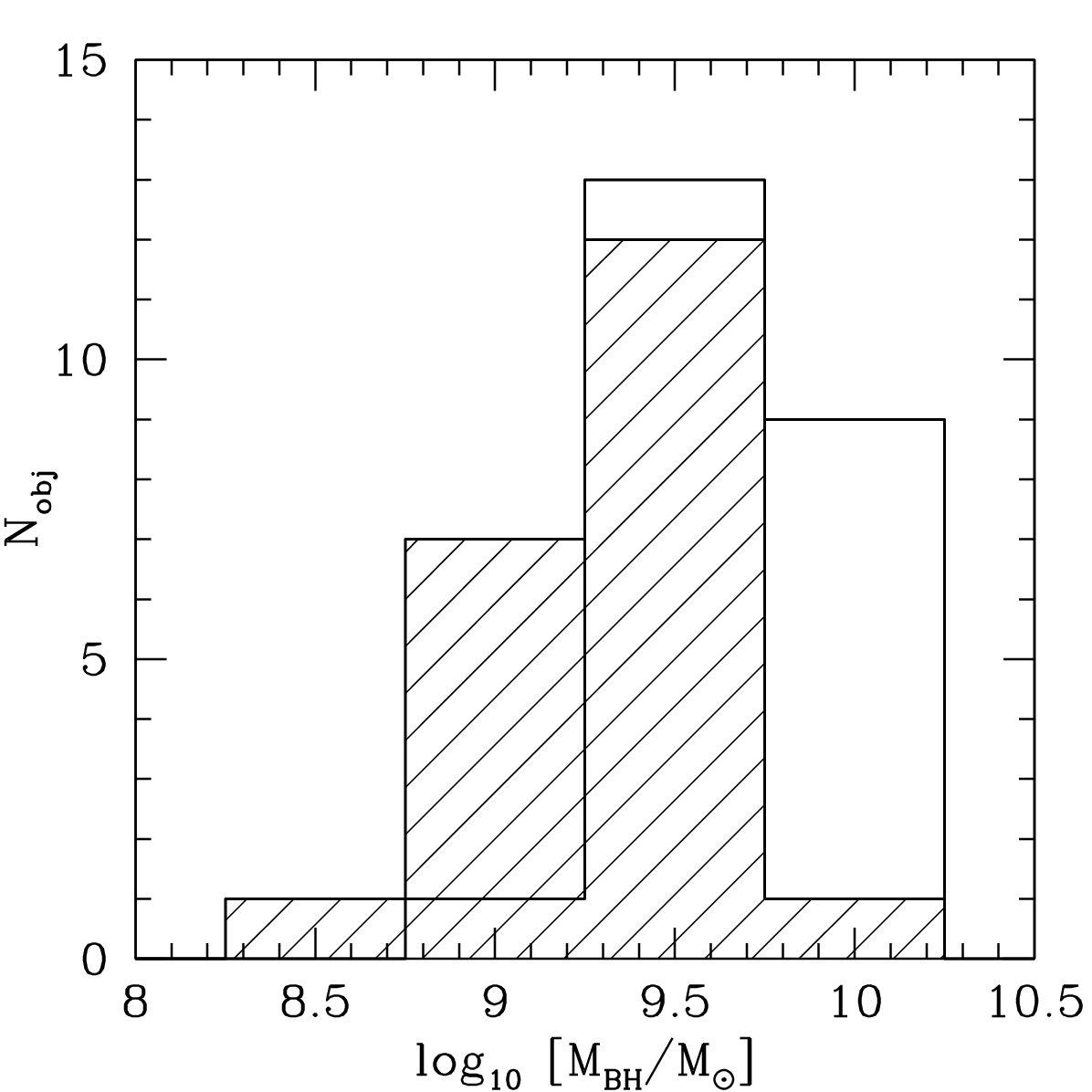}
\includegraphics[width=9.0cm,height=9cm, angle=0]{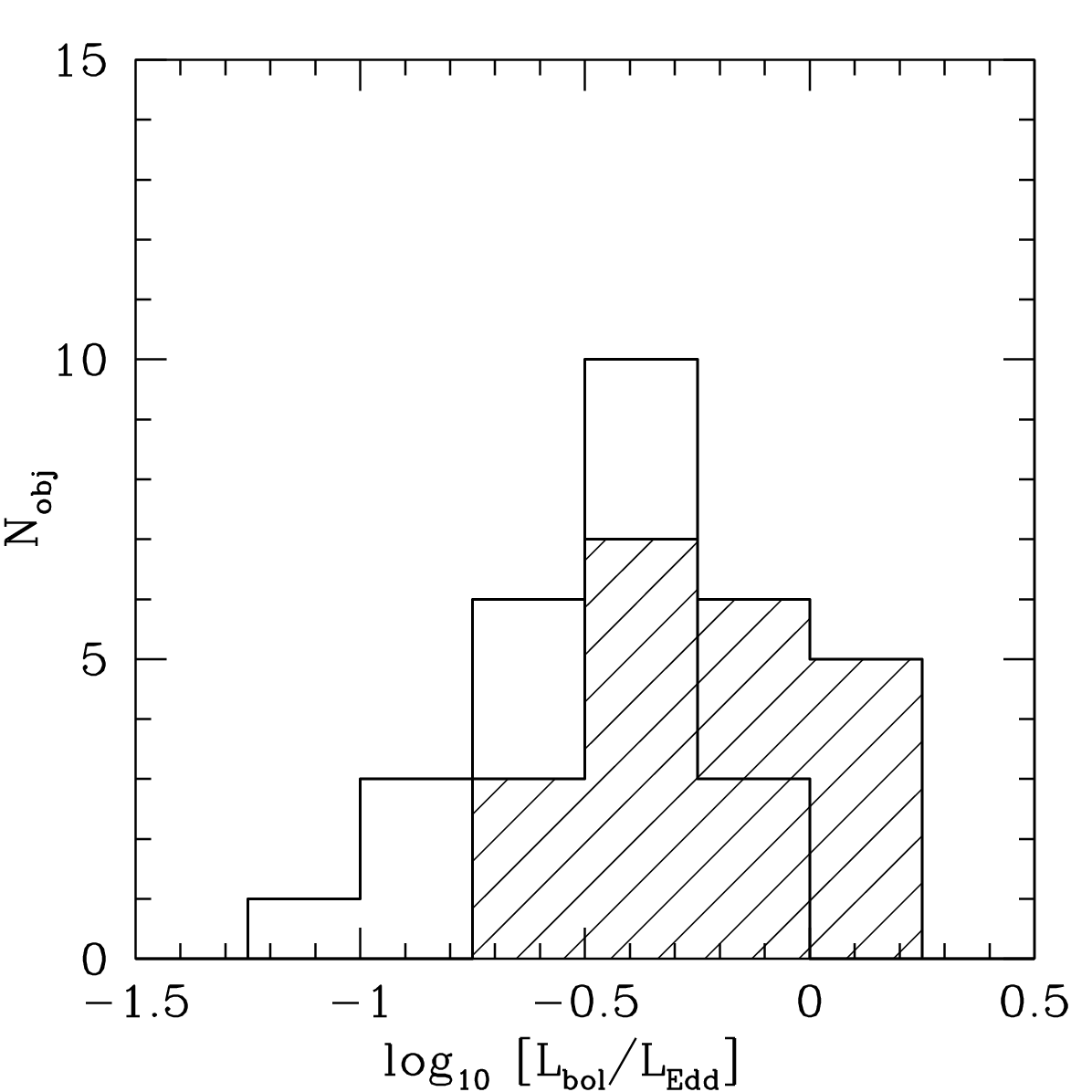}
%\vspace{22cm}
\caption[]{Distribution of sources whose \hbbc\ is of Lorentzian (hatched histogram) or Gaussian type  as a function of black hole mass in units of solar masses (left) and of Eddington ratio (right).} \label{fig:distrlm}
\end{figure*}

\begin{figure*}
\begin{center}
\includegraphics[width=9cm,height=9cm, angle=0]{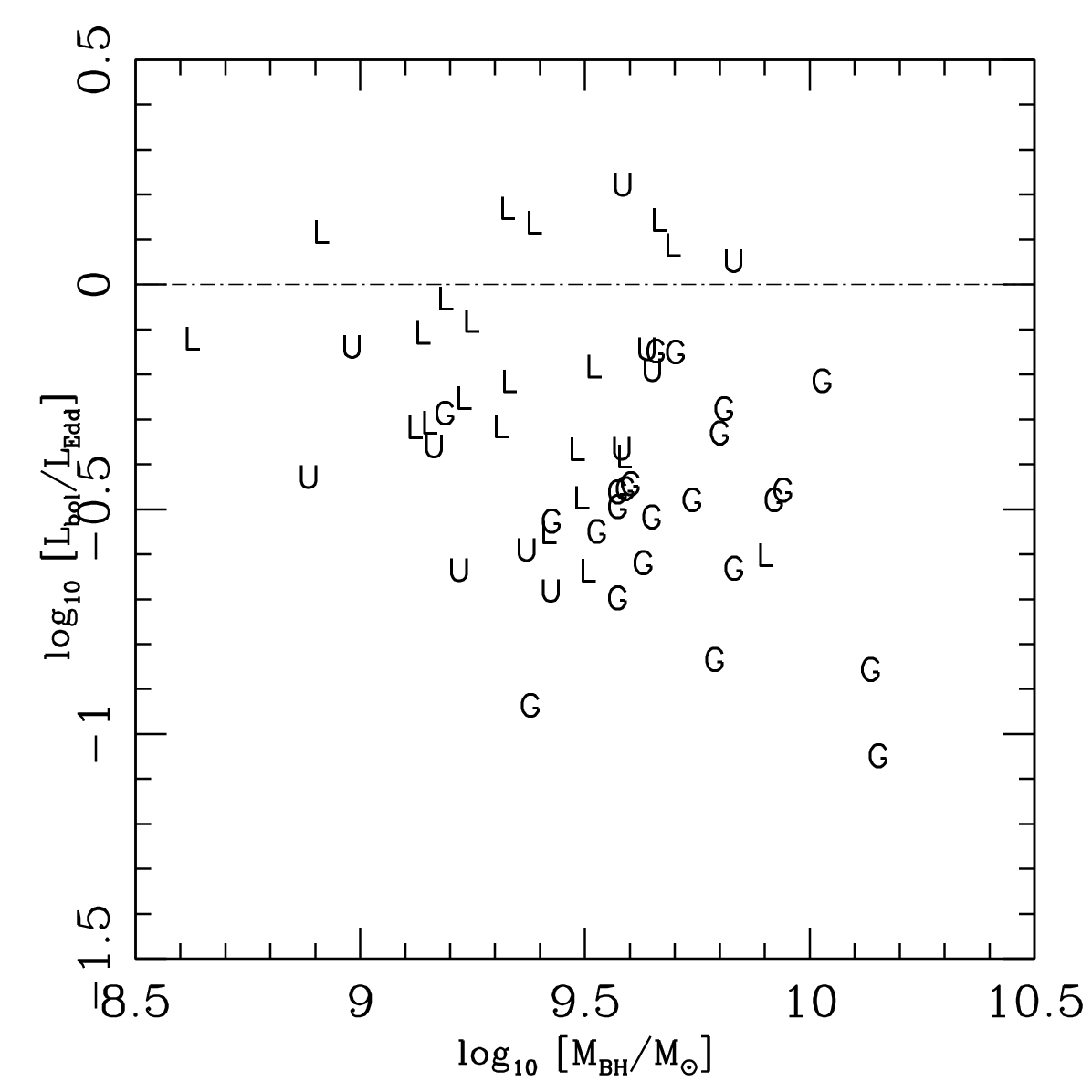}
\end{center}
%\vspace{22cm}
\caption[]{Distribution of the  ISAAC sources in the \lledd\ vs. \mbh\ (in solar units) plane. Each source is labeled as in Fig. \ref{fig:lb}.  } \label{fig:prof}
\end{figure*}
\vfill\clearpage

\begin{figure*}
\includegraphics[width=9cm,height=9cm, angle=0]{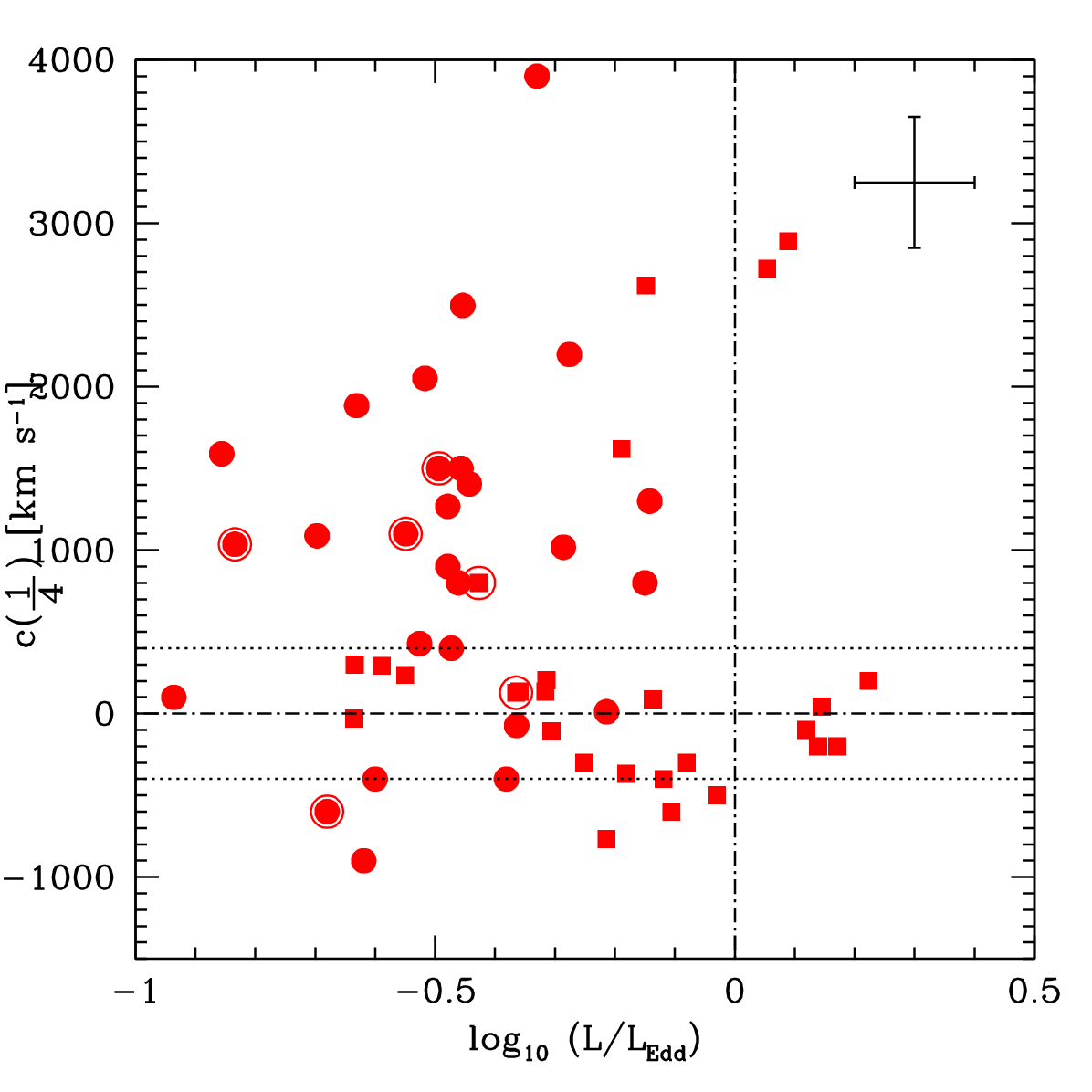}
\includegraphics[width=9cm,height=9cm, angle=0]{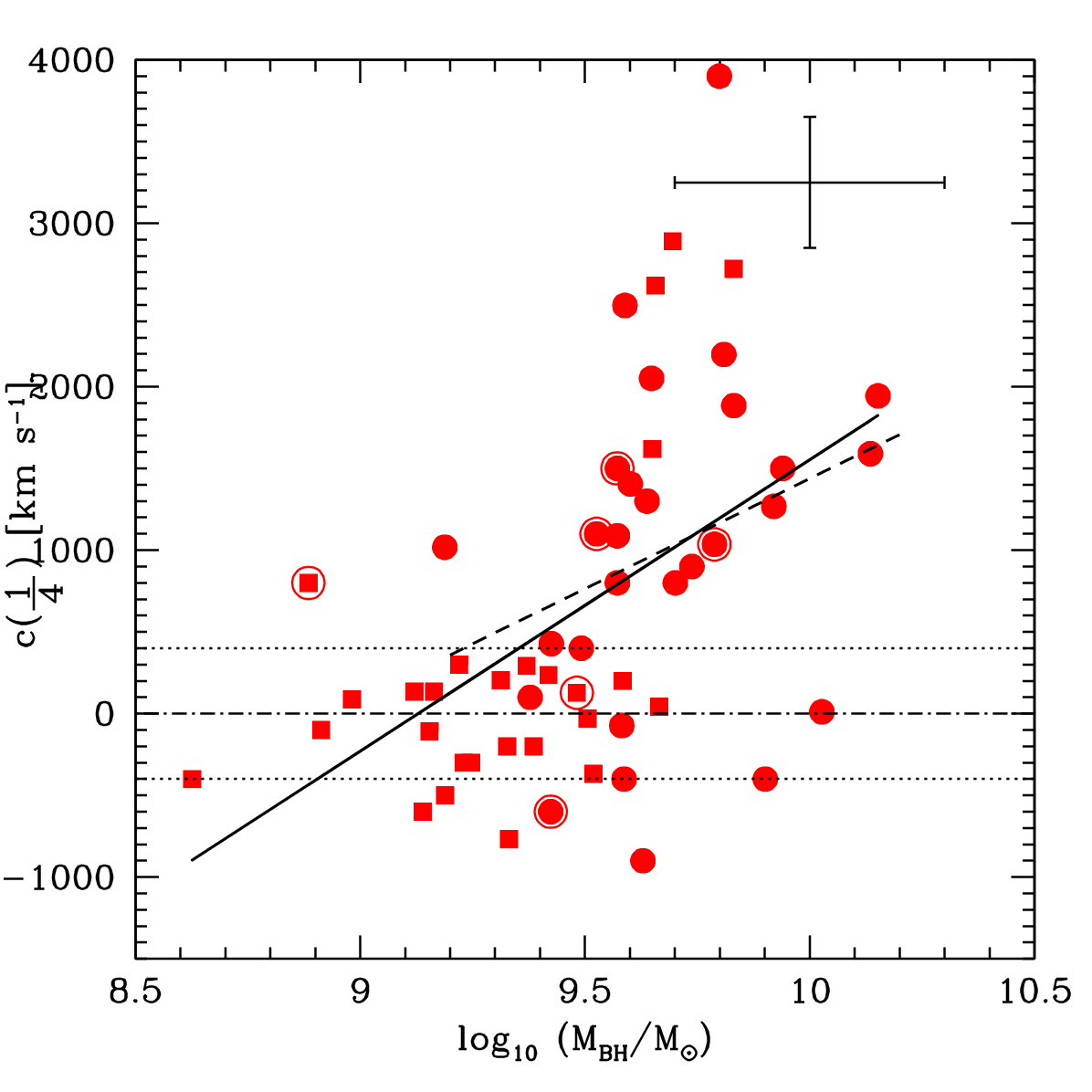}
\includegraphics[width=9cm,height=9cm, angle=0]{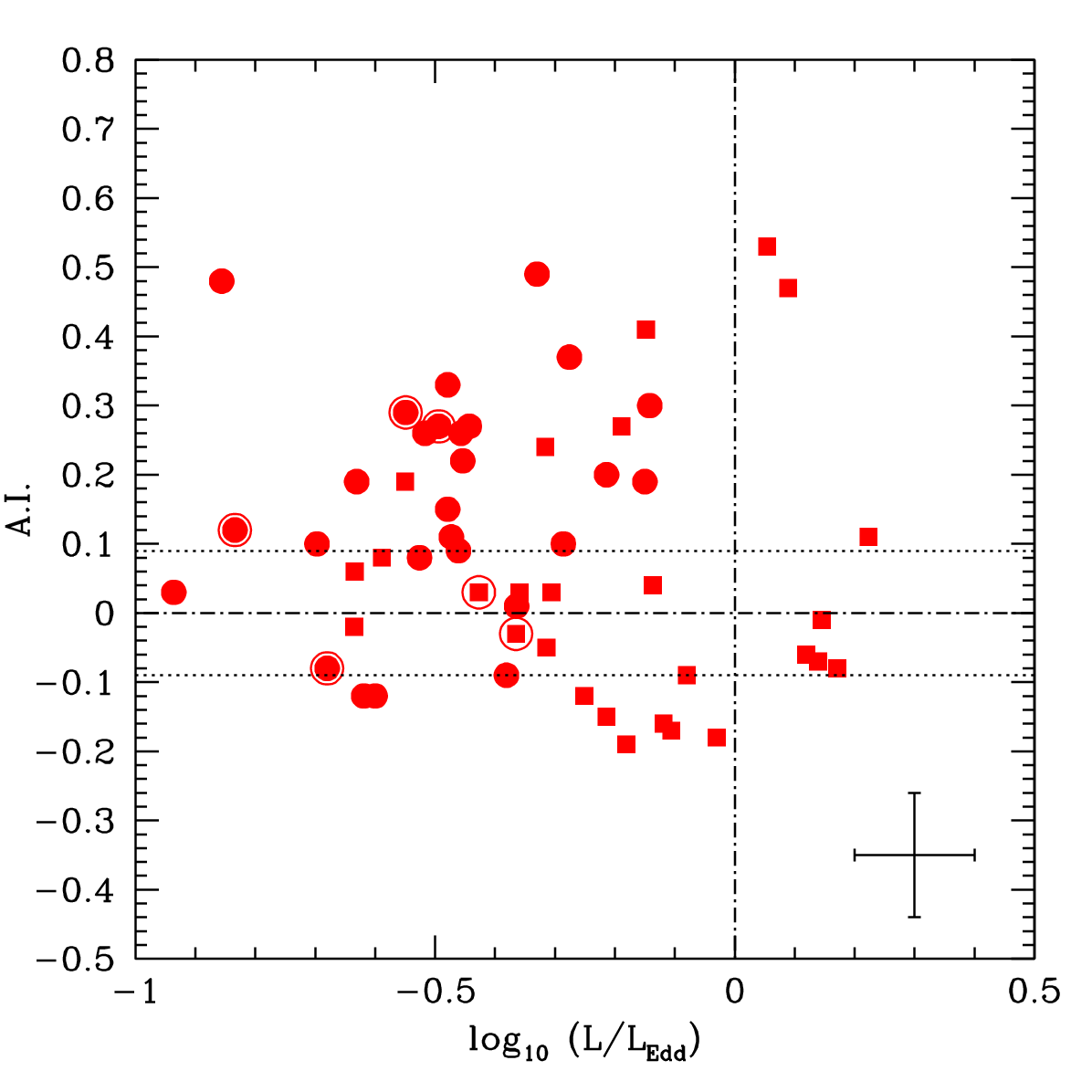}
\includegraphics[width=9cm,height=9cm, angle=0]{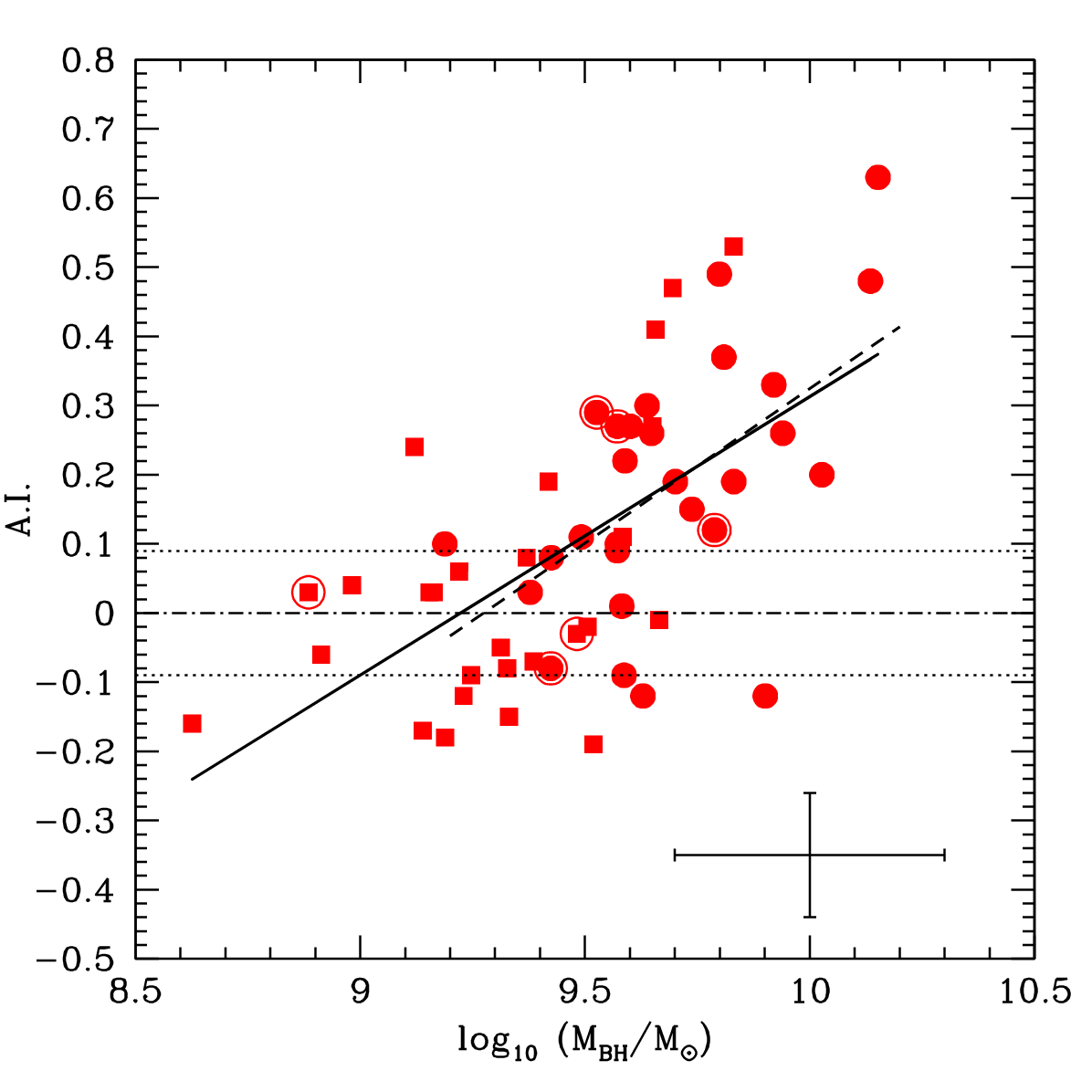}
%\vspace{22cm}
\caption[]{{ Dependence on black hole \mbh\ and
Eddington ratio for \hb\ \coneq\  (upper panels)  and A.I. (lower panels). \coneq\ is in units of \kms, \mbh\ in units of solar masses. Circled data points identify RL sources. Typical error bars at a 2$\sigma$ confindence levels are displayed in the  corners of each panel. Significant correlations are presents for \coneq\ and A.I. vs. $\log$ \mbh; the thick lines are best fits with the unweighted least squares method. The dashed lines represent unweighted least squares fits restricted to Pop. B.}} \label{fig:aic14}
\end{figure*}

\begin{figure*}
\includegraphics[width=9cm,height=9cm, angle=0]{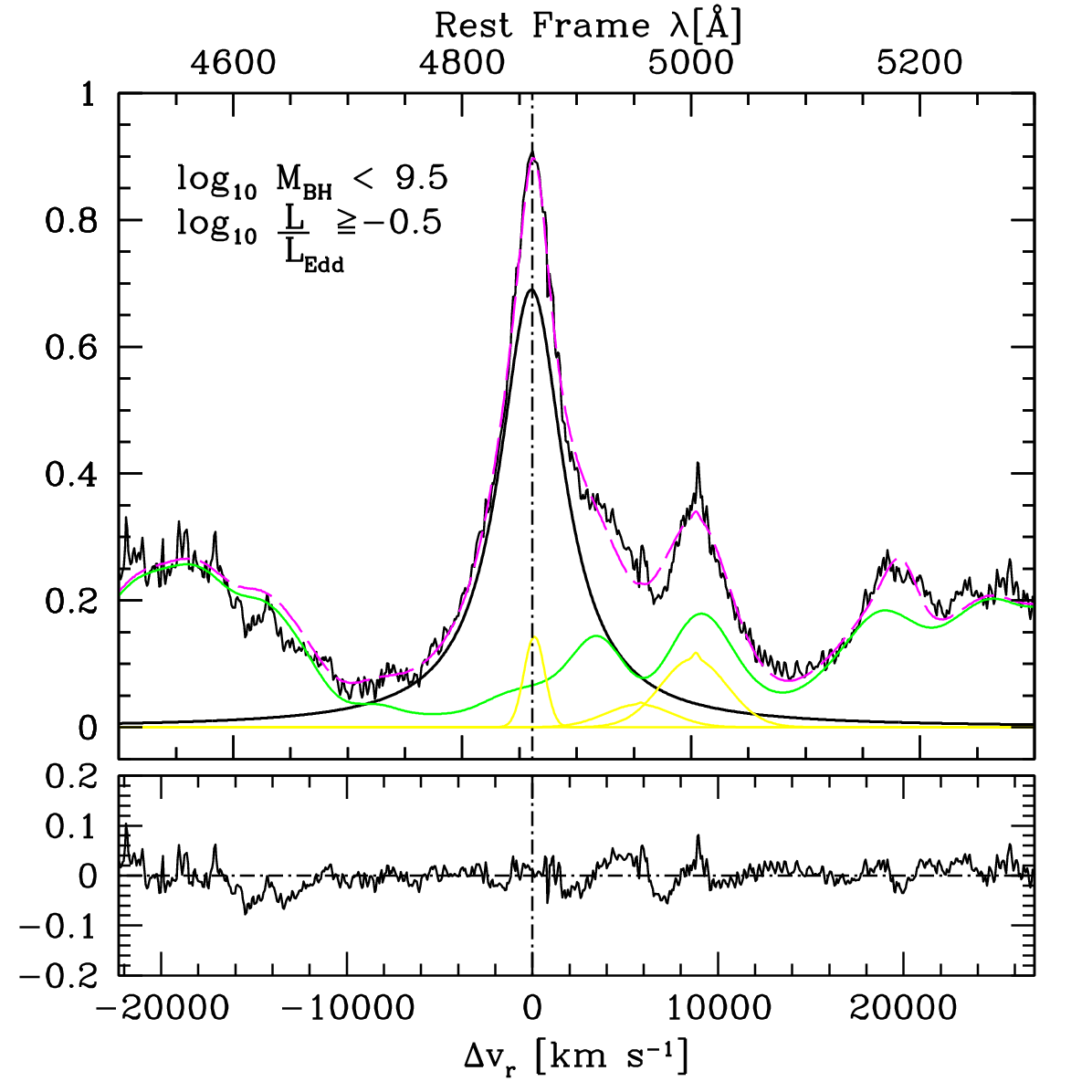}
\includegraphics[width=9cm,height=9cm, angle=0]{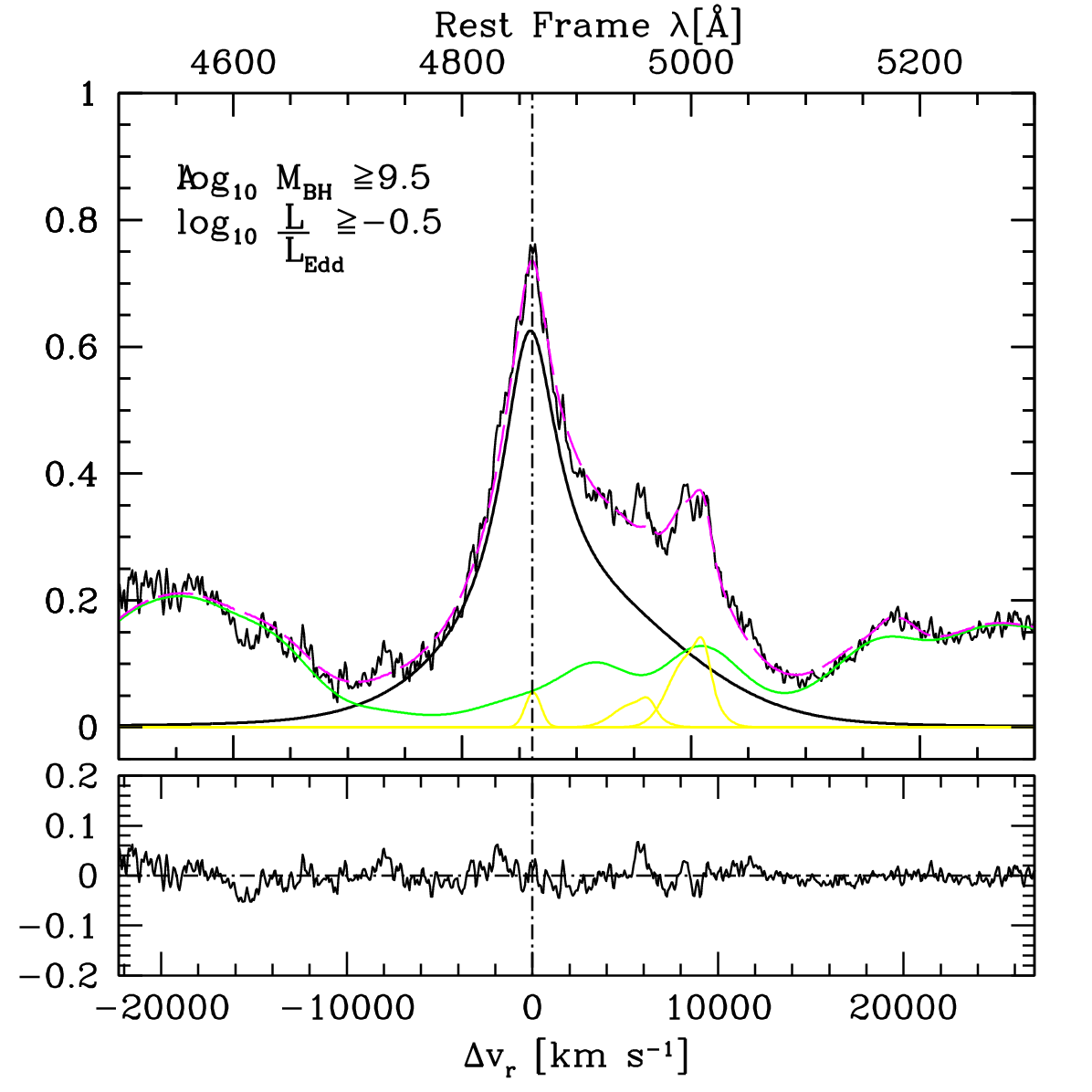}
\includegraphics[width=9cm,height=9cm, angle=0]{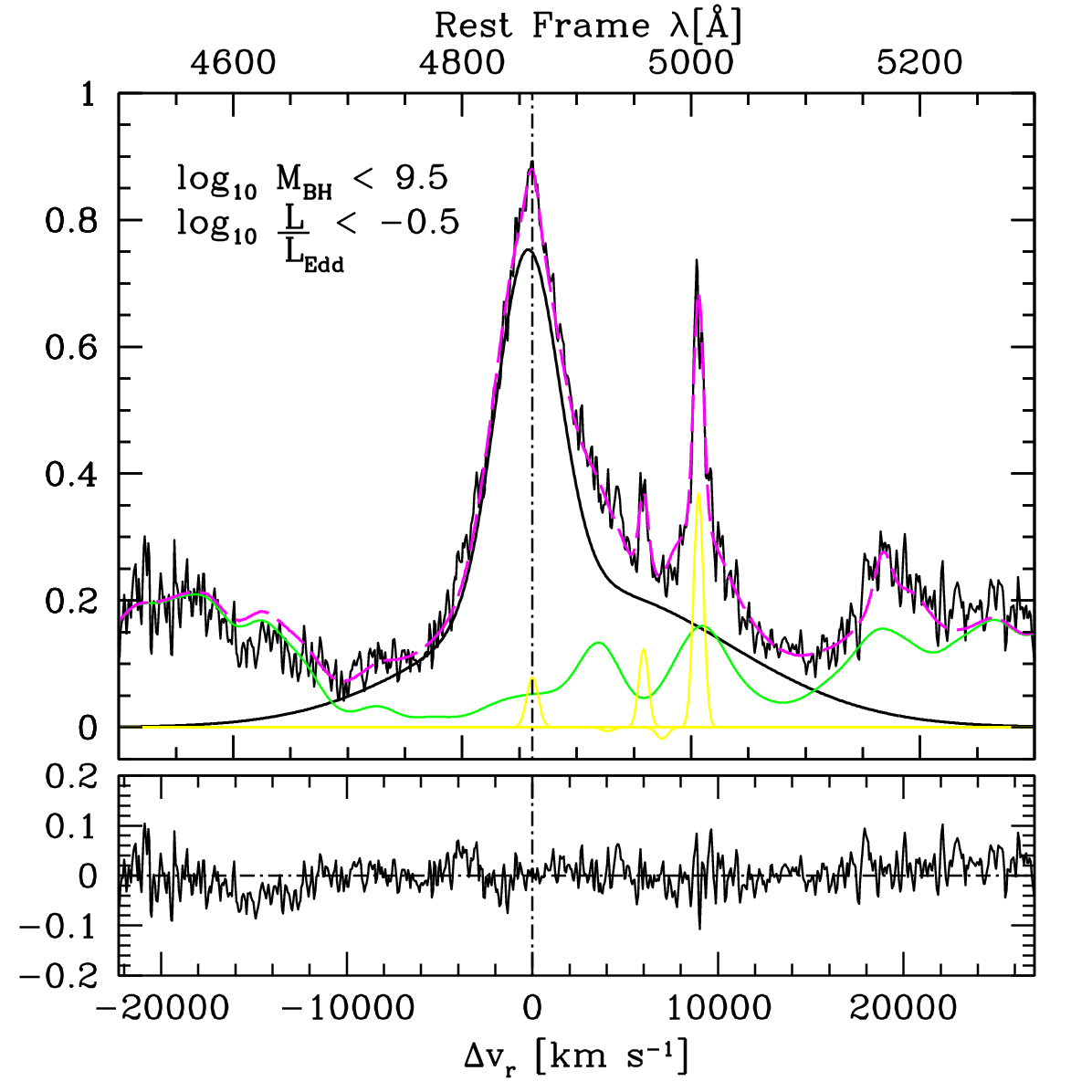}
\includegraphics[width=9cm,height=9cm, angle=0]{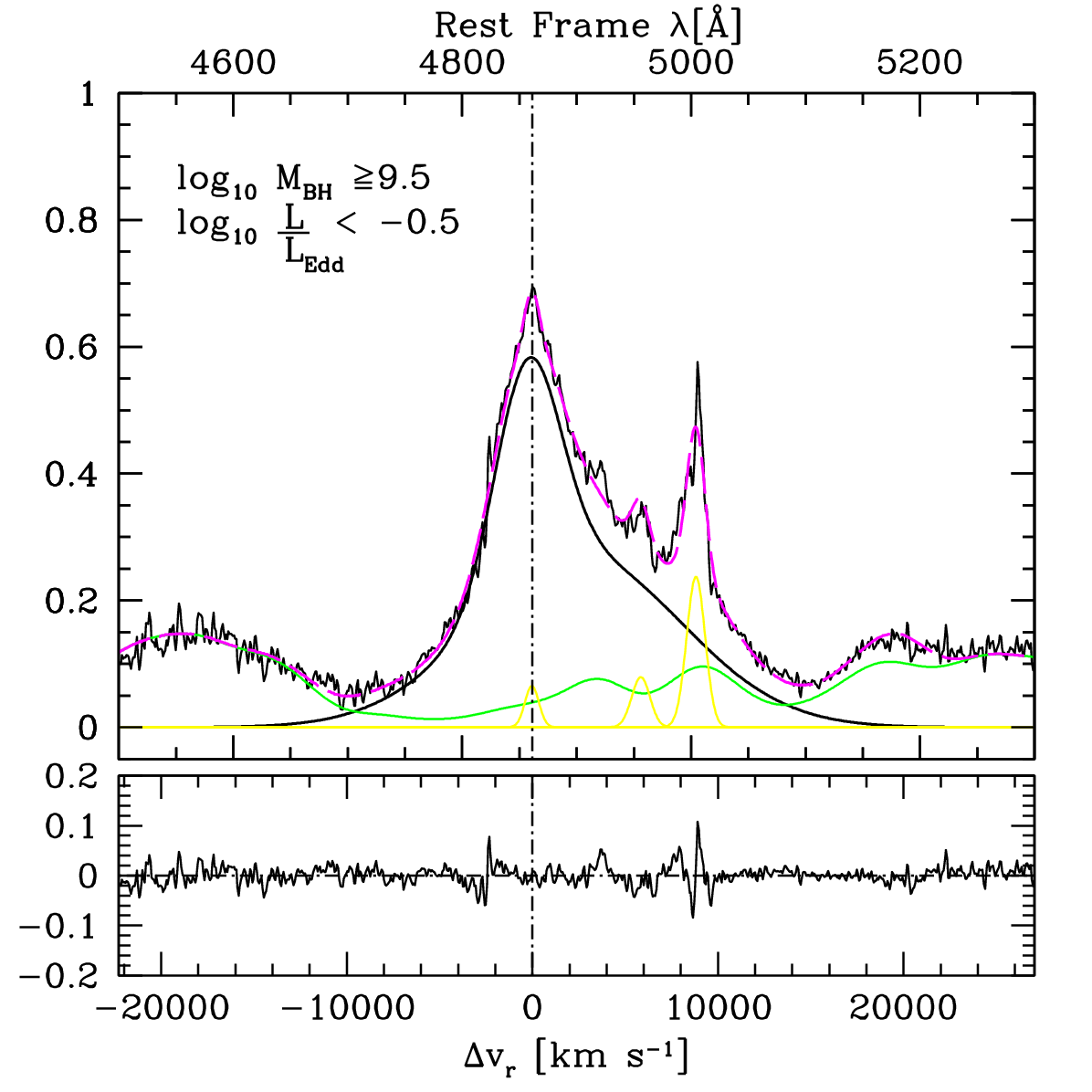}
%\vspace{22cm}
\caption[]{{ \hb\ median profiles for sources in 2 ranges of \mbh\ and of \lledd. The upper panels show the higher \lledd range ($\log $ \lledd $\ge -0.5$), while the bottom panels are   for $-1.1 \la   \log $ \lledd $< -0.5$. The lower mass range in the left panel is $8.5 \la \log$\mbh $< 9.5$. Meaning of lines is the same of Fig. \ref{fig:4binan} and \ref{fig:dcom}. See text for more details.}} \label{fig:ml}
\end{figure*}

\end{document}